\documentclass[aps,pra]{revtex4}
\usepackage{dcolumn}
\usepackage{graphicx}

\begin{document}

\title{Multipole (E1, M1, E2, M2, E3, M3) transition wavelengths and rates
 between $3l^{-1}5l'$ excited and ground states  in nickel-like
 ions}

\author{ U. I. Safronova,  A. S. Safronova }
 \affiliation{Physics Department, University of Nevada, Reno, NV
 89557}

\author{ P. Beiersdorfer}
 \affiliation{Lawrence Livermore National Laboratory, Livermore, CA
 94550}

\begin{abstract}
A relativistic many-body method is developed to calculate energy
and transition rates for multipole transitions in many-electron
ions. This method is based on relativistic many-body perturbation
theory (RMBPT), agrees with MCDF calculations in lowest-order,
includes all second-order correlation corrections and includes
corrections from negative energy states. Reduced matrix elements,
oscillator strengths, and transition rates are calculated for
electric-multipole (dipole (E1), quadrupole (E2), and octupole
(E3))  and magnetic-multipole (dipole (M1), quadrupole (M2), and
octupole (M3)) transitions between $3l^{-1}5l'$ excited and ground
states in Ni-like ions with nuclear charges ranging from $Z$ = 30
to 100. The calculations start from a
$1s^22s^22p^63s^23p^63d^{10}$ Dirac-Fock potential. First-order
perturbation theory is used to obtain intermediate-coupling
coefficients, and second-order RMBPT is used to determine the
matrix elements. A detailed discussion of the various
contributions to the dipole matrix elements and energy levels is
given for nickellike tungsten ($Z$ = 74). The contributions from
negative-energy states are included in the second-order E1, M1, E2
 M2, E3, and M3 matrix elements. The resulting transition energies and
transition rates are compared with experimental values and with
results from other recent calculations. These atomic data are
important in modeling of M-shell radiation spectra of heavy ions
generated in electron beam ion trap experiments and in M-shell
diagnostics of plasmas.

\end{abstract}


\maketitle

\section{Introduction}

The Ni-isoelectronic sequence has been studied extensively  in
connection with x-ray lasers
~\cite{pre-05a,prl-05a,pra-04a,pra-03a,pr-2003a,jpb-5f-93,scofield,chen,li,japan,nilsen}.
Recently, an investigation into the use of atomic databases in
simulation of Ni-like gadolinium x-ray laser was presented by King
{\it et al.\/} in Ref.~\cite{gd-04}. Measurements of $3d-5f$ and
$3d-6f$ transition energies in Ni-like ions (Ag$^{19+}$,
Sn$^{22+}$, Pr$^{31+}$, Gd$^{36+}$, Yb$^{42+}$, Ta$^{45+}$,
Ir$^{49+}$, Th$^{62+}$, and , U$^{64+}$) were reported by Elliott
{\it et al.\/} in Ref.~\cite{elliot}. Additional measurements of
the $3d_{5/2} - 6f_{7/2}$ transition energies in Ni-like
Tm$^{41+}$, Hf$^{44+}$, Re$^{47+}$, Pb$^{54+}$, and Th$^{62+}$
were carried out by Beiersdorfer in Ref.~\cite{ad-13b}.
 Recently, the x-ray spectral
measurements of the line emission of $n$ = 3--4, 3--5, 3--6, and
3--7  transitions in  Ni- to Kr-like Au ions in electron beam trap
(EBIT) plasma were reported by May {\it et al.\/} in
Ref.~\cite{au-03}. X-ray spectra of Ni-like W including 3-4, 5,
and 6 transitions recorded by a broadband microcalorimeter, were
analyzed in Ref.~\cite{alla,ad-15a}.
 A detailed analysis of 3-4
and 3-5 transitions in the x-ray spectrum by laser produced
plasmas of Ni-like highly-charged ions
 was presented by Doron
{\it et al.\/} \cite{doron} (Ba$^{28+}$), by  Zigler {\it et
al.\/} \cite{zigler} (La$^{29+}$ and Pr$^{31+}$), by Doron {\it et
al.\/} \cite{doron-58} (Ce$^{30+}$). Studies of Ni-like ions
(Gd$^{36+}$, W$^{46+}$ have also been carried out on tokamaks
~\cite{ad-18a,ad-18b}.

Various computer codes were employed to calculate transitions in
Ni-like ions. In particular, ab-initio calculations were performed
in Ref.~\cite{doron} using the {\sc relac} relativistic computer
code to identify $3d-nf$ ($n$=4 to 8) transitions of Ni-like Ba.
Atomic structure calculations for highly ionized tungsten (Co-like
W$^{47+}$ to Rb-like W$^{37+}$) were done by Fournier
\cite{fournier} with using the graphical angular momentum coupling
code ANGULAR and the fully relativistic parametric potential code
RELAC. The Hebrew University Lawrence Livermore Atomic Code {\sc
hullac}  is also based on a relativistic model potential
\cite{klapisch}. Ab-initio calculations with the {\sc hullac}
relativistic code was used for detailed analysis of spectral lines
by by  Zigler {\it et al.\/} \cite{zigler} and by May {\it et
al.\/} in Ref.~\cite{au-03}. Zhang {\it et al.\/} \cite{zhang},
using the Dirac-Fock-Slater (DFS) code evaluated excitation
energies and oscillator strengths of 3-4 and 3-5 transitions for
the 33 Ni-like ions with 60$\leq Z \geq$92. The multiconfiguration
Dirac=Fock calculations of the $3d_{3/2}-5f_{5/2}$,
$3d_{5/2}-5f_{7/2}$, $3d_{3/2}-6f_{5/2}$, and $3d_{5/2}-6f_{7/2}$
transitions were reported by Elliot {\it et al.\/} in
Ref.~\cite{elliot}.
 The wavelengths and transition rates
for $3l-nl'$ electric-dipole transitions in Ni-like xenon are
presented by Skobelev {\it et al.\/} in Ref.~\cite{skobelev}.
Results were obtained by three methods: the relativistic
Hartree-Fock (HFR) self-consistent-field method (Cowan code),
multiconfiguration Dirac=-Fock (MCDH) method (Grant code), and the
{\sc hullac} code. The contribution of lots of weak correlation on
transition wavelengths and probabilities  by including partly
single and double excitation from the $3l$ inner-shells into the
$4l$ and $5l$ orbital layers of highly-charged Ni-like ions were
discussed by Dong {\it et al.\/} in Ref.~\cite{dong-03}. Energy
levels, transition probabilities, and electron impact excitation
for possible x-ray line emissions of Ni-like tantalum ions were
recently calculated by Zhong {\it et al.\/} in Ref.~\cite{ta-05}.

   The relative magnitudes of
the electric-multipole (E1, E2, E3) and magnetic-multipole (M1,
M2, M3) radiative decay rates calculated by the MCDF approach,
were presented by Bi\'{e}mont \cite{biemont} for~ lowest 17 levels
of highly ionized nickel-like ions. Observation of
electric-quadrupole (E2) and magnetic-octupole (M3) decay in the
x-ray spectrum of highly charged Ni-like ions (Th$^{62+}$ and
U$^{64+}$) were reported by Beiersdorfer {\it et al.\/} in
Ref.~\cite{beier}. The lowest excited level in Ni-like ions,
$3d^94d\ ^3D_3$, decays only via an M3 decay. The radiative decay
of this line was measured recently in Xe$^{26+}$, Cs$^{27+}$, and
XBa$^{28+}$ by Tr{\"{a}}bert {\it et al.\/}
~\cite{trabert,ad-27a}. Calculated values of transitions
wavelengths and rates for Ni-like ions with 30$\leq Z \leq$100
presented in Ref.~\cite{trabert} were obtained by using
Relativistic Many-Body Perturbation Theory (RMBPT) method. Reduced
matrix elements, oscillator strengths, and transition rates into
the ground state for all allowed and forbidden electric- and
magnetic-dipole and electric- and magnetic-quadrupole transitions
(E1, M1, E2, M2) in Ni-like ions were presented by Hamasha {\it et
al.\/} in Ref.~\cite{ni-cjp}. Relativistic many-body calculations
of multipole (E1, M1, E2, M2, E3, M3) transition wavelengths and
rates between $3l^{-1}4l'$ excited and ground states  in
nickel-like ions were recently reported by Safronova {\it et
al.\/} in Ref.~\cite{ni-adndt}.

In the present paper, RMBPT is used for systematic study of atomic
characteristics of transitions in Ni=like ions in a broad range of
the nuclear charge $Z$ = 3--100. Specifically, we determine
energies of $3s^23p^63d^{9}5l(J)$, $3s^23p^53d^{10}5l(J)$, and
$3s3p^63d^{10}5l(J)$ states of Ni-like ions with nuclear charges
$Z$=30-100.  The calculations are carried out to second order in
perturbation theory.  We consider all possible $3l$ holes and $5l$
particles leading to
 the 67 odd-parity   $3d^{-1}5p(J)$, $3d^{-1}5f(J)$,
$3p^{-1}5s(J)$, $3p^{-1}5d(J)$, $3p^{-1}5g(J)$, $3s^{-1}5p(J)$,
and $3s^{-1}5f(J)$ excited states and  the 74 even-parity
$3d^{-1}5s(J)$, $3d^{-1}5d(J)$, $3d^{-1}5g(J)$, $3p^{-1}5p(J)$,
$3p^{-1}5f(J)$, $3s^{-1}5s(J)$, $3s^{-1}5d(J)$ and $3s^{-1}5g(J)$
excited states in  Ni-like ions with $Z$=30 to 100.

RMBPT is also used to determine line strengths, oscillator
strengths, and transition rates  for all  allowed and forbidden
electric-multipole and magnetic-multipole (E1, E2, E3, M1, M2, M3)
from $3s^23p^63d^{9}5l(J)$, $3s^23p^53d^{10}5l(J)$, and
$3s3p^63d^{10}5l(J)$ excited states into the ground state in
Ni-like ions. Retarded E1, E2, and E3 matrix elements are
evaluated in both length and velocity forms. A detailed discussion
of the various contributions to the dipole matrix elements and
energy levels is given for nickellike tungsten ($Z$ = 74), which
plays an important role in tokamaks ~\cite{ad-18a,ad-18b},
electron beam ion traps ~\cite{ad-15a,ad-29a,ad-29ab}, x-ray
lasers ~\cite{ad-29b}, and z-pinches ~\cite{alla}.

\begin{figure}[tbp]
\centerline{ \includegraphics[scale=0.35]{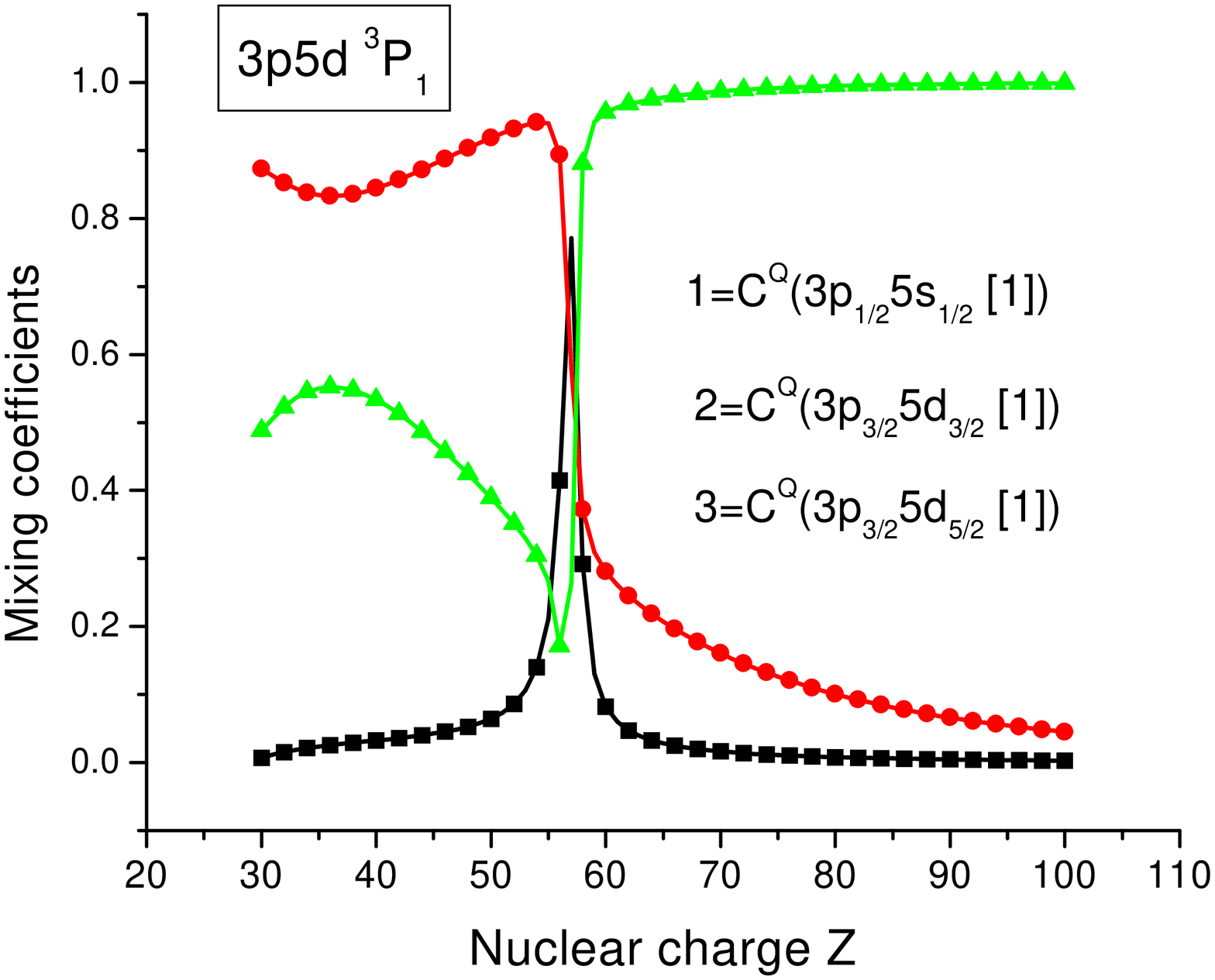}
             \includegraphics[scale=0.35]{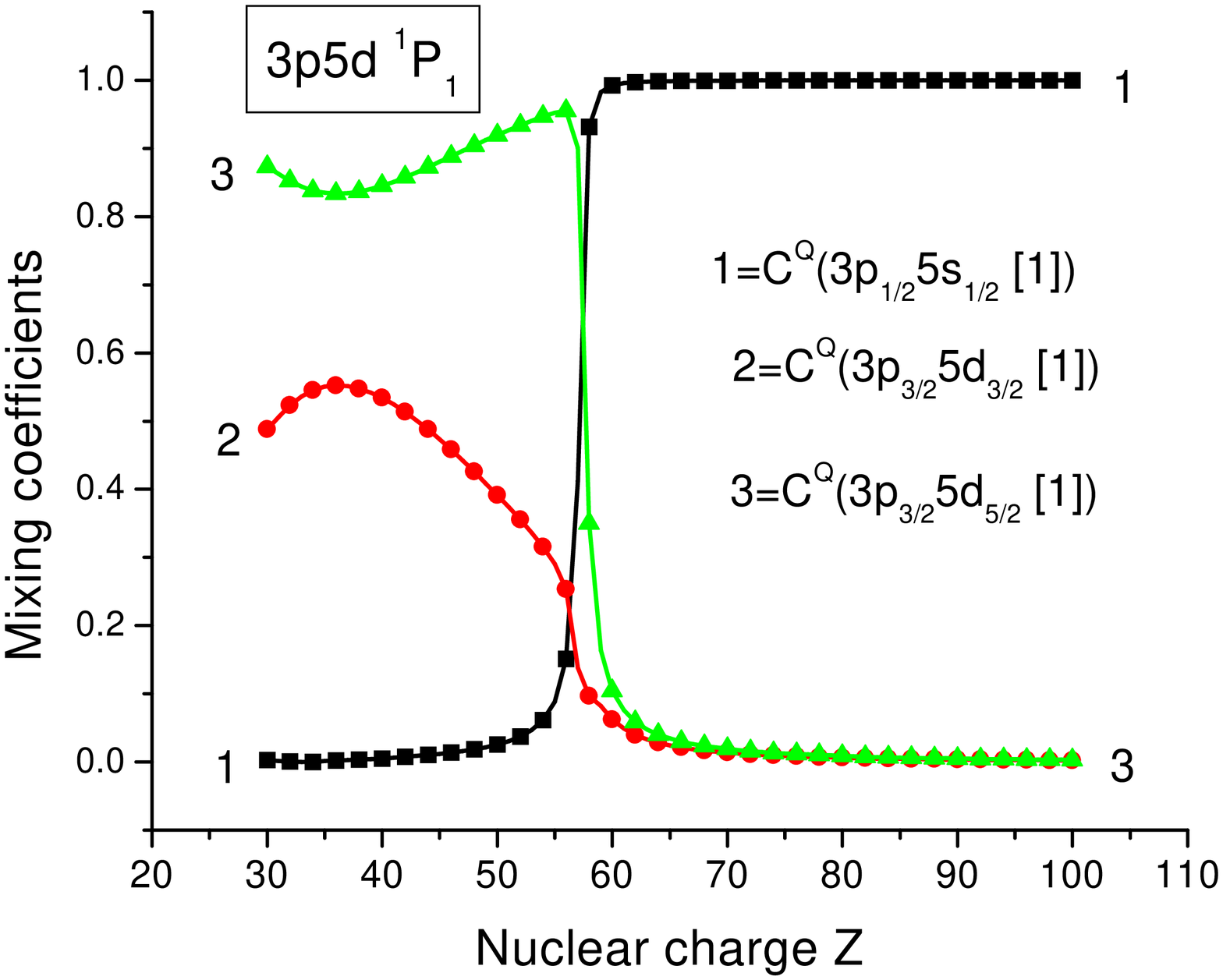}}
\caption{Mixing coefficients for the $3p5d\ ^{1,3}P_1$ levels as
functions of $Z$} \label{fig-ppf}
\end{figure}

\begin{figure}[tbp]
\centerline{\includegraphics[scale=0.35]{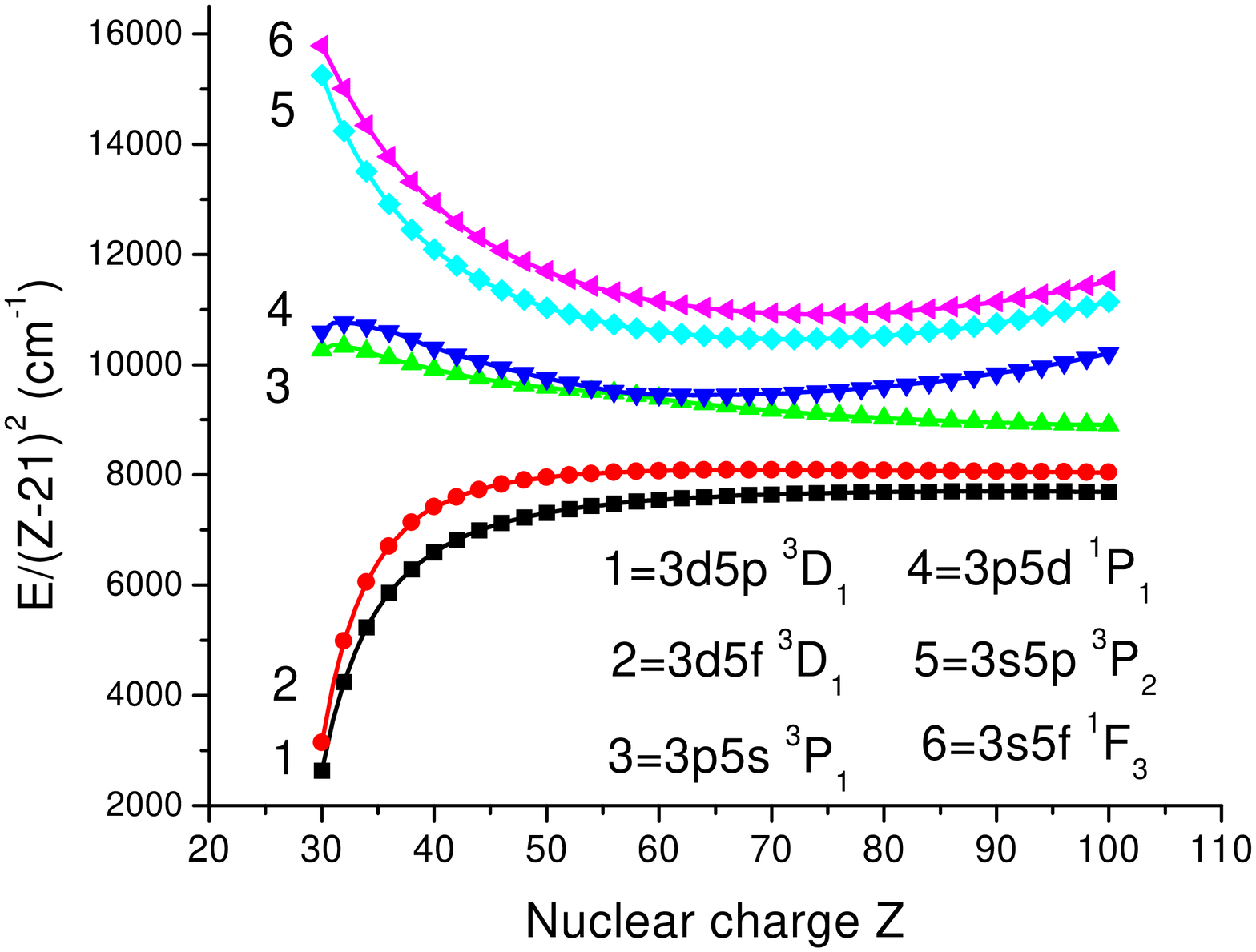}
            \includegraphics[scale=0.35]{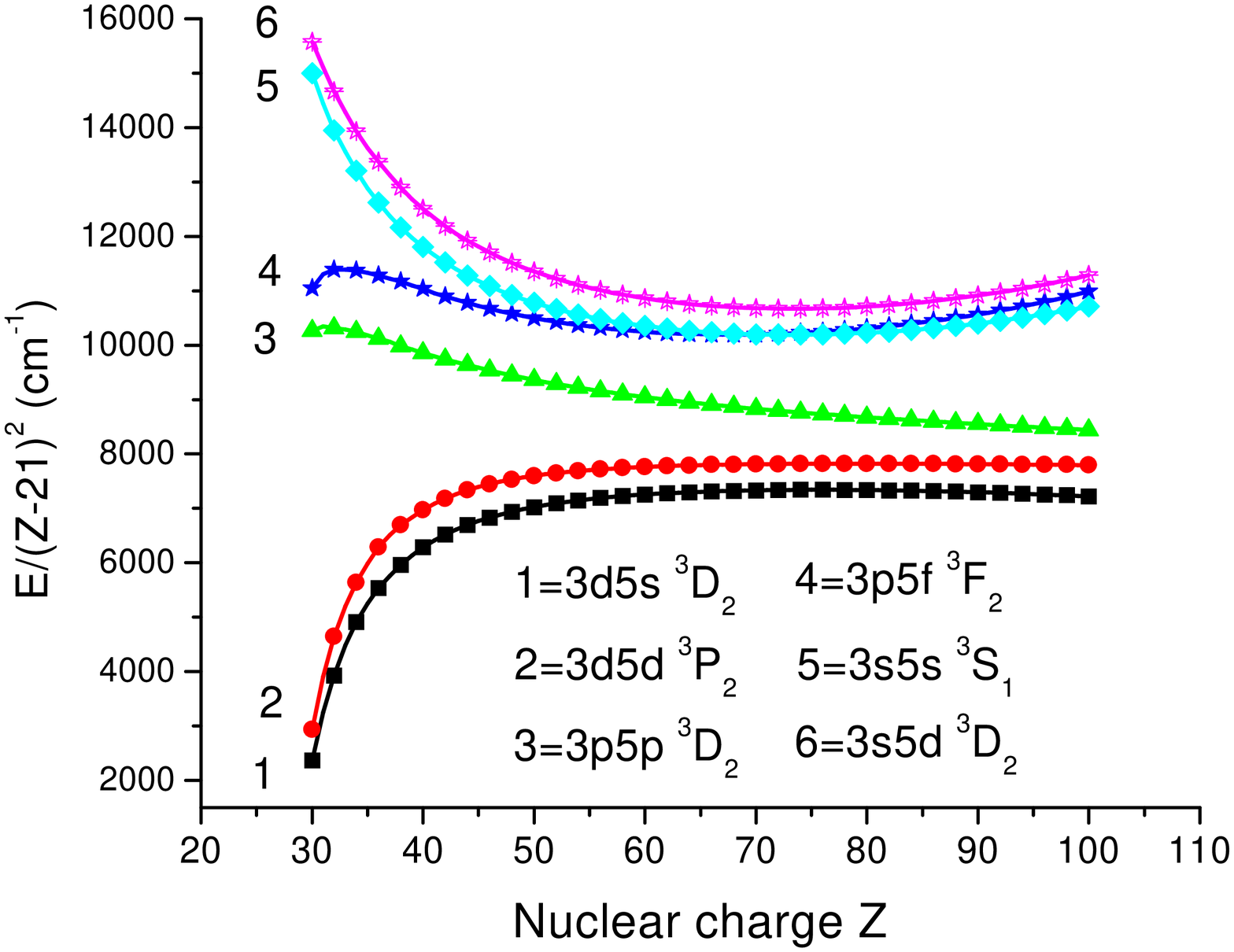}}
            \caption{
Energies ($E/(Z-21)^2$) in~ cm$^{-1}$ for  odd- and even-parity
states  in Ni-like ions as functions of $Z$} \label{fig-en}
\end{figure}

\begin{figure}[tbp]
\centerline{\includegraphics[scale=0.35]{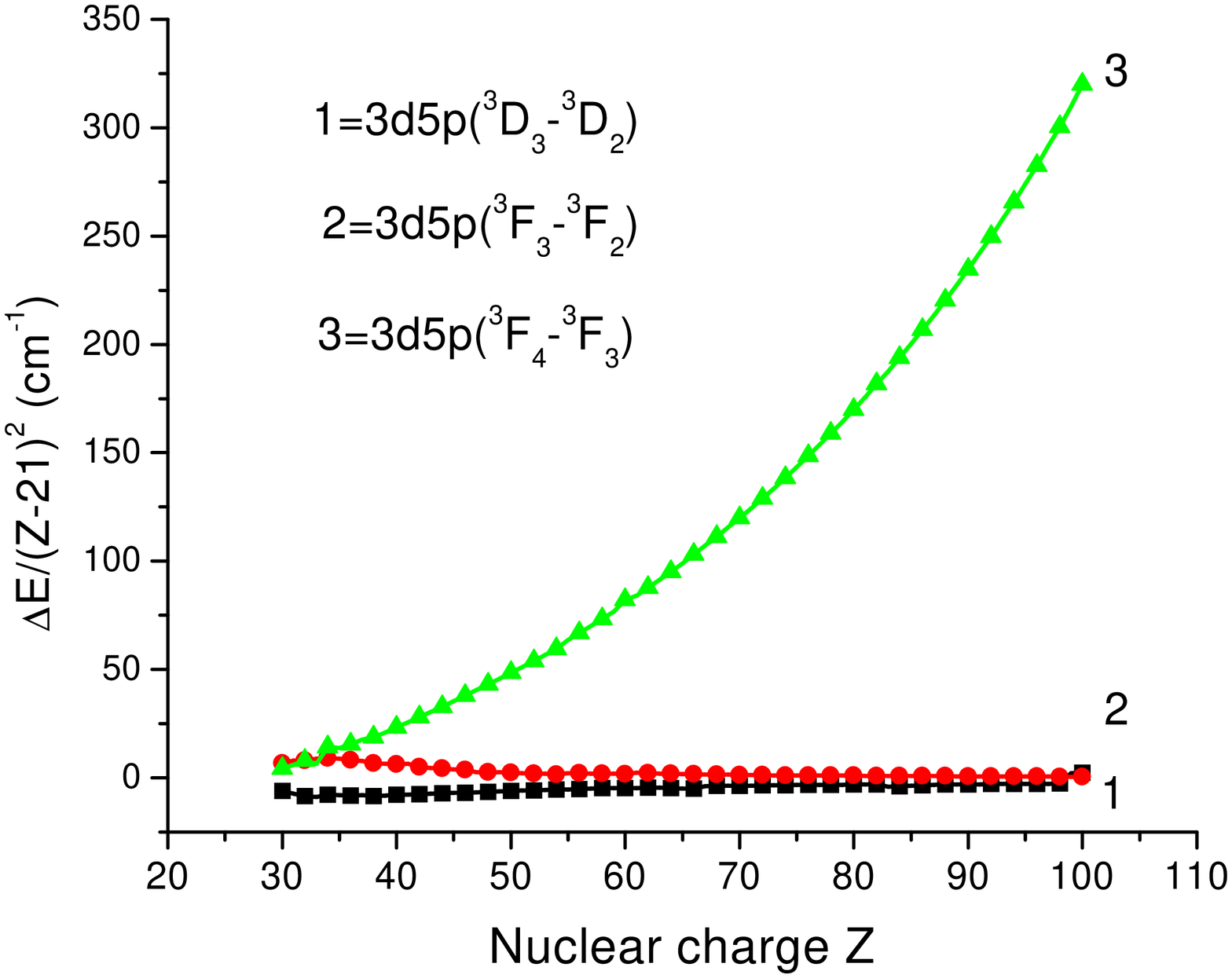}
            \includegraphics[scale=0.35]{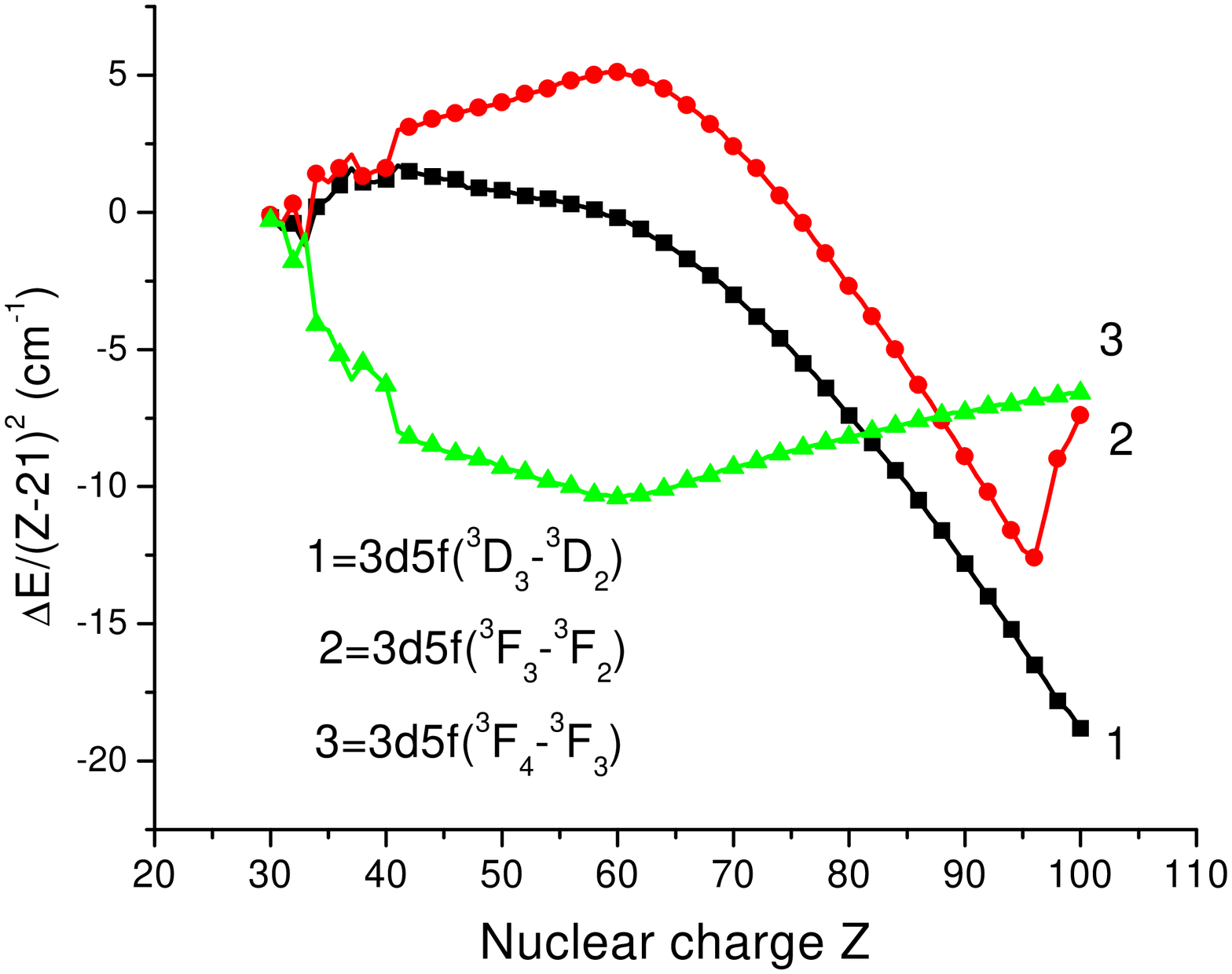}}
 \centerline{\includegraphics[scale=0.35]{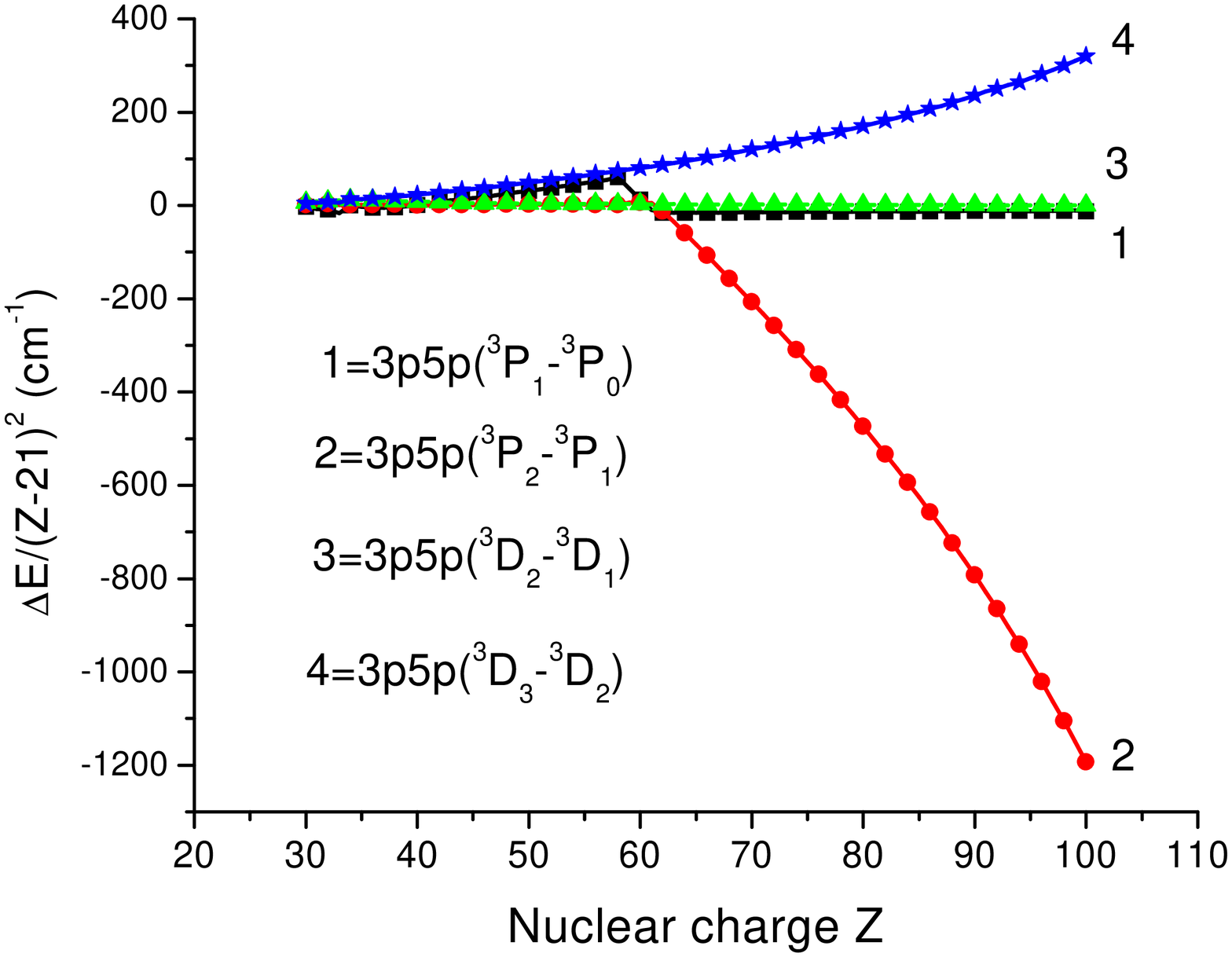}
            \includegraphics[scale=0.35]{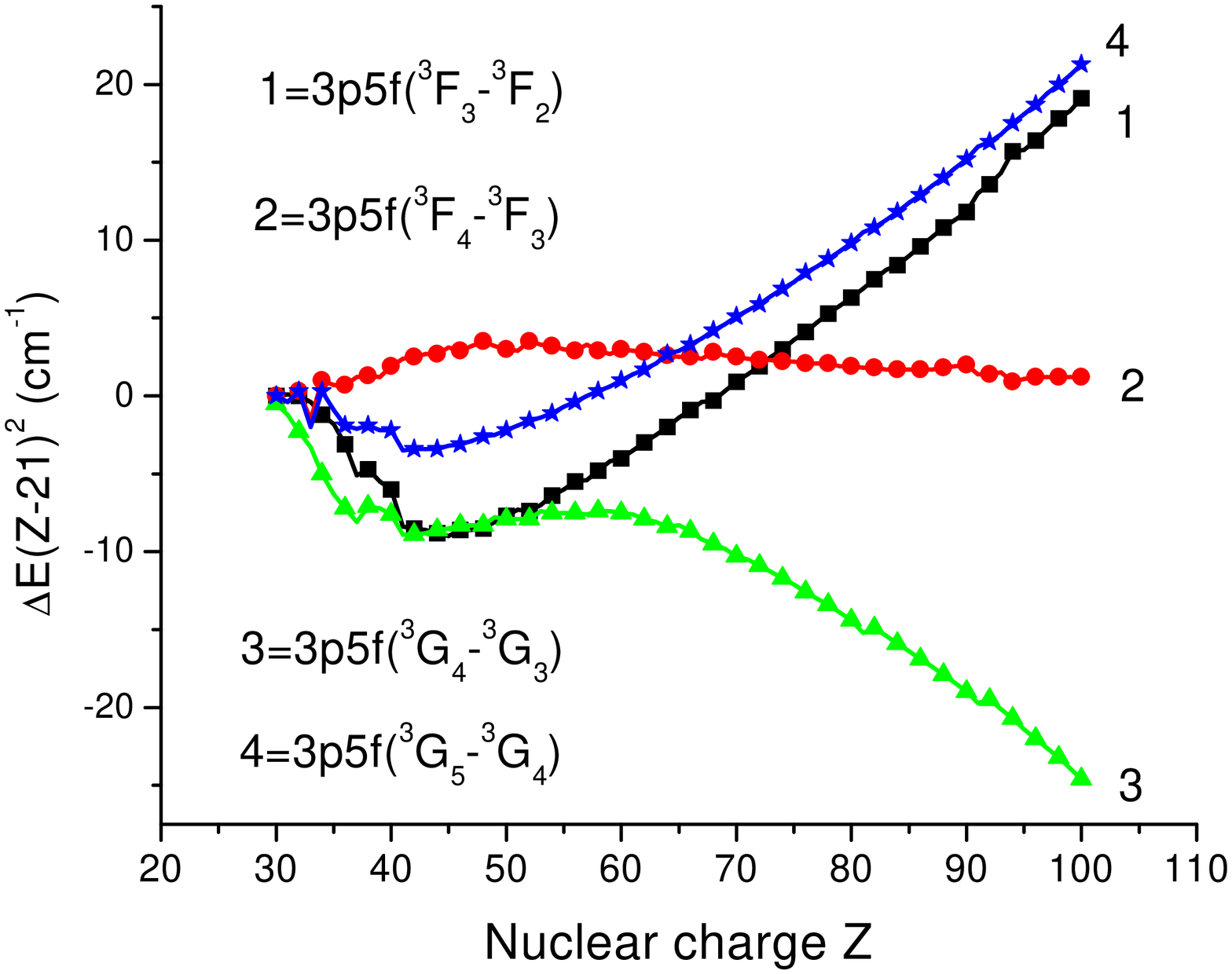}}
\caption{ Energy splitting  ($\Delta E/(Z-21)^2$) in~ cm$^{-1}$
for terms of odd- and even-parity states with  in Ni-like ions as
function of $Z$} \label{fig-del1}
\end{figure}

\begin{figure}[tbp]
\centerline{\includegraphics[scale=0.35]{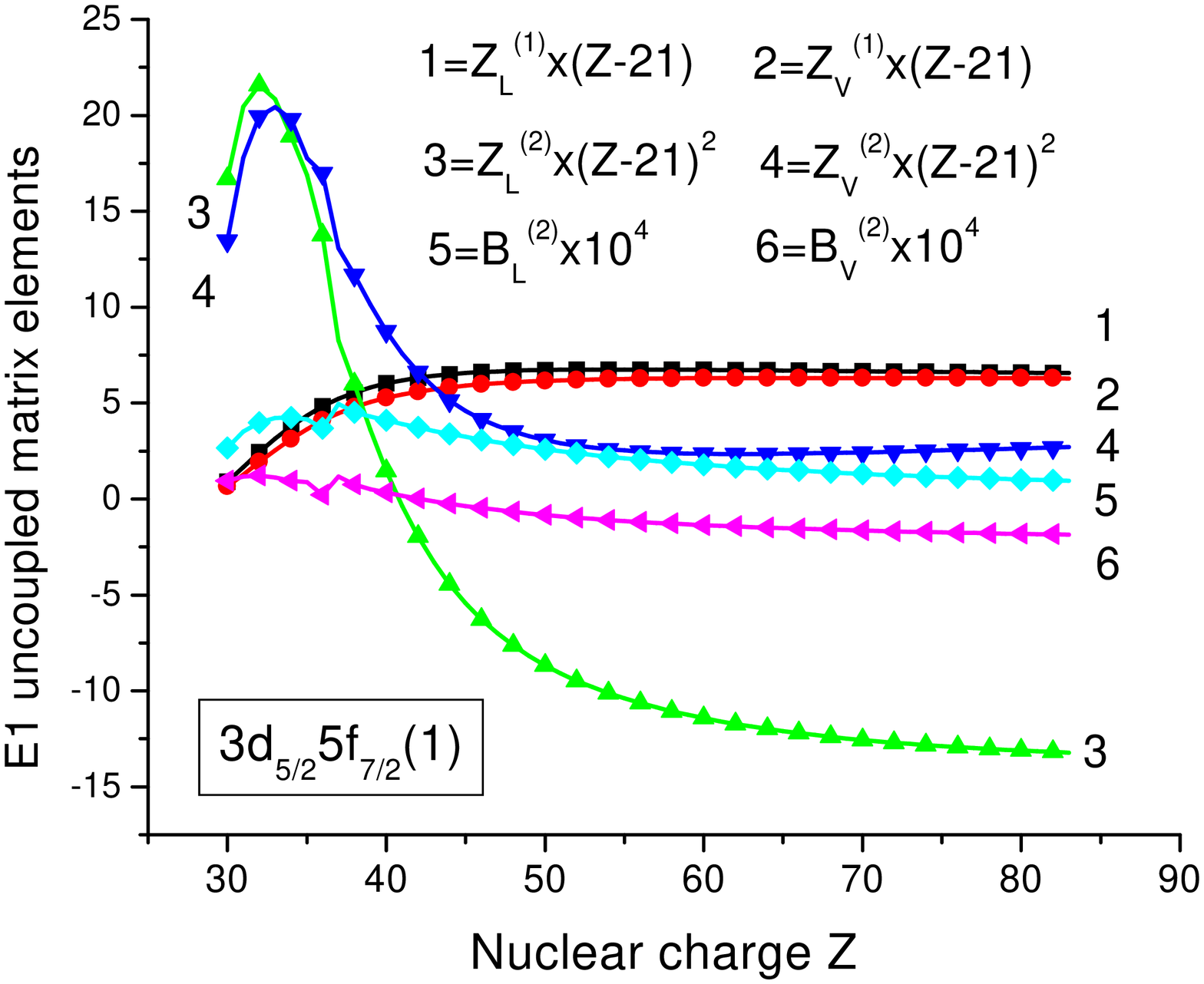}
            \includegraphics[scale=0.35]{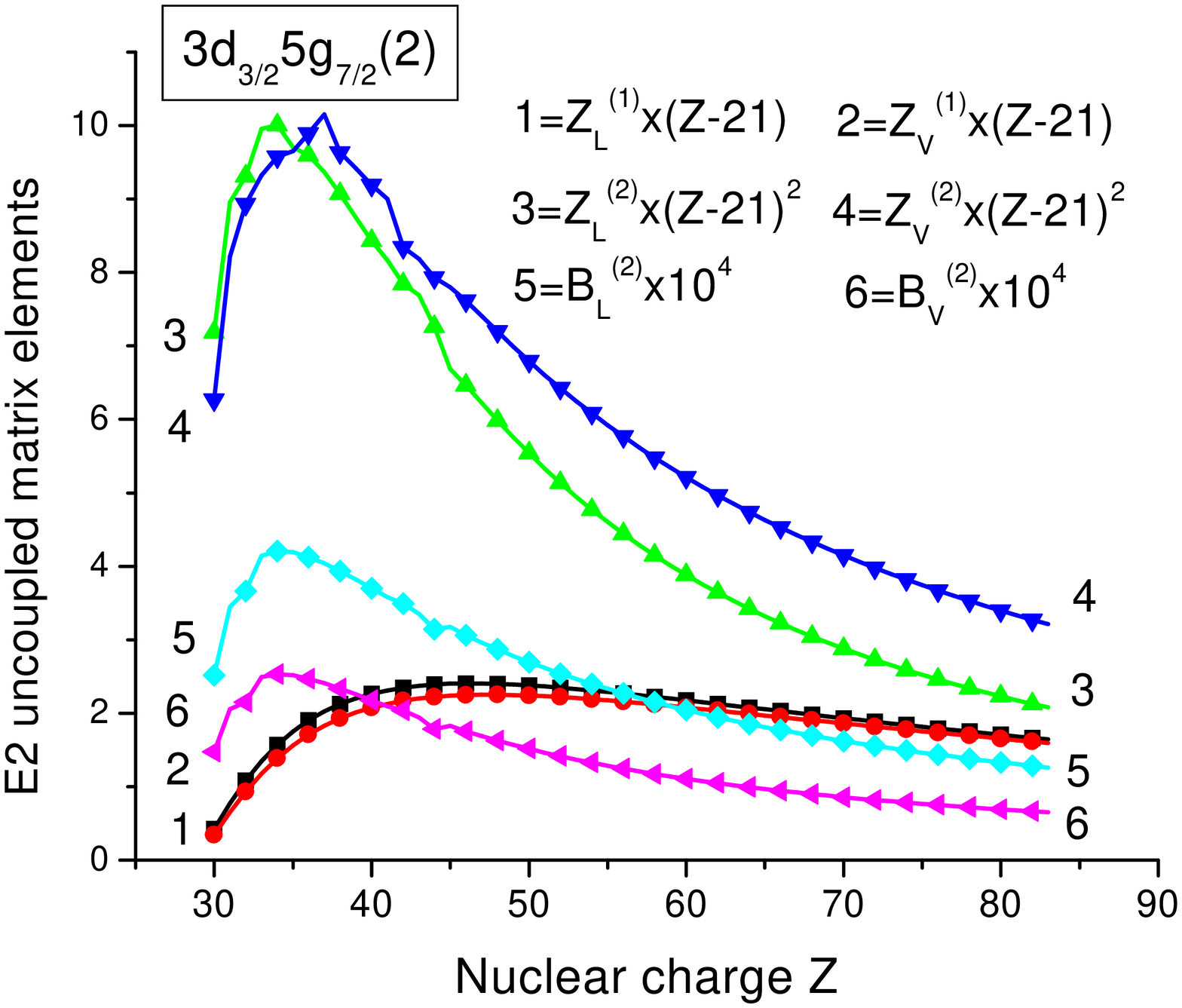}}
\centerline{\includegraphics[scale=0.35]{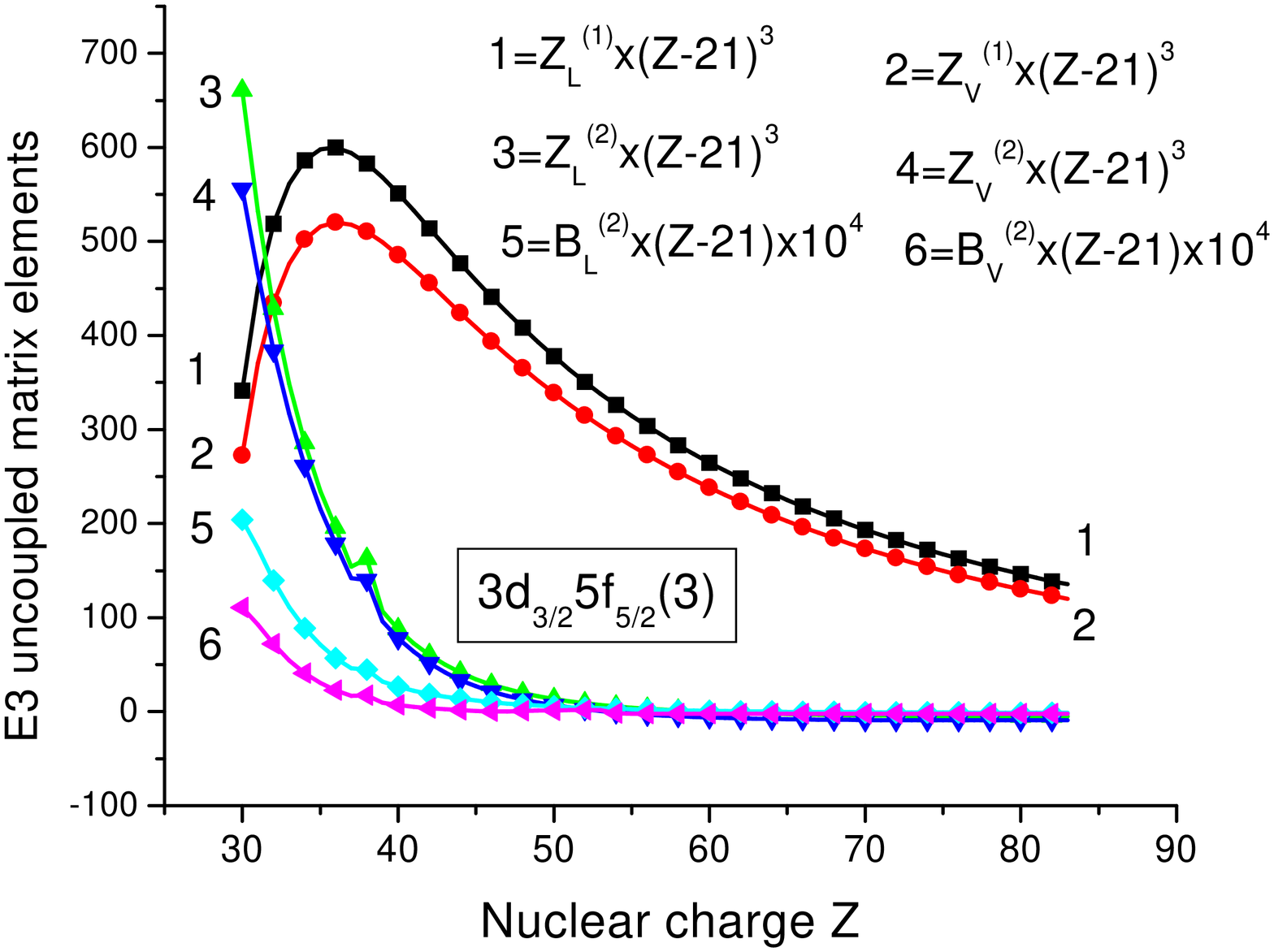}} \caption{The
first- and second-order Coulomb corrections ($Z^{(1)}$,
$Z^{(2)}$), and second-order Breit-Coulomb corrections ($B^{(2)}$)
for E1,  E2, and E3  uncoupled matrix elements for transitions
between excited and  ground states calculated in length ($L$)and
velocity ($V$) forms in Ni-like ions.  } \label{e1-uncop}
\end{figure}

\begin{figure}[tbp]
\centerline{\includegraphics[scale=0.35]{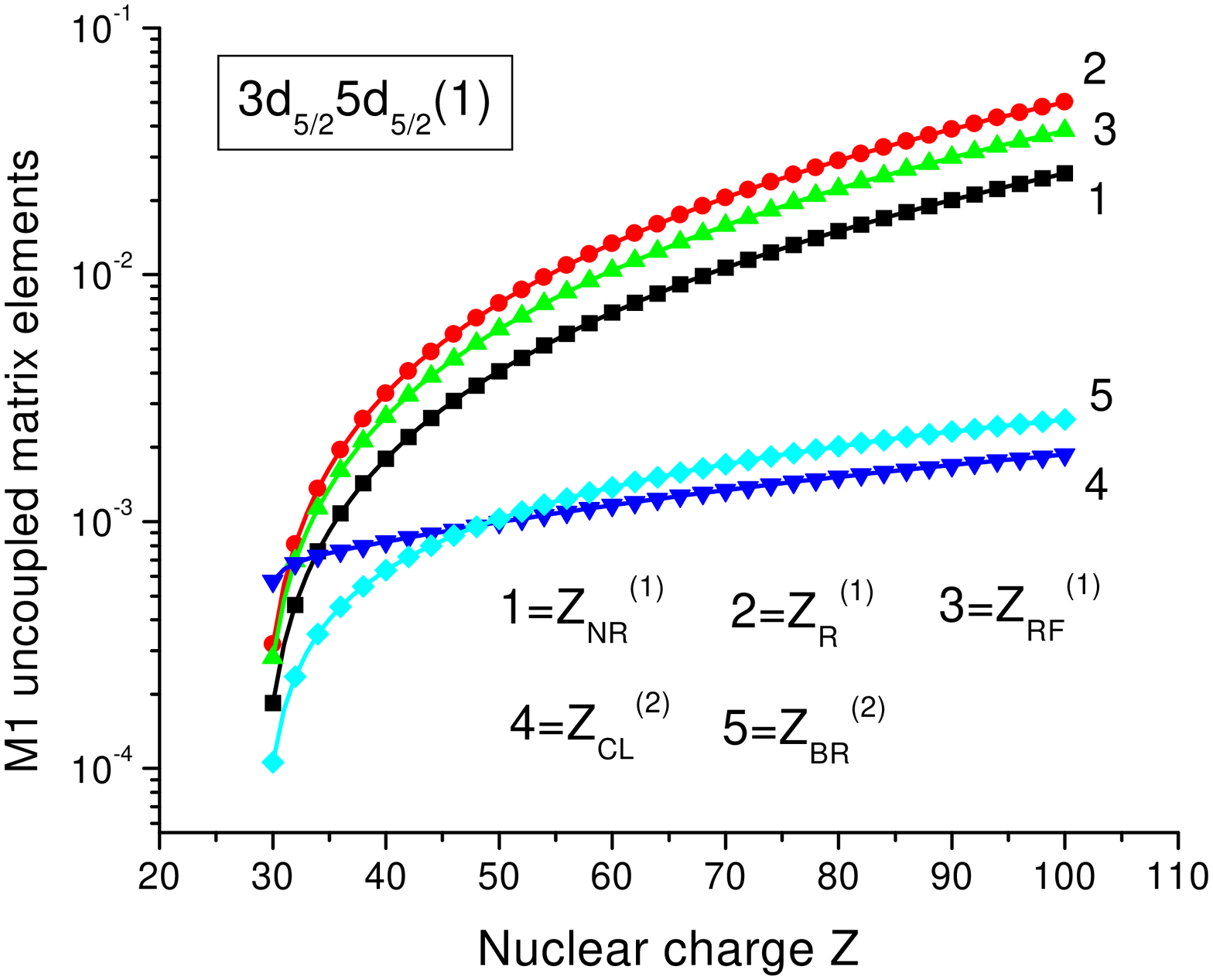}
            \includegraphics[scale=0.35]{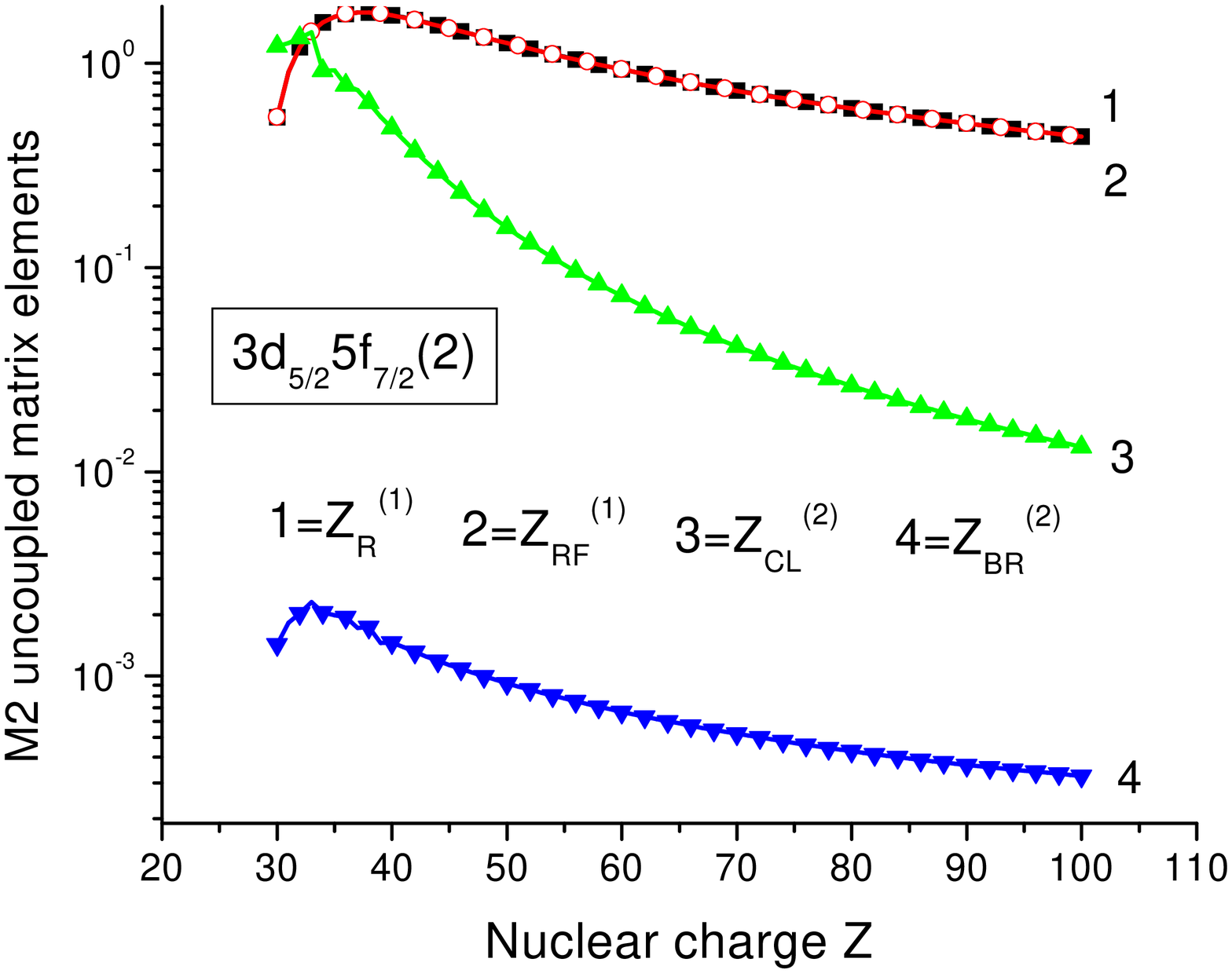}}
\centerline{\includegraphics[scale=0.35]{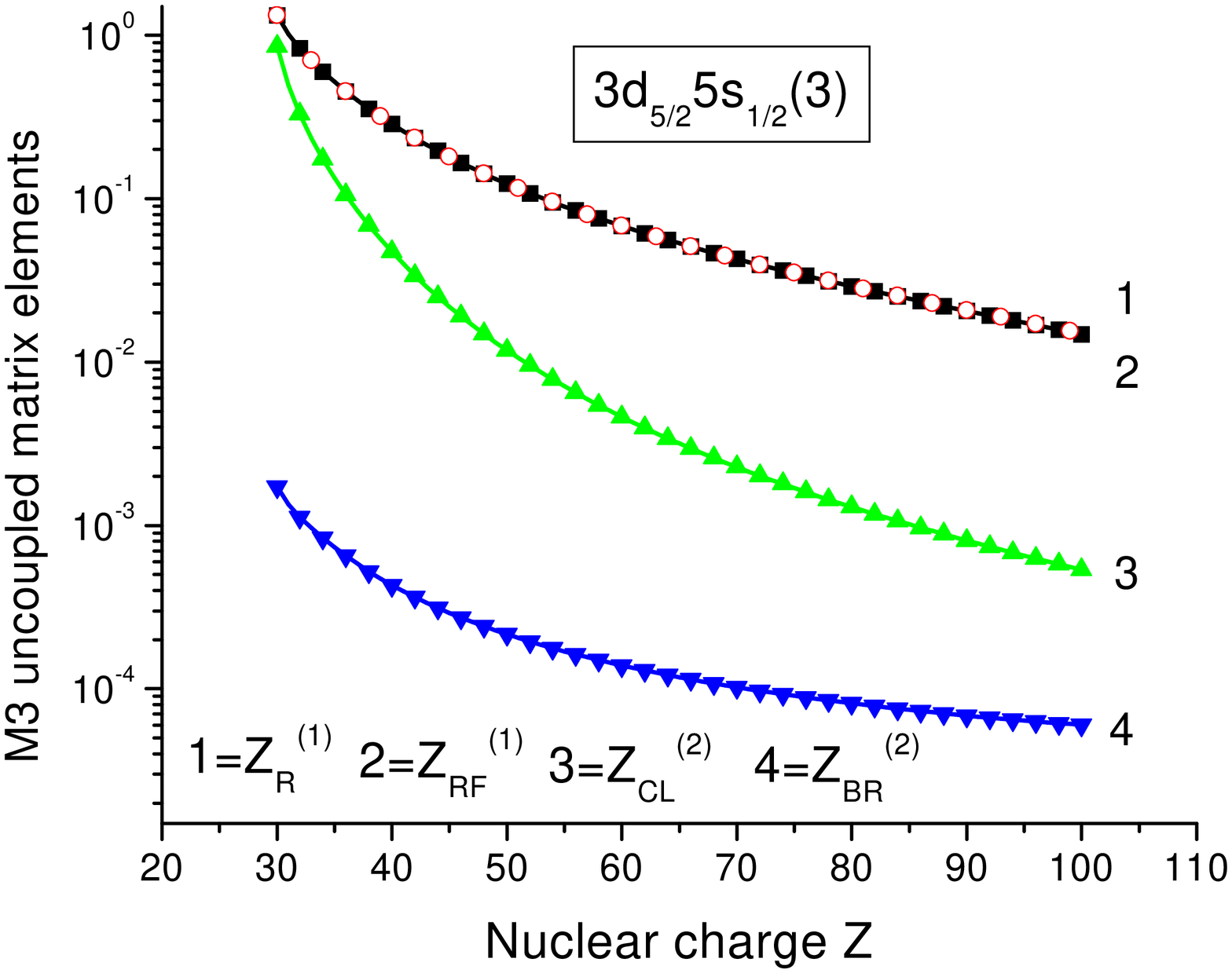}} \caption{The
first- and second-order  corrections ($Z^{(1)}$, $Z^{(2)}$) for
M1, M2 and M3 uncoupled matrix elements for transition between
excited and  ground states in Ni-like ions. The first-order
($Z^{(1)}$) matrix elements calculated in nonrelativistic
($Z^{(1)}_{\rm NR}$), relativistic
frequency-independent($Z^{(1)}_{\rm R}$), and relativistic
frequency-dependent ($Z^{(1)}_{\rm RF}$)  approximations  are
presented. The second-order Coulomb  ($Z^{(2)}_{\rm CL}$) and
Breit-Coulomb corrections ($Z^{(2)}_{\rm BR}$)) are compared.}
\label{m1-uncop}
\end{figure}

\begin{figure}[tbp]
\centerline{\includegraphics[scale=0.45]{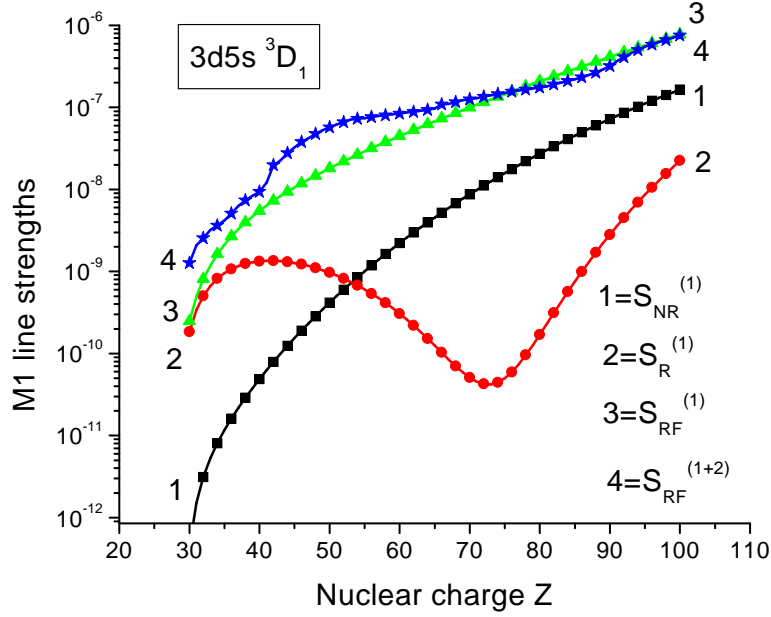}} \caption{ The
total M1 line strengths ($S^{(1+2)}_{\rm RF}$) between the $3d5s\
^1P_1$ and ground states in Ni-like ions as function of $Z$. The
first-order ($S^{(1)}$) line strengths calculated in
nonrelativistic ($S^{(1)}_{\rm NR}$), relativistic
frequency-independent ($S^{(1)}_{\rm R}$), and relativistic
frequency-dependent approximations ($S^{(1)}_{\rm RF}$)  are
presented.} \label{s-m1}
\end{figure}

\begin{figure}[tbp]
\centerline{\includegraphics[scale=0.35]{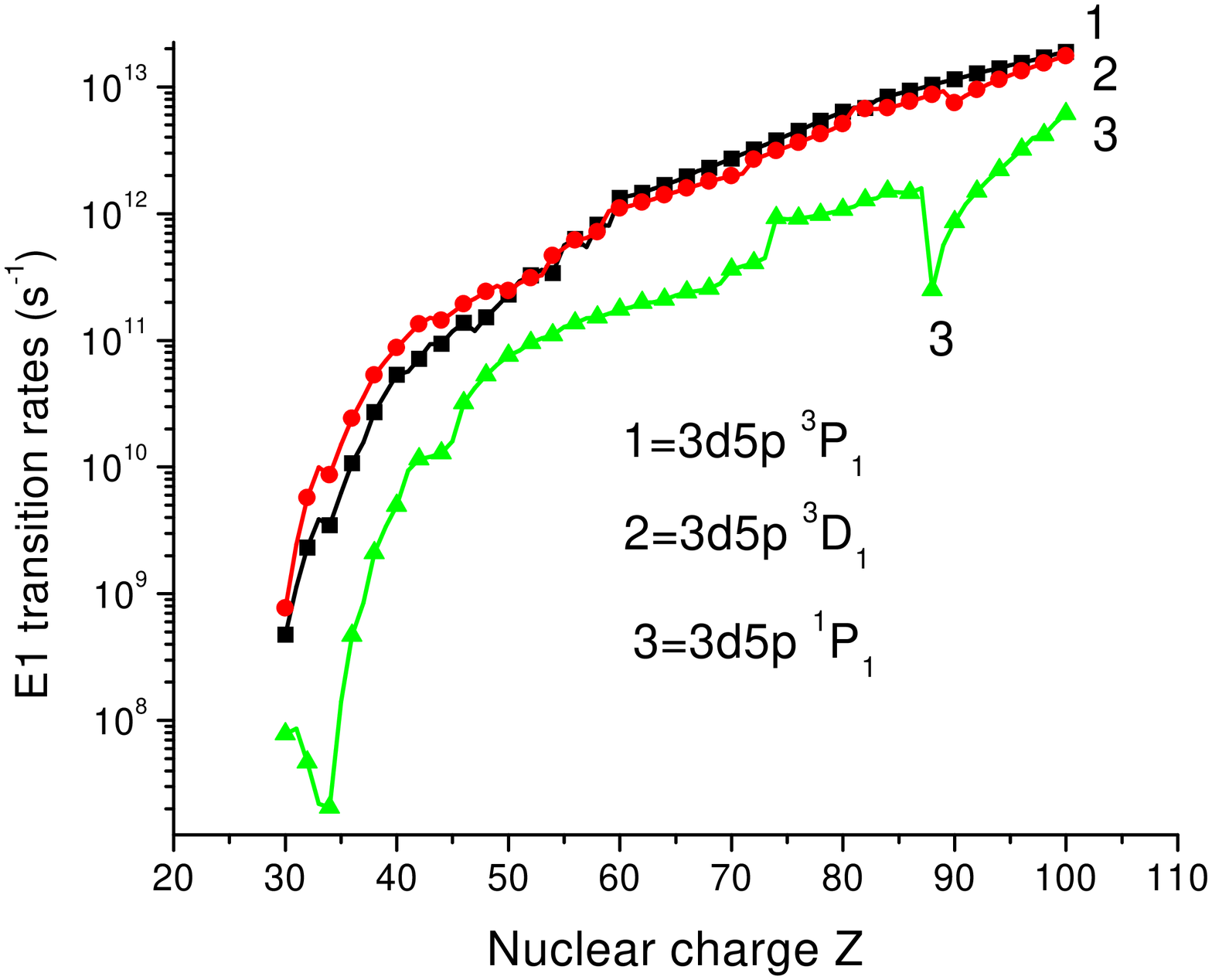}
            \includegraphics[scale=0.35]{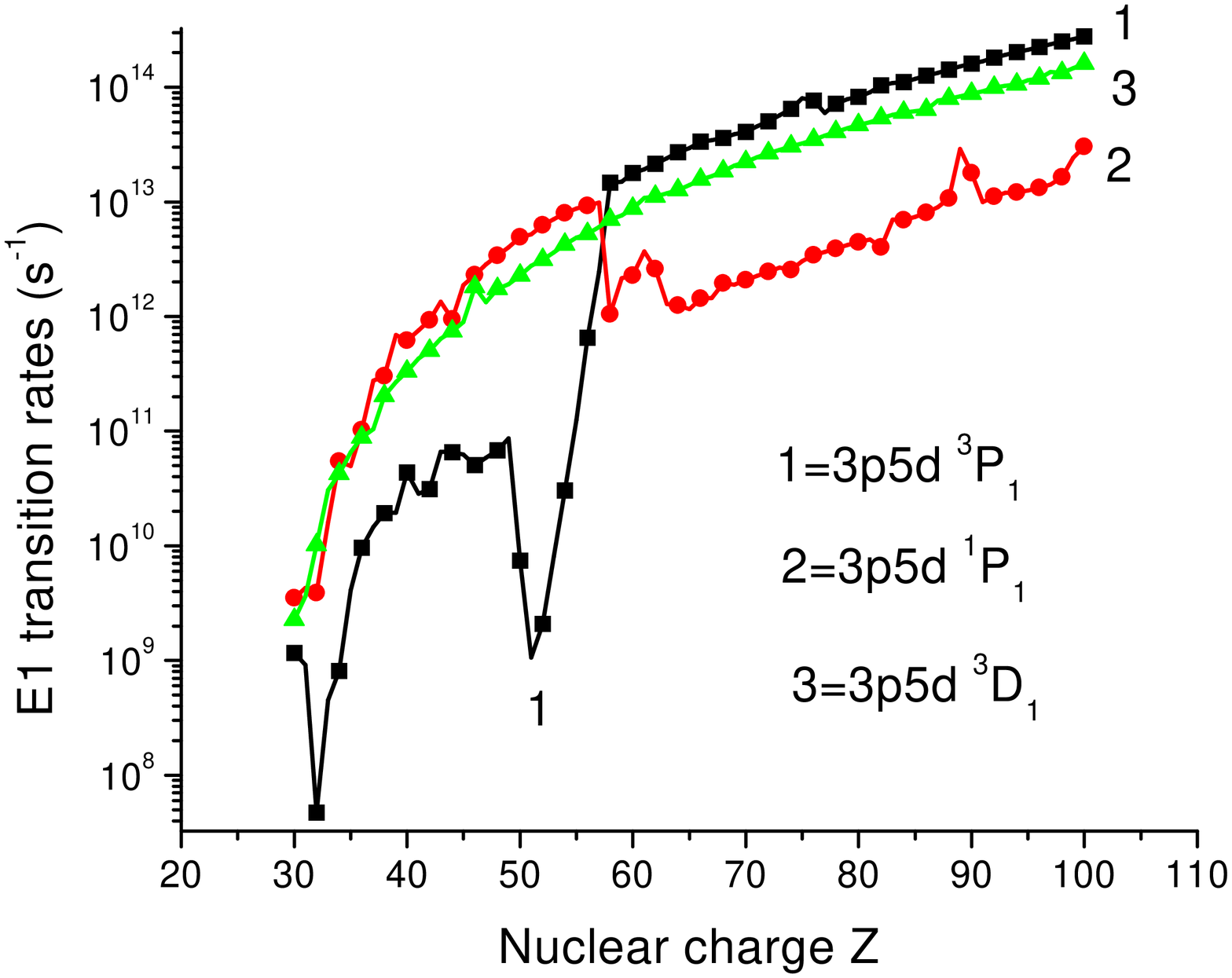}}
\centerline{\includegraphics[scale=0.35]{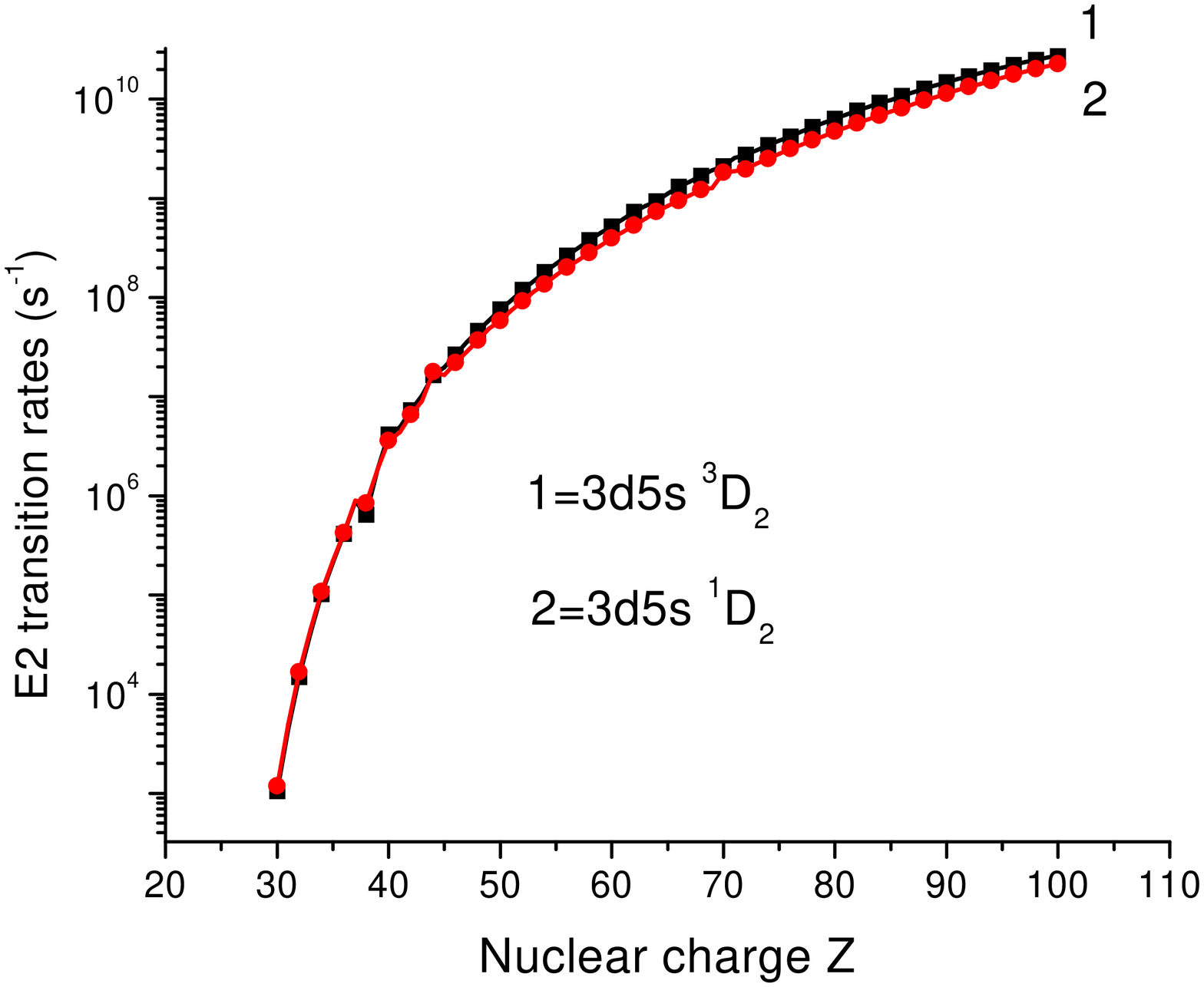}
            \includegraphics[scale=0.35]{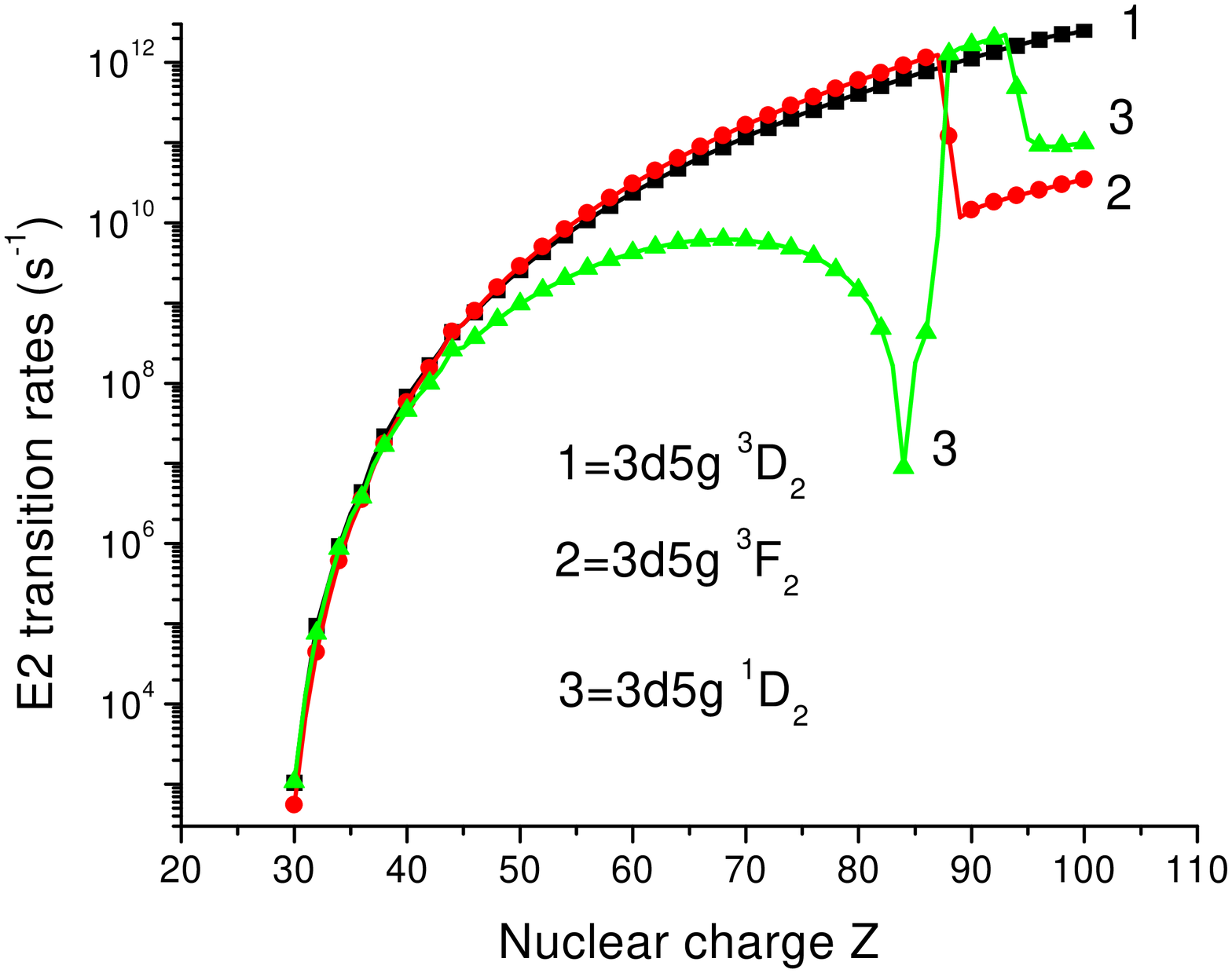}}
\centerline{\includegraphics[scale=0.35]{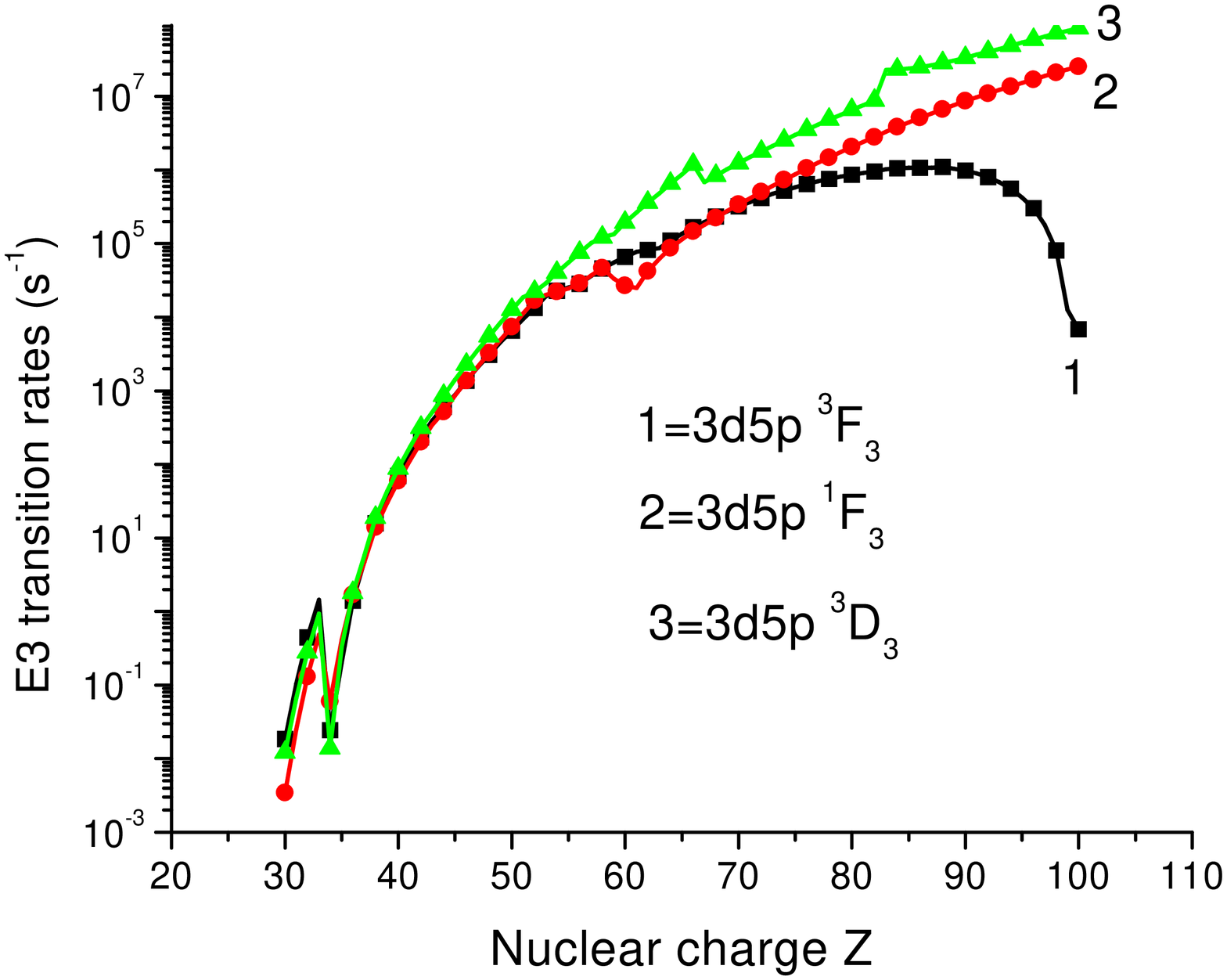}
            \includegraphics[scale=0.35]{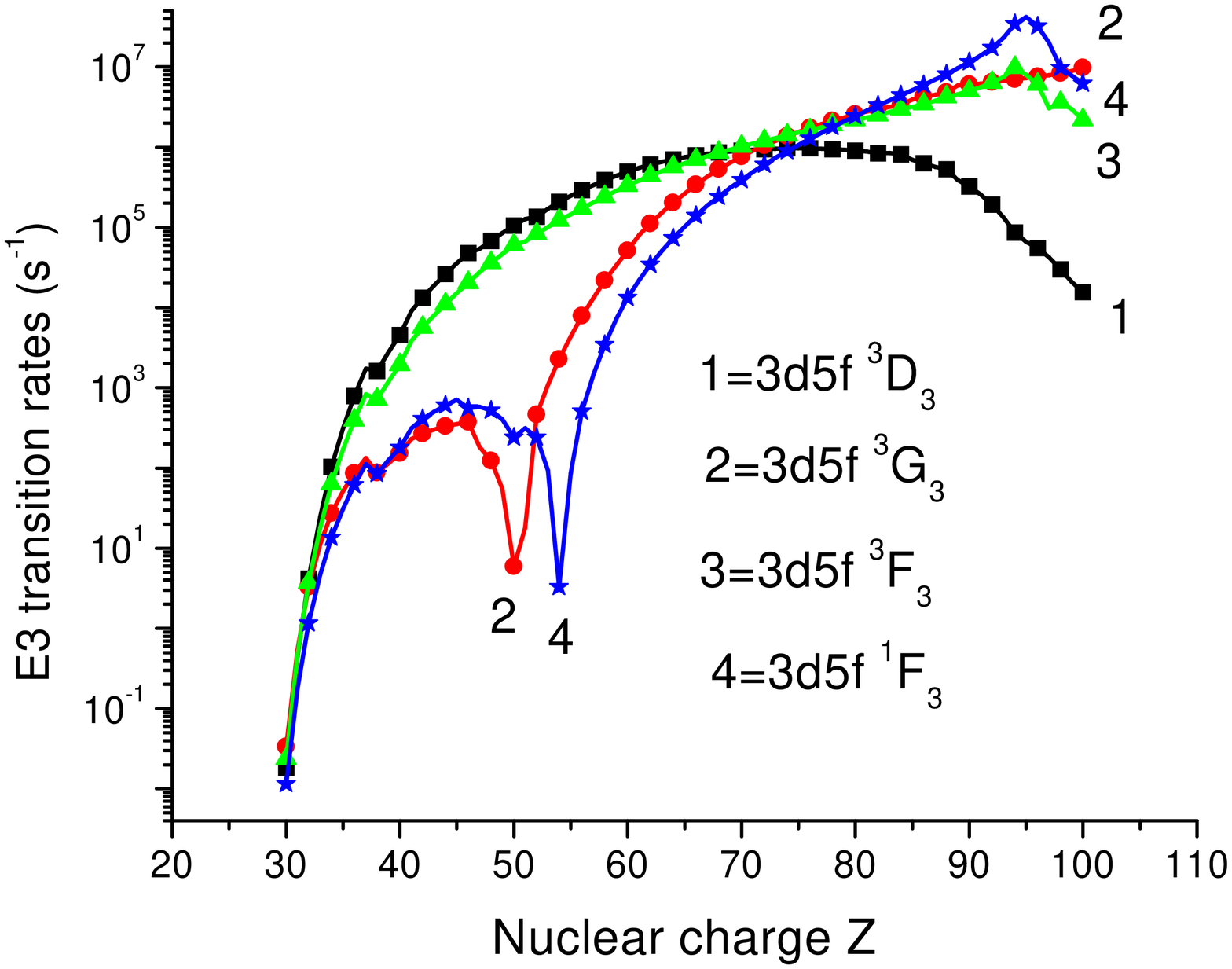}}
            \caption{
E1,E2, and E3 transition rates  for transitions between the
$3l5l'$  states with $J$=1--3 and ground state in Ni-like ions as
function of $Z$} \label{tr-e1}
\end{figure}

\begin{figure}[tbp]
\centerline{\includegraphics[scale=0.35]{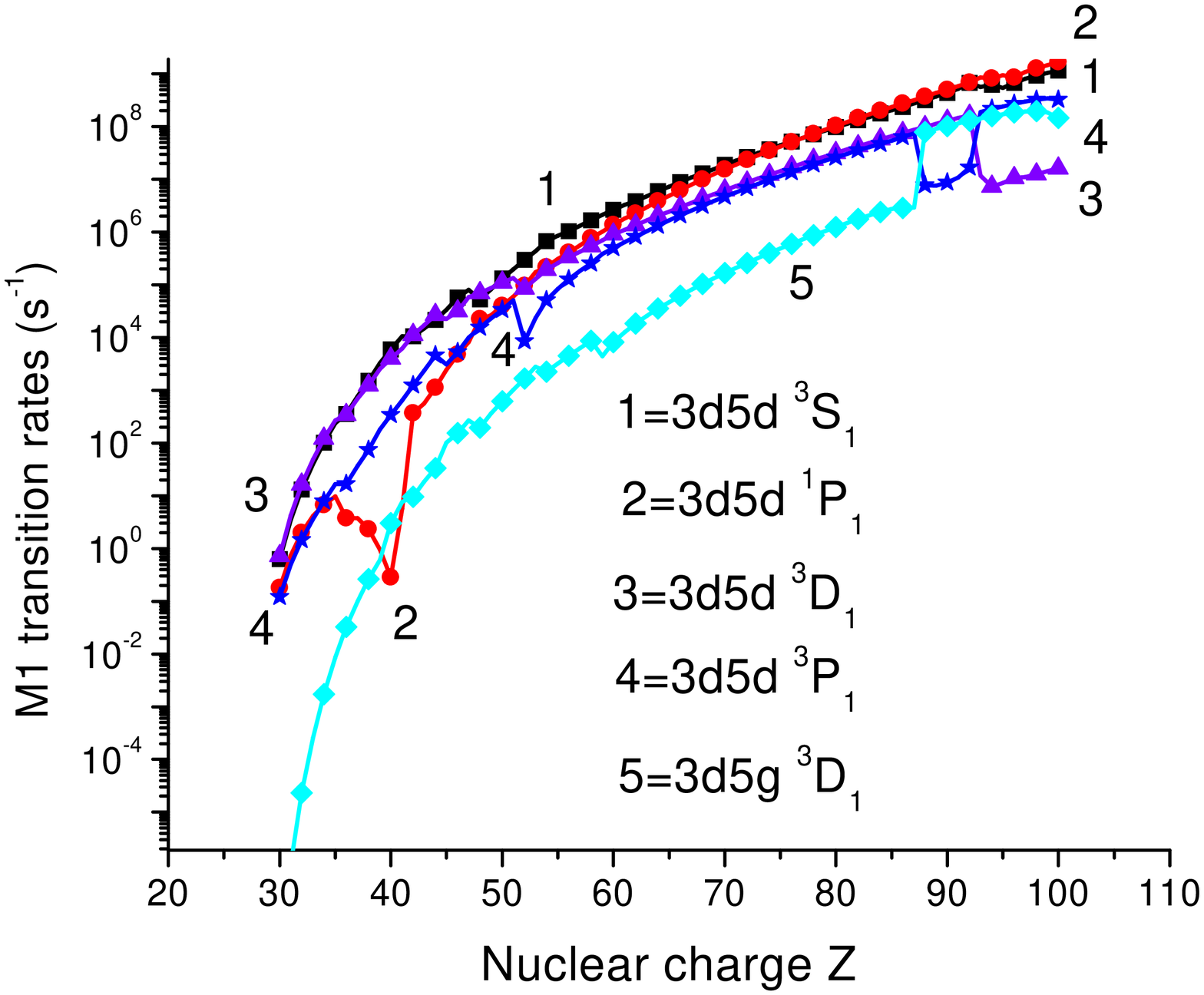}
            \includegraphics[scale=0.35]{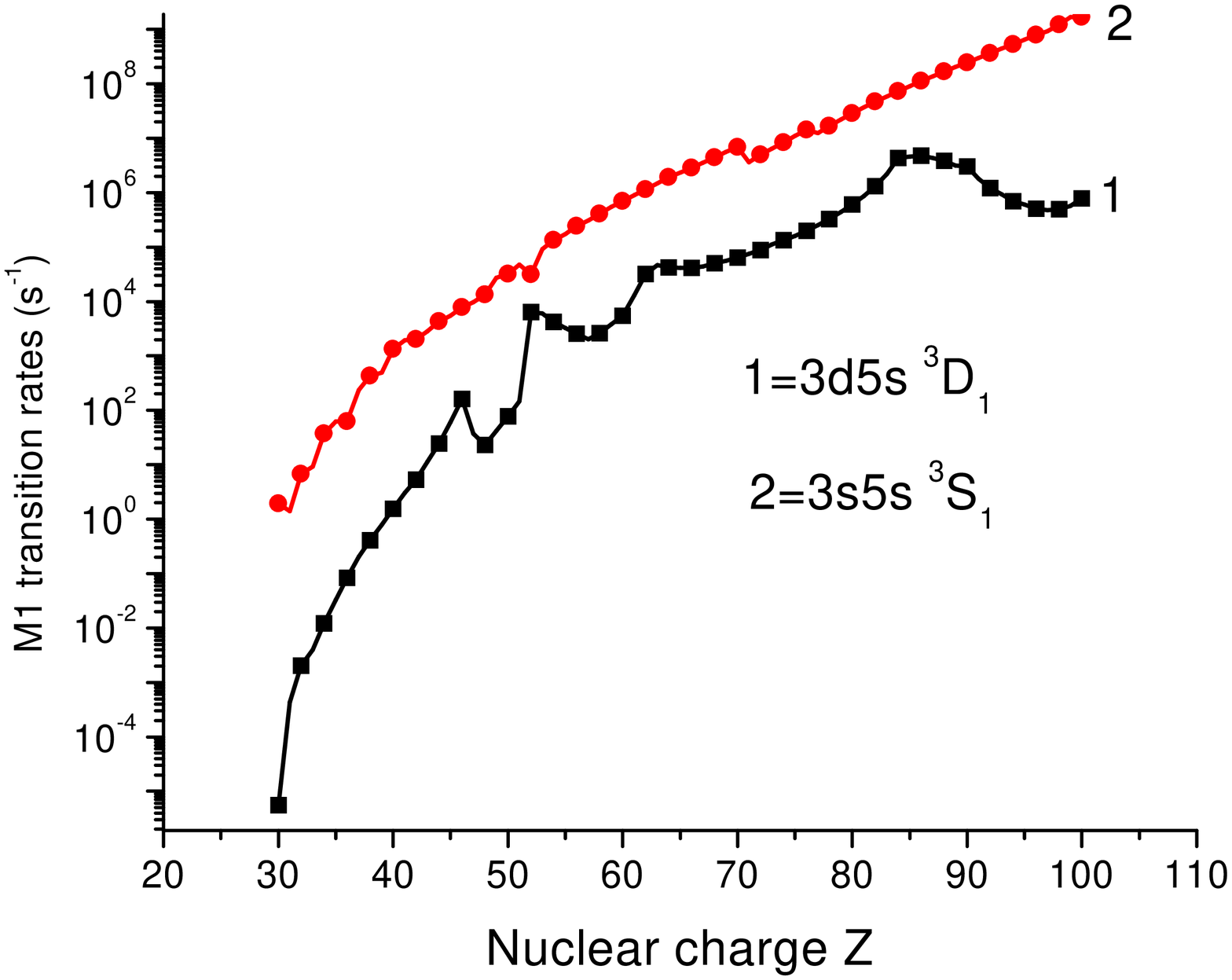}}
\centerline{\includegraphics[scale=0.35]{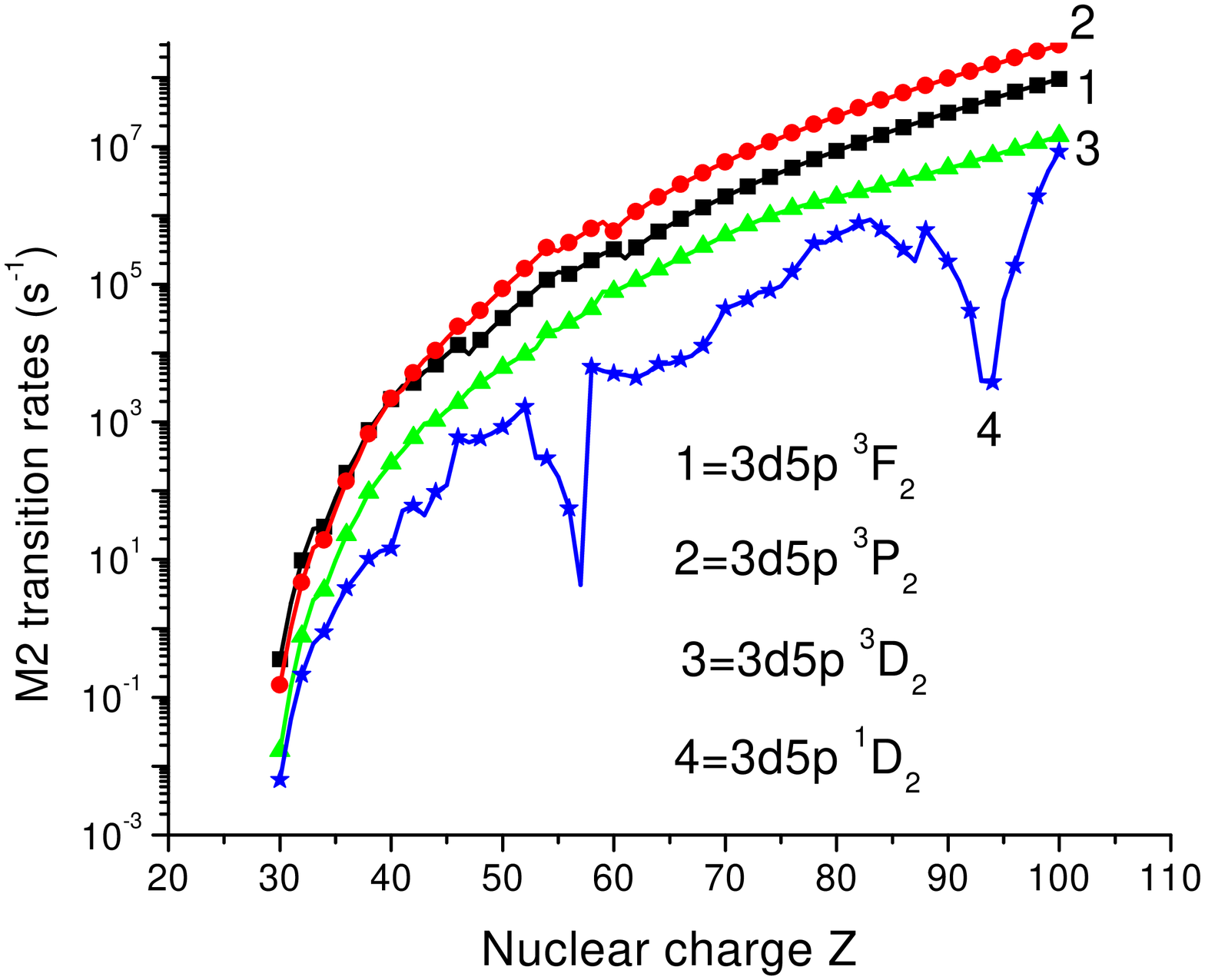}
            \includegraphics[scale=0.35]{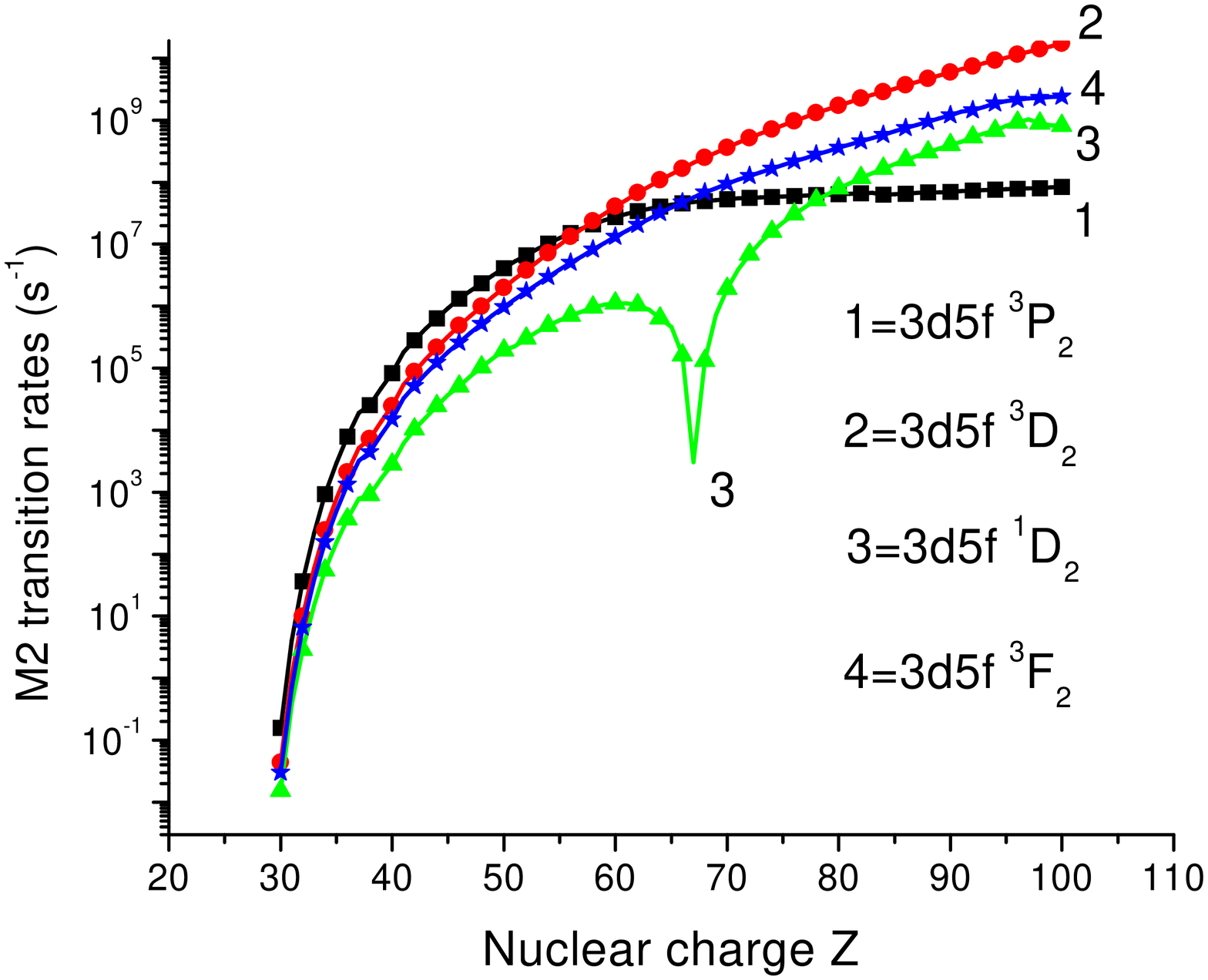}}
\centerline{\includegraphics[scale=0.35]{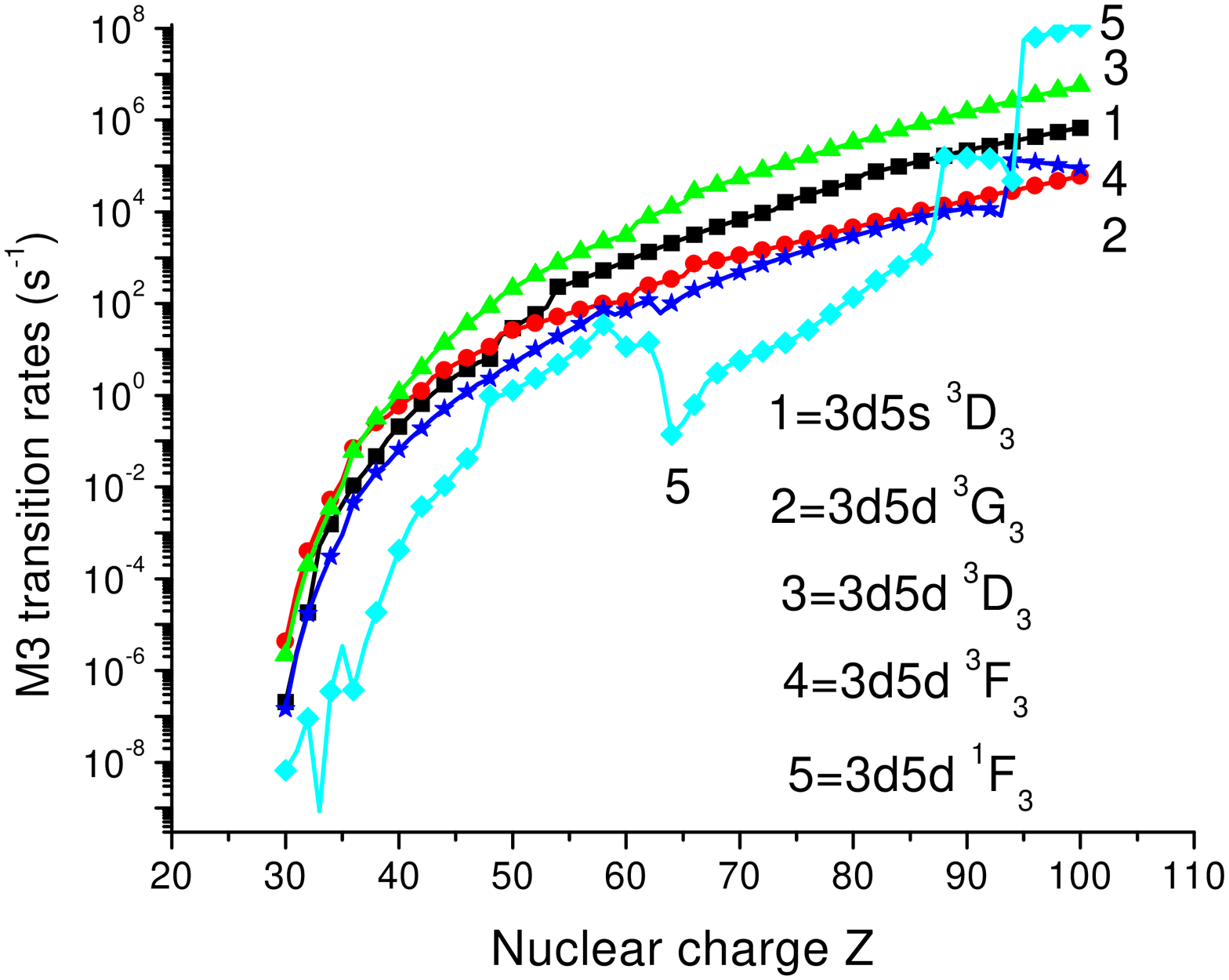}
            \includegraphics[scale=0.35]{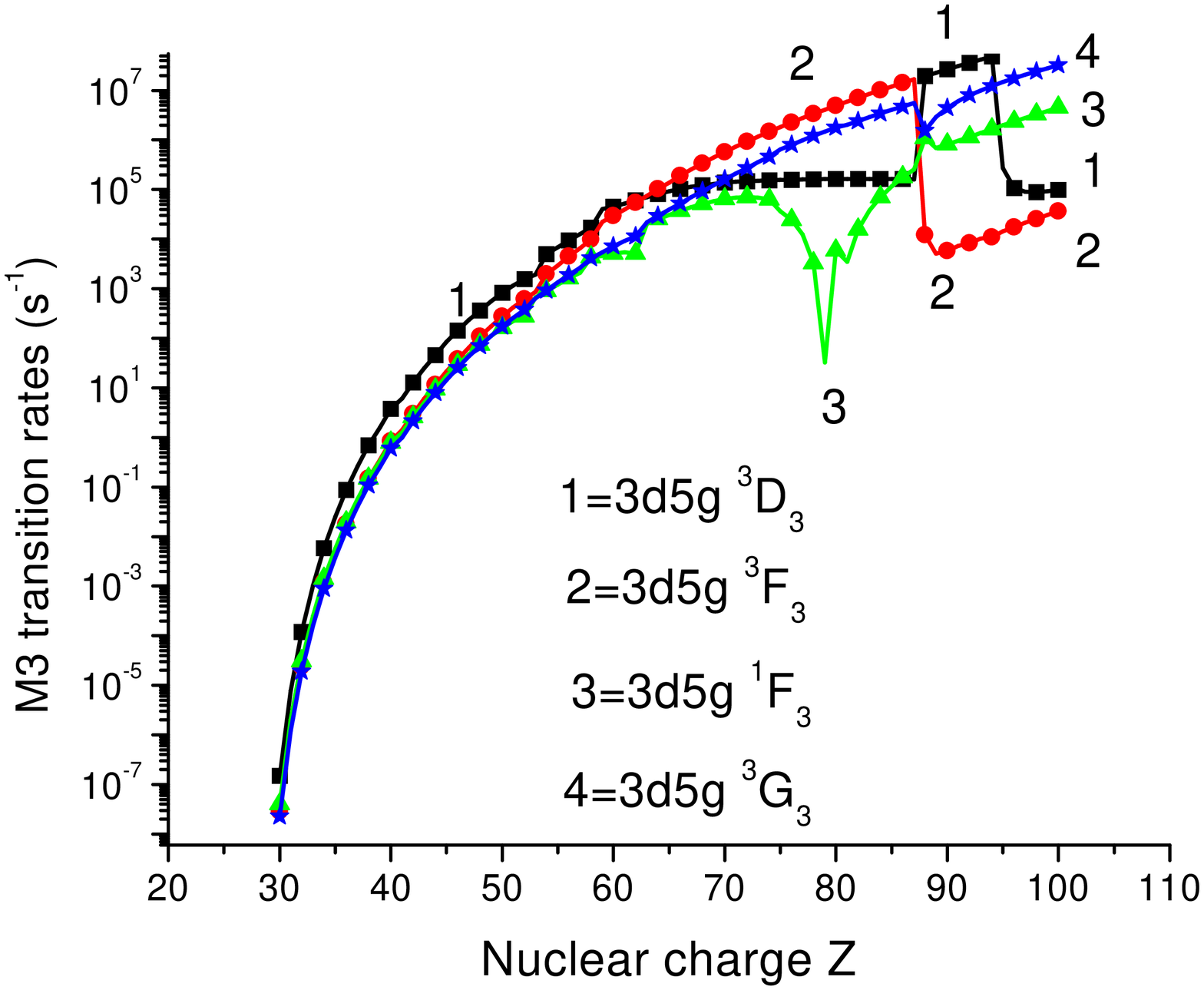}}
            \caption{
M1, M2, and M3 transition rates between the $3l5l'$ states with
$J$=1--3 and ground state in Ni-like ions as function of $Z$}
\label{tr-m1}
\end{figure}

\section{METHOD}
 Details of the RMBPT
method were presented in Refs.~\cite{avg-en1,ni} for calculation
of energies of hole-particle states, in Ref.~\cite{be-en} for
calculation of energies of particle-particle states, in
Ref.~\cite{be-tr} for calculation of radiative electric-dipole
rates in two-particle states, and in Ref.~\cite{ne,ni-cjp} for
calculation of radiative   electric-dipole, electric-quadrupole,
magnetic-dipole, and magnetic-quadrupole rates in Ne- and Ni-like
systems. We will present here only the model space for Ni-like
ions  without repeating the detailed discussions given in
\cite{avg-en1,ni}, \cite{be-en},\cite{be-tr}, and
\cite{ne,ni-cjp}. The calculations are carried out using  sets of
basis Dirac-Hartree-Fock (DHF) orbitals. The orbitals used in the
present calculation are obtained as linear combinations of
B-splines. These B-spline basis orbitals are determined using the
method described in Ref.~\cite{2wrj}. We use 50 B-splines of order
10 for each single-particle angular momentum state and we include
all orbitals with orbital angular momentum $l \leq 9$ in our basis
set.

For atoms with one hole in closed shells and one electron above
closed
shells, the model space is formed from hole-particle states of the type $%
a_{v}^{+}a_{a}|0\rangle \,,$ where $|0\rangle $ is the
closed-shell
 $1s_{1/2}^{2}2s_{1/2}^{2}2p_{1/2}^{2}2p_{3/2}^{4}3s_{1/2}^{2}3p_{1/2}^{2}3p_{3/2}^{4}3d_{3/2}^{4}3d_{3/2}^{6}$
ground state. The single-particle indices $v$ range over states in
the valence shell and the single-hole indices $a$ range over the
closed core. For our study of low-lying states $3l^{-1}4l^{\prime
}$ states of Ni-like ions, values of $a$
are $3s_{1/2}$, $3p_{1/2}$, $3p_{3/2}$, $3d_{3/2}$, and $3d_{5/2}$,while values of $v$ are $%
5s_{1/2} $, $5p_{1/2}$, $5p_{3/2}$, $5d_{3/2}$, $5d_{5/2}$
$5f_{5/2}$,  $5f_{7/2}$,  $5g_{7/2}$, and $5g_{9/2}$. To obtain an
orthonormal model states, we consider the coupled states $\Phi
_{JM}(av)$ defined by
\begin{eqnarray}
\Phi _{JM}(av)
&=&\sqrt{(2J+1)}\sum_{m_{a}m_{v}}(-1)^{j_{v}-m_{v}}\left(
\begin{array}{ccc}
j_{v} & J & j_{a} \\
-m_{v} & M & m_{a}
\end{array}
\right)   a_{vm_{v}}^{\dagger }a_{am_{a}}|0\rangle \,.
\label{uncoup}
\end{eqnarray}
Combining the $n=3$ hole orbitals and the $n=4$ particle orbitals
in nickel, we obtain 68 odd-parity states consisting of
 5 $J=0$ states,
13 $J=1$ states, 16 $J=2$ states, 15 $J=3$ states, 11 $J=4$
states, six $J=5$ states, and two $J=6$ states. Additionally,
there are 74 even-parity states consisting of 5 $J=0$ states, 13
$J=1$ states, 17 $J=2$ states, 16 $J=3$ states, 12 $J=4$ states,
seven $J=5$ states, three $J=6$ states, and one $J=7$ state. The
distribution of the 142 states in the model space is summarized in
Table~\ref{tab0}. In this table, we give both $jj$ and $LS$
designations for hole-particle states.
 Instead of using the  $3l_{j}^{-1}5l'_{j'}$ or $3l^{-1}5l'$ designations,
 we use  simpler
designations  $3l_{j}5l'_{j'}$ or $3l5l'$ in this table and in all
following tables and the text  below.

\section{ Excitation energies}

\subsection{Example: Energy matrix for W$^{46+}$ }

In Table~\ref{tab2}, we give various contributions to the
second-order energies for the special case of Ni-like tungsten,
$Z=74$. In this table, we show the one-body and two-body
second-order Coulomb contributions to the energy matrix labeled
$E^{(2)}_{1}$ and $E^{(2)}_{2}$, respectively. The corresponding
Breit-Coulomb contributions are given in columns headed
$B^{(2)}_{1}$ and $B^{(2)}_{2}$. The one-body second-order energy
is obtained as a sum  of the valence $E^{(2)}_v$ and hole
$E^{(2)}_a$ energies with the latter being the dominant
contribution. The values of $E^{(2)}_{1}$ and $B^{(2)}_{1}$ are
non-zero only for diagonal matrix elements. Although there are 142
diagonal and 1636 non-diagonal matrix elements for
$(3l_{j}5l'_{j'})\,(J)$ hole-particle states, we list only the
part of odd-parity subset with $J$=1 in Table~\ref{tab2}. It can
be seen from the table that second-order
  Breit-Coulomb corrections are relatively large and, therefore, must
 be included in accurate calculations.
 The values of non-diagonal matrix elements given in
 columns headed $E^{(2)}_{2}$ and $B^{(2)}_{2}$  are comparable with values of
diagonal two-body  matrix elements. However, the values of
one-body contributions, $E^{(2)}_{1}$ and $B^{(2)}_{1}$, are
larger than the values of two-body contributions, $E^{(2)}_{2}$
and $B^{(2)}_{2}$, respectively. As a result, total second-order
diagonal matrix elements are much larger than the non-diagonal
matrix elements, which are shown in Table~\ref{tab3}.

In Table~\ref{tab3}, we present results for the zeroth-, first-,
and second-order Coulomb contributions,  $E^{(0)}$,  $E^{(1)}$,
and $E^{(2)}$,
 and the first- and second-order Breit-Coulomb
corrections, $B_{hf}^{(1)}$ and  $B^{(2)}$. It should be noted
that corrections for the frequency-dependent Breit interaction
\cite{3wrj} are included in the first order only. The difference
between the first-order Breit-Coulomb corrections calculated with
and without frequency dependence is less than 1\%.  As one can see
from Table~\ref{tab3}, the ratio of non-diagonal and diagonal
matrix elements is larger for the first-order contributions than
for the second-order contributions. Another difference in the
first- and second-order contributions is the symmetry properties:
 the first-order non-diagonal matrix elements are symmetric and
 the second-order non-diagonal matrix elements are not symmetric.
 The values of
$E^{(2)}[a'v'(J),av(J)]$ and $E^{(2)}[av(J),a'v'(J)]$ matrix
elements differ in some cases by a factor 2--3 and occasionally
have opposite signs.

We now discuss how the final  energy levels are obtained from the
above contributions. To determine the first-order energies of the
states under consideration, we diagonalize the symmetric
first-order effective Hamiltonian, including both the Coulomb and
Breit interactions. The first-order expansion coefficient
$C^{N}[av(J)]$ (often a mixing coefficient)is the $N$-th
eigenvector of the first-order effective Hamiltonian and
$E^{(1)}[N]$ is the corresponding eigenvalue. The resulting
eigenvectors are used to determine the second-order Coulomb
correction $E^{(2)}[N]$, the second-order Breit-Coulomb correction
$B^{(2)}[N]$ and the QED correction $E_{\rm {Lamb}}[N]$.

In Table \ref{tab4}, we list the following contributions to the
energies of 13 excited states in W$^{46+}$: the sum of the zeroth
and first-order energies $E^{(0+1)}$ = $E^{(0)}$
+$E^{(1)}$+$B_{hf}^{(1)}$, the second-order Coulomb energy
$E^{(2)}$, the second-order Breit-Coulomb correction $B^{(2)}$,
the QED correction $E_{\rm {LAMB}}$,  and the sum of the above
contributions
 $E_{\rm {tot}}$. The QED correction is
approximated as the sum of the one-electron self energy and the
first-order vacuum-polarization energy. The screened self-energy
and vacuum polarization data given by Kim {\it et al.\/}
\cite{kim}, which are in close agreement with screened self-energy
calculations by Blundell~ \cite{blundell} are used to determine
the QED correction $E_{\rm {LAMB}}$.

When starting calculations from relativistic DHF wavefunctions, it
is natural to use $jj$ designations for uncoupled transition and
energy matrix elements; however, neither $jj$ nor $LS$ coupling
describes the {\em physical} states properly, except for the
single-configuration state $3d_{5/2}5g_{9/2}(7) \equiv 3d5g\ ^3\!
I_7$. Both designations are used in Table \ref{tab4}.

\subsection{$Z$-dependence of eigenvectors and eigenvalues in Ni-like ions}

In Figs.~\ref{fig-ppf} - \ref{fig-en}, we illustrate the
$Z$-dependence of the eigenvectors and eigenvalues of the $3l_{j}\
5l'_{j'}\ (J)$ hole-particle states. We refer to a set of states
of the same parity and the same $J$ as
 a complex of states.
Strong mixing for the $3l_{j}\ 4l'_{j'}\ (J)$ hole-particle states
was discussed in many papers (see, for example, \cite
{ni,ni-cjp}).  It should be noted that the $3l_{j}\ 5l'_{j'}\ (J)$
states are less mixed than the $3l_{j}\ 4l'_{j'}\ (J)$  states.
For odd-parity complex with $J$=1, we found strong mixing only for
states with $3p$-hole, $3p_{j}5s_{1/2}\ (1)$ and $3p_{j}\ 5d_{j'}\
(1)$ states.  In Fig.~\ref{fig-ppf},
 we show the dependence of the eigenvectors using the
example of odd-parity states with $J$=1. This particular $J$=1
odd-parity complex includes  13 states which are listed in
Table~\ref{tab0}. Using
 the first-order expansion
coefficients $C^{N}[av(J)]$ defined in the previous section, we
can present the resulting eigenvectors as
\begin{eqnarray}
\Phi (N) &=&C^{N}[3d_{3/2}5p_{1/2}(1)]\Phi
[3d_{3/2}5p_{1/2}(1)]+C^{N}[3d_{5/2}5p_{3/2}(2)]\Phi
[3d_{5/2}5p_{3/2}(1)]+
\nonumber \\
&&C^{N}[3d_{3/2}5p_{3/2}(1)]\Phi
[3d_{3/2}5p_{3/2}(1)]+C^{N}[3d_{5/2}5f_{5/2}(2)]\Phi
[3d_{5/2}5f_{5/2}(1)]+
\nonumber \\
&&C^{N}[3d_{5/2}5f_{7/2}(2)]\Phi
[3d_{5/2}5f_{7/2}(1)]+C^{N}[3d_{3/2}5f_{5/2}(2)]\Phi
[3d_{3/2}5f_{5/2}(1)]+
\nonumber\\
&&C^{N}[3p_{3/2}5s_{1/2}(1)]\Phi
[3p_{3/2}5s_{1/2}(1)]+C^{N}[3p_{1/2}5s_{1/2}(1)]\Phi
[3p_{1/2}5s_{1/2}(1)]+
\nonumber\\
&&C^{N}[3p_{3/2}5d_{3/2}(1)]\Phi
[3p_{3/2}5d_{3/2}(1)]+C^{N}[3p_{3/2}5d_{5/2}(1)]\Phi
[3p_{3/2}5d_{5/2}(1)]+
\nonumber\\
&&C^{N}[3p_{1/2}5d_{3/2}(1)]\Phi [3p_{1/2}5d_{3/2}(1)]+ \nonumber\\
&&C^{N}[3s_{1/2}5p_{1/2}(1)]\Phi
[3s_{3/2}5p_{1/2}(1)]+C^{N}[3s_{1/2}5p_{3/2}(1)]\Phi
[3s_{3/2}5p_{3/2}(1)]
\end{eqnarray}
As a result,  169 $C^{N}[av(J)]$ coefficients are needed to
describe the 13 eigenvalues.  For simplicity, we plot only three
of the 13 mixing coefficients for the level  $N$=$3p5d\ ^{1,3}P_1$
in Fig.~\ref{fig-ppf}. These coefficients are chosen to illustrate
the mixing of the states; the remaining mixing coefficients give
very small contributions to this level. We observe strong mixing
between $[3p_{1/2}5s_{1/2}(1)]$ + $[3p_{3/2}5d_{3/2}(1)]$ +
$[3p_{3/2}5d_{5/2}(1)]$ states for $Z$=57 - 58.

Energies, relative to the ground state, of odd- and even-parity
states with
 $J$=1, 2 and 3, divided by $(Z-21)^2$, are shown in Fig.~\ref{fig-en}.
It should be noted that $Z$ was decreased by 21 to provide better
presentation of the energy diagrams. We plot the limited number of
energy levels to illustrate $Z$ dependence choosing one
representative from a configuration. As a result, we show  6
levels instead of 68 odd-parity states, and 6 levels instead of 74
for the even-parity states in Fig.~\ref{fig-en}.  $LS$
designations are chosen by  small values of
 multiplet splitting for low-$Z$ ions. To
confirm  those $LS$ designations we obtain the fine structure
splitting for the $3d5s\ ^3D$, $3d5p[\ ^3P,\ ^3D,\ ^3F]$, $3d5d[\
^3P,\ ^3D,\ ^3F,\ ^3G]$, $3d5f[\ ^3P,\ ^3D,\ ^3F,\ ^3G,\ ^3H]$,
$3d5g[\ ^3D,\ ^3F,\ ^3G,\ ^3H,\ ^3I]$,
 $3p5s[\ ^3P$], $3p5p[\ ^3P,\ ^3D]$, $3p5d[\ ^3P,\ ^3D,\ ^3F]$,
$3p5f[\ ^3D,\ ^3F,\ ^3G]$, $3p5g[\ ^3F,\ ^3G,\ ^3H]$, $3s5p\ ^3P$,
$3s5f\ ^3F$, and $3s5g\ ^3G$
 triplets.

Energy differences between levels of odd- and even-parity triplet
terms, divided by $(Z-21)^2$, are illustrated in
Fig.~\ref{fig-del1}.
 The  energy intervals for the $3d5p(\ ^3P_{2}-\ ^3P_{1})$,
$3d5p(\ ^3D_{3}-\ ^3D_{2})$, $3d5p(\ ^3F_{3}-\ ^3F_{2})$, $3d5f(\
^3P_{2}-\ ^3P_{1})$,  $3d5f(\ ^3G_{4}-\ ^3G_{3})$, $3d5f(\
^3G_{5}-\ ^3G_{4})$, $3d5f(\ ^3H_{5}-\ ^3H_{4})$, and  $3p5s(\
^3P_{1}-\ ^3P_{0})$ states are very small and almost do not change
with $Z$ as can be seen from Fig.~\ref{fig-del1}. It is the very
sharp change of
 splitting with $Z$ for the $3d5f\ ^3F$ and $3p5f\ ^3F$ terms but the
 energies $\Delta E/(Z-21)^2$ change by small values, from 5~cm$^{-1}$ to
 -20~cm$^{-1}$ and 20~cm$^{-1}$ to
 -30~cm$^{-1}$, respectively.
The  energy intervals vary strongly with the nuclear charge for
the $3d5p(\ ^3P_{1}-\ ^3P_{0})$, $3d5p(\ ^3D_{2}-\ ^3D_{1})$,
$3d5p(\ ^3F_{4}-\ ^3F_{3})$, $3d5f(\ ^3P_{1}-\ ^3P_{0})$, and
 $3d5f(\ ^3H_{6}-\ ^35_{4})$ states.  Our calculations show that the fine
structures of almost all levels illustrated in
Figs.~\ref{fig-del1} do not follow the Land\'{e} rules even for
small $Z$. The unusual splittings may be caused by  changes from
$LS$ to $jj$ coupling, with mixing from other triplet and singlet
states.
  The different J states are mixed differently.
  Further experimental confirmation would be very helpful
  in verifying the correctness of these sometimes sensitive mixing parameters.

\section{\bf Electric-dipole, electric-quadrupole, and electric-octupole  matrix elements}
We calculate  electric-dipole (E1) matrix elements for the
transitions between
 the 13 odd-parity $3d_{j}5p_{j'}(1)$, $3d_{j}5f_{j'}(1)$,
$3p_{j}5s_{1/2}(1)$, $3p_{j}5d_{j'}(1)$, and $3s_{1/2}5p_{j'}(1)$
 excited states and
the ground state,  electric-quadrupole (E2)  matrix elements
between the 17 even-parity $3d_{j}5s_{1/2}(2)$,
$3d_{j}5d_{j'}(2)$, $3d_{j}5g_{j'}(2)$, $3p_{j}5p_{j'}(2)$, $3p_{j}5f_{j'}(2)$, and
$3s_{1/2}5d_{j'}(2)$ excited states and the ground state, and
electric-quadrupole (E2)  matrix elements
between the 15 odd-parity $3d_{j}5p_{j'}(3)$,
$3d_{j}5f_{j'}(3)$,  $3p_{j}5s_{1/2}(3)$, $3p_{j}5d_{j'}(3)$, $3p_{j}5g_{j'}(3)$, and
$3s_{1/2}5f_{j'}(2)$ excited states and the ground state
 for Ni-like ions with nuclear charges $Z=30-100$. Analytical expressions
  for multipole matrix elements
in the first and the second order RMBPT are given by Eqs.~(2.12)-(2.17)  of
Ref.~\cite{ni}.

The first- and second-order Coulomb corrections and second-order
Breit-Coulomb corrections to
  reduced E1 and E2
matrix elements will be referred to as $Z^{(1)}$, $Z^{(2)}$, and
$B^{(2)}$, respectively, throughout the text. These contributions
are calculated in both length and velocity gauges. In this
section, we show the importance of the different contributions and
discuss the gauge dependence of the E1, E2, and E3
 matrix elements.

\subsection{Example: E1, E2, and E3 matrix elements for W$^{46+}$ }
In Table~\ref{tab-euncop}, we list values of {\em uncoupled}
first- and second-order E1, E2, and E3 matrix elements $Z^{(1)}$,
$Z^{(2)}$, $B^{(2)}$, together with derivative terms $P^{({\rm
derv})}$, for Ni-like tungsten, $Z$=74. We list values for the
 E1
 transitions between  odd-parity  states with $J$=1,  the
ground state and the  E2 transitions between even-parity states
with $J$=2 and the ground state, the  E3
 transitions between  odd-parity  states with $J$=3, respectively. Matrix
elements in both  length ($L$) and velocity ($V$) forms are given.
We can see that the first-order matrix elements, $Z^{(1)}_L$ and
$Z^{(1)}_V$, differ by 5--10\%; however, the $L$--$V$ differences
between second-order matrix elements are much larger for some
transitions.  Also  for the E1 transitions, the derivative term in
length form, $P^{({\rm derv)}}_L$, is almost equal to $Z^{(1)}_L$
but  the derivative term in velocity form, $P^{({\rm derv)}}_V$,
 is smaller than $Z^{(1)}_V$  by three to four orders of magnitude.
For the E2 transitions, the value of $P^{({\rm derv)}}$ in velocity
form almost equals $Z^{(1)}$ in velocity  form and the $P^{({\rm
derv)}}$
 in length form is larger by factor of two than $Z^{(1)}$ in length
 form.
 For the E3 transitions, the value of $P^{({\rm derv)}}$ in velocity
form is larger by factor of two than $Z^{(1)}$ in velocity  form and the $P^{({\rm
derv)}}$
 in length form is larger by factor of three than $Z^{(1)}$ in length
 form.

Values of E1, E2, and E3 {\em coupled} reduced matrix elements in
length and velocity forms are illustrated in Table~\ref{tab-ecop}
for the limited set of transitions.  Although we use an
intermediate-coupling scheme, it is nevertheless convenient to
label the physical states using the $jj$ labelling for high-$Z$
and the $LS$ labelling for low-$Z$; both designations are used in
Table~\ref{tab-ecop}. The first two columns in Table
\ref{tab-ecop} show $L$ and $V$ values of {\em coupled} reduced
matrix elements calculated in first order. The $L-V$ difference is
about 5--10\%. Including the second-order contributions (columns
headed RMBPT in Table \ref{tab-ecop}) decreases the $L-V$
difference to 0.02--2\%. This non-zeroth $L-V$ difference
 arises because we start our RMBPT
calculations using a non-local Dirac-Fock (DF) potential.  If we
were to replace the DF potential by a local potential, the
differences would disappear completely. It should be emphasized
that we include the negative energy state (NES) contributions to
sums over intermediate states (see Ref.~\cite{be-tr} for details).
Neglecting the NES contributions leads to small changes in the
$L$-form matrix elements but to substantial changes in some of the
$V$-form matrix elements with a consequent loss of gauge
independence.

\subsection{$Z$-dependences of E1 and E2 matrix elements  in Ni-like ions}

In Fig.~\ref{e1-uncop}, differences between length and velocity
forms are illustrated for the various contributions to uncoupled
$0-3d_{5/2}5f_{7/2}(1)$, $0-3d_{3/2}5g_{7/2}(2)$, and
$0-3d_{3/2}5f_{5/2}(3)$  matrix elements, where $0$ is the ground
state. In the case of E1 transitions, the first-order matrix
element $Z^{(1)} $ is proportional to $1/Z$, the second-order
Coulomb matrix element $Z^{(2)} $  is proportional to $1/Z^2$, and
the second-order Breit-Coulomb matrix element $B^{(2)} $ is almost
independent of $Z$ (see \cite {be-tr}) for high $Z$. Therefore, we
plot $Z^{(1)}\times (Z-21)$, $Z^{(2)}\times (Z-21)^2$, and
$B^{(2)}\times 10^4$  for the $0-3d_{5/2}5f_{7/2}(1)$ transition.
  All these contributions  are positive, except for the
second-order Coulomb matrix elements $Z^{(2)}$ in lengths form.

The difference between length- and velocity-forms for  E2
transitions is illustrated  in Fig.~\ref{e1-uncop} for the
uncoupled $0-3d_{3/2}5g_{7/2}(2)$ matrix element.  In the case of
E2 transitions, the first-order matrix element $Z^{(1)} $ is
proportional to $1/Z^2$, the second-order Coulomb matrix element
$Z^{(2)} $  is proportional to $1/Z^3$, and the second-order
Breit-Coulomb matrix element $B^{(2)} $ is proportional to $1/Z$
for high $Z$. We plot $Z^{(1)}\times (Z-21)$, $Z^{(2)}\times
(Z-21)^2$, $B^{(2)}\times 10^4$ for the better illustration of
those contributions in the right panel of Fig.~\ref{e1-uncop}. All
these contributions are positive.

The difference between length- and velocity-forms for  E3
transitions is illustrated  in Fig.~\ref{e1-uncop} for the
uncoupled $0-3d_{3/2}5f_{5/2}(3)$ matrix element.
  We plot $Z^{(1)}\times (Z-21)^2$,
$Z^{(2)}\times (Z-21)^3$, and $B^{(2)}\times (Z-21) \times 10^4$
in the bottom panel of Fig.~\ref{e1-uncop}.   The second-order
Breit-Coulomb correction to the E3 matrix element $B^{(2)} $ is
much smaller in velocity form than in length form, as seen in the
figure.

The differences between results in length- and velocity-forms
shown in Fig.~\ref{e1-uncop} are compensated by additional
second-order terms called  ``derivative terms'' $P^{({\rm
derv})}$; they are defined by Eq.~(2.16) of Ref.~\cite{ni} (see,
also Tables~\ref{tab-euncop} and ~\ref{tab-ecop}).
 The
derivative terms arise because transition amplitudes depend on the
energy, and the transition energy changes order-by-order in RMBPT
calculations.

\section{\bf Magnetic-dipole,  magnetic-quadrupole, and magnetic-octupole matrix elements}
We calculate  magnetic-dipole (M1)  matrix elements for the
transitions between the 13 even-parity $3d_{j}5s_{1/2}(1)$,
$3d_{j}5d_{j'}(1)$, $3d_{j}5d_{j'}(1)$,  $3p_{j}5p_{j'}(1)$,
$3p_{j}5f_{j'}(1)$, $3s_{1/2}5s_{1/2}(1)$, and
$3s_{1/2}5d_{j'}(1)$ excited states and the ground state,
magnetic-quadrupole (M2)  matrix elements between the 16
odd-parity $3d_{j}5p_{j'}(2)$, $3d_{j}5f_{j'}(2)$,
$3p_{j}5s_{1/2}(2)$, $3p_{j}5d_{j'}(2)$, $3p_{j}5g_{j'}(2)$,
$3s_{1/2}5p_{j'}(2)$, and $3s_{1/2}5f_{j'}(2)$ excited states and
the ground state, and magnetic-octupole (M3)  matrix elements for
the transitions between the 16 even-parity $3d_{j}5s_{1/2}(3)$,
$3d_{j}5d_{j'}(3)$, $3d_{j}5d_{j'}(3)$, $3d_{j}5g_{j'}(3)$,
$3p_{j}5p_{j'}(3)$, $3p_{j}5f_{j'}(3)$, $3s_{1/2}5s_{1/2}(3)$, and
$3s_{1/2}5d_{j'}(3)$ excited states and the ground state for
Ni-like ions with nuclear charges $Z=30-100$.

We calculate first- and second-order Coulomb, second-order
Breit-Coulomb corrections, and second-order derivative term  to
  reduced M1 and M2
matrix elements  $Z^{(1)} $, $Z^{(2)}$, $B^{(2)}$, and $P^{({\rm
derv})}$, respectively, using the method described in Eqs.~(2.13)
- (2.18) of Ref.\cite{ni} and  Eqs.~(A3--A5) of
Ref.~\cite{safr2}, respectively.
 In this section, we
illustrate the importance of the relativistic and
frequency-dependent
 contributions to the first-order M1 and M2 matrix
elements. We also show the importance of the taking into account
the second-order RMBPT contributions to M1 and M2 matrix elements
and we subsequently discuss the necessity of including the
negative-energy contributions to sums over intermediate states.
{\em Ab initio} relativistic calculations require careful
treatment of negative-energy states (virtual electron-positron
pairs).  In second-order matrix elements,
 such contributions explicitly arise from those terms in
the sum over states for which $\varepsilon _{i}<-mc^{2}$. The
effect of the NES contributions to M1-amplitudes has been studied
recently in Ref.~\cite{be-m1}. The NES contributions drastically
change  the second-order Breit-Coulomb matrix elements $B^{(2)}$.
However, the second-order Breit-Coulomb correction  contributes
only 2--5\% to uncoupled M1  matrix elements and, as a result,
negative-energy states change the total values of M1 matrix
elements by a few percent only.

\subsection{$Z$-dependences of M1, M2, and M3 matrix elements  in Ni-like ions}
 The differences between first-order M1 uncoupled
matrix elements, calculated in nonrelativistic, relativistic
frequency-independent, and relativistic frequency-dependent
approximations are illustrated in the left panel of
Fig.~\ref{m1-uncop} for the $0-3d_{5/2}5d_{5/2}(1)$ matrix
element. The corresponding matrix elements are labeled
$Z^{(1)}_{\rm RF}$, $Z^{(1)}_{\rm R}$, and $Z^{(1)}_{\rm NR}$.
Formulas for relativistic frequency-dependent and non-relativistic
first-order M1 matrix elements are given by Eqs.~(3--6) of
Ref.~\cite{be-m1}.
 We also plot the second-order Coulomb  contributions, $Z^{(2)}_{\rm CL}$, and the
second-order Breit-Coulomb contributions, $Z^{(2)}_{\rm BR}$, in
the same figure.
  As we observe from the left panel of
Fig.~\ref{m1-uncop},  the values of $Z^{(1)}_{\rm NR}$ are twice
as small as  the values of $Z^{(1)}_{\rm R}$ and $Z^{(1)}_{\rm
RF}$. Therefore, relativistic effects are very large for M1
transitions. The frequency-dependent relativistic matrix elements
$Z^{(1)}_{\rm RF}$  differ from the relativistic
frequency-independent matrix elements $Z^{(1)}_{\rm R}$ by
10--40\%. The differences between other first-order matrix
elements calculated with and without frequency dependence are also
 the order of a few percent. Uncoupled second-order  M1 matrix
elements $Z^{(2)}_{\rm CL}$ are comparable to first-order matrix
elements $Z^{(1)}_{\rm RF}$  for small $Z$ but the relative size
of the second-order contribution decreases for high $Z$. This is
expected since second-order Coulomb matrix elements $Z^{(2)}_{\rm
CL}$ are proportional to $Z$ for high $Z$ while first-order matrix
elements $Z^{(1)}_{\rm RF}$ grow as $Z^2$. The second-order
Breit-Coulomb matrix elements $Z^{(2)}_{\rm BR}$ are proportional
to $Z^3$ and become larger than $Z^{(2)}_{\rm CL}$ for  high $Z$.

 The differences between first-order M2 uncoupled
matrix elements, calculated in  relativistic frequency-dependent,
and relativistic frequency-independent approximations are
illustrated for the $0-3d_{5/2}5f_{7/2}(2)$ matrix element in the
right panel of  Fig.~\ref{m1-uncop}. The corresponding  matrix
elements are labeled $Z^{(1)}_{\rm RF}$ and $Z^{(1)}_{\rm R}$.
Formulas for relativistic frequency-dependent and
frequency-independent first-order M2 matrix elements are given by
Eqs.~(A3--A5) of Ref.~\cite{safr2}.
 We also plot the second-order Coulomb  contributions, $Z^{(2)}_{\rm CL}$, and the
second-order Breit-Coulomb contributions, $Z^{(2)}_{\rm BR}$, in
the same figure.

 The differences between first-order M3 uncoupled
matrix elements, calculated in  relativistic frequency-dependent,
and relativistic frequency-independent approximations are
illustrated for the $0-3d_{5/2}5s_{1/2}(3)$ matrix element in the
bottom panel of Fig.~\ref{m1-uncop}. The corresponding  matrix
elements are labeled $Z^{(1)}_{\rm RF}$ and $Z^{(1)}_{\rm R}$.
 We also plot the second-order Coulomb  contributions, $Z^{(2)}_{\rm CL}$, and the
second-order Breit-Coulomb contributions, $Z^{(2)}_{\rm BR}$, in
the same figure.

In Fig.~\ref{s-m1}, we illustrate the $Z$-dependence of the line
strengths of M1 transition from the $3d5s\ ^3D_1$ excited state to
the ground state. In this figure, we plot the values of the
first-order line strengths $S^{(1)}_{\rm NR}$ , $S^{(1)}_{\rm R}$,
and $S^{(1)}_{\rm RF}$ calculated in the same approximations as
the M1 uncoupled matrix elements: nonrelativistic, relativistic
frequency-independent, and relativistic frequency-dependent
approximations, respectively. The total line strengths
$S^{(1+2)}$, which include second-order corrections,   are also
plotted. It should be noted that the value of nonrelativistic
matrix element, $Z^{(1)}_{\rm NR}(0-3d_{5/2}5s_{1/2}(1))$ equal to
zero. Small mixing inside of the even-parity complex with $J$=1
between $3d_{5/2}5s_{1/2}$, $3d_{5/2}5d_{3/2}$ $3d_{5/2}5d_{5/2}$,
and $3p_{1/2}5p_{1/2}$ states gives non-zero value even for $Z$ =
30 of the first-order line strengths $S^{(1)}_{\rm NR}$. Non-zero
value of $Z^{(1)}_{\rm R}(0-3d_{5/2}5s_{1/2}(1))$ increases the
first-order line strengths by three order of magnitude for $Z$ =
30. The difference between the values of $S^{(1)}_{\rm R}$, and
$S^{(1)}_{\rm RF}$ is 26~\% for $Z$ = 30. The second-order
contribution gives additional contribution for the value of the
line strengths, the ratio of $S^{(1+2)}_{\rm RF}$ and
$S^{(1)}_{\rm RF}$ is about 5 for $Z$ = 30. The ratios between
$S^{(1)}_{\rm NR}$ , $S^{(1)}_{\rm R}$,  $S^{(1)}_{\rm RF}$, and
$S^{(1+2)}_{\rm RF}$ are changed with $Z$ as can be seen from
Fig.~\ref{s-m1} by increasing relativistic effects.

\subsection{Example: E1, E2, E3,  M1, M2, and M3 transition rates for W$^{46+}$ }
The E1, E2, E3,  M1, M2, and M3 transition probabilities $A$
(s$^{-1}$) for the transitions between the ground state and
$3lj5l^{\prime }j^{\prime }(J)$ states  are obtained in terms of
line strengths $S$~(a.u.) and wavelength $\lambda$(\AA) as
\begin{eqnarray}
A(E1) &=&\frac{2.02613\times 10^{18}}{(2J+1)\lambda ^{3}}\ S(E1),\ \ A(M1)=%
\frac{2.69735\times 10^{13}}{(2J+1)\lambda ^{3}}S(M1)  \nonumber \\
A(E2) &=&\frac{1.11995\times 10^{18}}{(2J+1)\lambda ^{5}}\ S(E2),\ \ A(M2)=%
\frac{1.49097\times 10^{13}}{(2J+1)\lambda ^{5}}\ S(M2)  \nonumber \\
A(E3) &=&\frac{3.14441\times 10^{17}}{(2J+1)\lambda ^{7}}\ S(E3),\ \ A(M3)=%
\frac{4.18610\times 10^{12}}{(2J+1)\lambda ^{7}}\ S(M3)
\end{eqnarray}
In Table~\ref{tab-ar}, we present our RMBPT calculations for E1, E2, E3, M1, M2,
 and M3 transition rates and wavelengths in the case of
Ni-like tungsten, $Z$=74.

\section{ Comparison of results with other theory and experiment}
We calculate energies of the 74 even-parity $3d_{j}5s_{1/2}(J)$,
$3d_{j}5d_{j'}(J)$, $3d_{j}5g_{j'}(J)$,  $3p_{j}5p_{j'}(J)$,   $3p_{j}5f_{j'}(J)$, \\
$3s_{1/2}5s_{1/2'}(J)$, $3s_{1/2}5d_{j'}(J)$,  and $3s_{1/2}5g_{j'}(J)$  excited states
and 68 odd-parity $3d_{j}5p_{j'}(J)$, $3d_{j}5f_{j'}(J)$,\\
$3p_{j}5s_{1/2}(J)$, $3p_{j}5d_{j'}(J)$, $3p_{j}5g_{j'}(J)$, $3s_{1/2}5p_{j'}(J)$, and
$3s_{1/2}5f_{j'}(J)$ excited states
 for Ni-like ions with
nuclear charges  $Z$=30-100. Reduced matrix elements, oscillator
strengths, and transition rates are determined for E1, E2, E3, M1,
M2, and M3 allowed and forbidden transitions into the ground state
for each ions. Comparisons are also given with other theoretical
results and with experimental data. Our results are presented in
two parts:   wavelengths and transition probabilities.

\subsection{ Transition energies}

In Table~\ref{tab-com1}, we compare our RMBPT results for the
excitation energies
 of the odd-parity states  in Ni-like tungsten  with
theoretical results obtained by different codes:  DFS code by
Zhang {\it et al.} \cite{zhang},  and  COWAN code \cite{web}. The
difference in results is about 0.1--0.2~\%. It should be noted
that the RMBPT and DFS codes used $jj$-coupling, however, the
COWAN code used $LS$-coupling for uncoupled matrix elements. To
compare results obtained after diagonalization of energy matrixes
in Table~\ref{tab-com1}, we use $LS$ designations. We found that
resulting $LS$ designations in three codes differ for some states.
 Those two labelling are different for some
levels. In the COWAN code, a label for every level was chosen by
maximum value among eigenvectors. It is not convenient sometimes
when two levels have the same label. In the present paper, we use
RMBPT code to evaluate energies for whole isoelectronic sequence.
It is known that  the crossing energy levels inside the  one
complex with fixed $J$ is forbidden by Wigner and Neumann theorem
(see, for example, in Ref.~\cite{Landau}). As a result, we can use
only numbering of the levels by the ordering of  energies. We
already mentioned, either $LS$ or $jj$ designations are used to
label the resulting eigenvectors and eigenvalues rather than
simply enumerating with an index $N$. We choose the $LS$
designations here since the $jj$ designations are used for
uncouple matrix elements. The  $LSJ$ labels are chosen by small
values  of multiplet splitting for low-$Z$ ions.

\subsection{E1, E2, E3, M1, M2, and M3 transition probabilities}

We present the resulting transition probabilities ($A_r$) in
Figs.~\ref{tr-e1} and \ref{tr-m1}. Transition rates  for the six
E1 lines from $3d5p\ ^3P_1,\ ^3D_1,\ ^1P_1$ and  $3p5d\ ^3P_1,\
^1P_1,\ ^3D_1$ levels to the  ground state  are plotted in the top
panel of Fig.~\ref{tr-e1}. The sharp features in the curves shown
in these figures  can be explained in many cases by strong mixing
of states inside  the odd-parity complex with $J$=1. The double
cusp in the interval $Z$=57-59 and deep minimum in the $Z$=51-53
range for the curve with the $3p5d\ ^3P_1$ label  can be explained
by mixing of the $3p_{3/2}5d_{3/2}\ (1)$,  $3p_{3/2}5d_{5/2}\
(1)$, and $3p_{1/2}5s_{1/2}\ (1)$ states. The  deep minimum in the
$Z$=86-87 range for the curve with the $3d5dp\ ^1P$ label  can be
explained by decreasing of the second-order contribution to the
$0-3d_{3/2}5p_{3/2}\ (1)$ dipole matrix element. This matrix
element gives the main part of contribution to the transition rate
for the $3d5p\ ^1P_1$ state.

Transition rates  for the five E2 lines from $3d5s\ ^3D_2,\ ^1D_2$
and $3d5g\ ^3D_2,\ ^3F_2,\ ^1D_2$ levels to the ground state are
plotted in the central panel of Fig.~\ref{tr-e1}. The curves
describing $3d5s\ ^3D_2,\ ^1D_2$ transition rates smoothly
increase  with $Z$ without any sharp features.  The difference in
values of $A_r$ for $3d5s\ ^3D_2$ and $3d5s\ ^1D_2$ lines is about
20--50\%. It is so small difference in the values of of $A_r$ for
$3d5g\ ^3D_2$ and $3d5g\ ^3F_2$ up to $Z$ = 60.  The double cusp
in the interval $Z$=88-89 range and deep minimum in the $Z$= 84
for the curve with the $3d5g\ ^1D_2$ label  can be explained by
mixing of the $3d_{5/2}5g_{7/2}\ (2)$ and $3d_{5/2}5g_{9/2}\ (2)$
states.

Transition rates  for the seven E3 lines from $3d5p\ ^3F_{3},\
^1F_{3},\ ^3D_3$ and $3d5f\ ^3D_{3},\ ^3G_{3},\ ^3F_{3},\ ^1F_3$
levels to the ground state are plotted in the bottom panel of
Fig.~\ref{tr-e1}. The  deep minimum in the $Z$= 50  for the curve
with the $3d5f\ ^3G_3$ label  can be explained by mixing of the
$3d_{5/2}5f_{5/2}\ (3)$ and $3d_{5/2}5f_{7/2}\ (3)$ states,
however the deep minimum in the $Z$= 54  for the curve with the
$3d5f\ ^1F_3$ label  can be explained by mixing of the
$3d_{3/2}5f_{5/2}\ (3)$ and $3d_{3/2}5f_{7/2}\ (3)$ states.

Transition rates  for the seven M1 lines from $3d5d\ ^3S_1,\
^3P_1,\ ^3D_1,\ ^1P_1$, $3d5g\ ^3D_1$, $3d5s\ ^3D_1$, and $3s5s\
^3S_1$ levels to the ground state are plotted in the top panel of
Fig.~\ref{tr-m1}. The deep minima in the $Z$= 41  for the curve
with the $3d5d\ ^1P_1$ label can be explained by strong mixing
between $3d_{5/2}5d_{3/2}$ and $3d_{5/2}5d_{5/2}$ states.  The
value of $A_r$ for $3s5s\ ^3S_1$ line is smaller than the value of
$A_r$ for $3d5s\ ^3D_1$ line by factor 10$^2$ - 10$^4$.

Transition rates for the eight M2 lines from $3d5p\ ^3F_2,\
^3P_2,\ ^3D_2,\ ^1D_2$ and $3d5f\ ^3P_2,\ ^3D_2,\ ^1D_2,\ ^3F_2$
levels to the ground state are plotted in central panel of
Fig.~\ref{tr-m1}. We can see from these figures that  the curves
describing M2 transition rates, except curves with $3d5p\ ^1D_2$
and  $3d5f\ ^1D_2$ labels, smoothly increase with $Z$ without any
sharp features. It should be noted that the main part of
contribution to the transition rate of the $3d5p\ ^1D_2$ state
gives $0-3d_{3/2}5p_{3/2}\ (2)$ dipole matrix element. This matrix
element has zero value in the first-order approximation. The small
non-zero value for the transition rate of the $3d5p\ ^1D_2$ state
is due to the correlation second-order contribution and mixing
inside of the  $3d_{j}5p_{j'}\ (2)$ complex.

Transition rates for the nine M3 lines from $3d5s\ ^3D_3$, $3d5d\
^3G_3,\ ^3D_3,\ ^3F_3,\ ^1F_3$ and $3d5g\ ^3D_3,\ ^3F_3,\ ^1F_3,\
^3G_3$ levels to the ground state are plotted in bottom panel of
Fig.~\ref{tr-m1}. The sharp features in the curves shown in these
figures  can be explained in many cases by strong mixing of states
inside of the odd-parity $3d_{j}5d_{j'}\ (3)$ and $3d_{j}5g_{j'}\
(3)$ complexes.

In Table~\ref{tab-com3}, wavelengths ($\lambda$ in $\AA$) and
oscillator strengths $f$ for odd-parity states with $J$=1 are
illustrated for Ni-like ions. We limit the table to those
transitions given in Ref.~\cite{zhang}. Comparison of $f$ obtained
by RMBPT  and DFS codes are given. We use $LSJ$ labelling for data
with RMBPT headings and the M17 -- M22 from Table~I and Table~III
of Ref.~\cite{zhang} for data with DFS headings.
 As can be seen from Table~\ref{tab-com3},
the difference between both results is about 5 - 20\%.  This
difference  can be explained by the second order contribution
included in our RMBPT calculations since results in
Refs.~\cite{zhang} were obtained in MCDF approximations. To
support this conclusion, we include values for oscillator
strengths   calculated in the first-order approximation in
Table~\ref{tab-com3} (column ''RMBPT1''). We can see from this
table that DFS data  better agree with results of the first-order
approximation  (RMBPT1) than with RMBPT results.

In Table~\ref{tab-com4}, wavelengths ($\lambda$ in $\AA$) and
transition rates ($A$ in s$^{-1}$) for odd-parity states with
$J$=1 are listed for Ni-like xenon. We compare our results with
theoretical results obtained by Skobelev {\it et al.}  in
Ref.~\cite{skobelev}. We already mentioned that results obtained
by three methods (HFR, MCDF, and HULLAC)  were compared in
~\cite{skobelev}. Our results  better agree with results obtained
by the HULLAC code,  as is seen from Table~\ref{tab-com4}. It
should be noted that HULLAC results are between our RMBPT results
and results of the first-order approximation  (RMBPT1) (see
columns with headings 'RMBPT' and 'RMBPT1' in
Table~\ref{tab-com4}).

Transition energies  and transition rates for  the $3d_{5/2}\
-5f_{7/2}$ and $3d_{3/2}\ -5f_{5/2}$ transitions in Ni-like ions
with $Z$ = 56--92  are given in Table~\ref{tab-com5}. We limit the
table to those ions with available experimental measurements. We
compare our RMBPT  calculations with experimental measurements
presented in Refs.\cite{doron,zigler,doron-58,elliot,au-03}.
 It should be noted that our RMBPT data are in excellent agreement
 with experimental measurements presented by Elliot  {\it et al.}  in
 Ref.~\cite{elliot}.

\section{\bf Conclusion}

We have presented a systematic second-order relativistic RMBPT
study of excitation energies, reduced matrix elements, line
strengths, and transition rates for $\Delta n$=1 electric- and
magnetic-dipole,  electric- and magnetic-quadrupole, and electric-
and magnetic-octupole transitions in  Ni-like ions with nuclear
charges $Z$=30--100. Our calculations of the retarded E1, E2, E3,
M1, M2, and M3 matrix elements include correlation corrections
from both Coulomb and Breit interactions. Contributions from
virtual electron-positron pairs were also included in the
second-order matrix elements. Both length and velocity forms of
the E1, E2, and E2 matrix elements were evaluated and small
differences, caused by the non-locality of the starting DF
potential, were found between the two forms. Second-order RMBPT
transition energies were used to evaluate oscillator strengths and
transition rates. Good agreement of our RMBPT data with other
accurate theoretical results
 leads  us to conclude that the
RMBPT method provides  accurate data for Ni-like ions. Results
from the present calculations provide benchmark values for future
theoretical and experimental studies of the nickel isoelectronic
sequence.

\section*{\bf Acknowledgments}
 The work  was supported in part by DOE-NNSA/NV
Cooperative Agreement DE-FC52-01NV14050.
  Work at the Lawrence Livermore
National Laboratory was performed under the auspices of the U.S.
Department of Energy  under Contract No. W-7405-Eng-48.

\begin{table}
\caption{Possible  hole-particle states in the $3l_{j}5l'_{j'}$
complex;
 $jj$~ and $LS$~coupling schemes}
\begin{ruledtabular}
\begin{tabular}{llllllllll}
\multicolumn{10}{c}{Odd-parity states}\\
\multicolumn{2}{c}{$J$=0,5,6}& \multicolumn{2}{c}{$J$=1}&
\multicolumn{2}{c}{$J$=2}& \multicolumn{2}{c}{$J$=3}&
\multicolumn{2}{c}{$J$=4\rule{0ex}{2.3ex}}\\
\multicolumn{1}{c}{$jj$~coupl.}&
\multicolumn{1}{c}{$LS$~coupl.}&
\multicolumn{1}{c}{$jj$~coupl.}&
\multicolumn{1}{c}{$LS$~coupl.}&
\multicolumn{1}{c}{$jj$~coupl.}&
\multicolumn{1}{c}{$LS$~coupl.}&
\multicolumn{1}{c}{$jj$~coupl.}&
\multicolumn{1}{c}{$LS$~coupl.}&
\multicolumn{1}{c}{$jj$~coupl.}&
\multicolumn{1}{c}{$LS$~coupl.\rule{0ex}{2.3ex}}\\
\hline
$3d_{3/2}5p_{3/2}(0)$&$3d5p\ ^3P_0$& $3d_{3/2}5p_{1/2}$&$3d5p\ ^3P$& $3d_{5/2}5p_{1/2}$&$3d5p\ ^3F$& $3d_{5/2}5p_{1/2}$&$3d5p\ ^3F$& $3d_{5/2}5p_{3/2}$&$3d5p\ ^3F$\\
$3d_{5/2}5f_{5/2}(0)$&$3d5f\ ^3P_0$& $3d_{5/2}5p_{3/2}$&$3d5p\ ^3D$& $3d_{5/2}5p_{3/2}$&$3d5p\ ^3P$& $3d_{5/2}5p_{3/2}$&$3d5p\ ^1F$& $3d_{5/2}5f_{5/2}$&$3d5f\ ^3H$\\
$3p_{1/2}5s_{1/2}(0)$&$3p5s\ ^3P_0$& $3d_{3/2}5p_{3/2}$&$3d5p\ ^1P$& $3d_{3/2}5p_{1/2}$&$3d5p\ ^1D$& $3d_{3/2}5p_{3/2}$&$3d5p\ ^3D$& $3d_{5/2}5f_{7/2}$&$3d5f\ ^3G$\\
$3p_{3/2}5d_{3/2}(0)$&$3p5d\ ^3P_0$& $3d_{5/2}5f_{5/2}$&$3d5f\ ^3P$& $3d_{3/2}5p_{3/2}$&$3d5p\ ^3D$& $3d_{5/2}5f_{5/2}$&$3d5f\ ^3D$& $3d_{3/2}5f_{5/2}$&$3d5f\ ^3F$\\
$3s_{1/2}5p_{1/2}(0)$&$3s5p\ ^3P_0$& $3d_{5/2}5f_{7/2}$&$3d5f\ ^3D$& $3d_{5/2}5f_{5/2}$&$3d5f\ ^3P$& $3d_{5/2}5f_{7/2}$&$3d5f\ ^3G$& $3d_{3/2}5f_{7/2}$&$3d5f\ ^1G$\\
$3d_{5/2}5f_{5/2}(5)$&$3d5f\ ^3H_5$& $3d_{3/2}5f_{5/2}$&$3d5f\ ^1P$& $3d_{5/2}5f_{7/2}$&$3d5f\ ^3D$& $3d_{3/2}5f_{5/2}$&$3d5f\ ^3F$& $3p_{3/2}5d_{5/2}$&$3p5d\ ^3F$\\
$3d_{5/2}5f_{7/2}(5)$&$3d5f\ ^3G_5$& $3p_{3/2}5s_{1/2}$&$3p5s\ ^1P$& $3d_{3/2}5f_{5/2}$&$3d5f\ ^1D$& $3d_{3/2}5f_{7/2}$&$3d5f\ ^1F$& $3p_{3/2}5g_{7/2}$&$3p5g\ ^3F$\\
$3d_{3/2}5f_{7/2}(5)$&$3d5f\ ^1H_5$& $3p_{1/2}5s_{1/2}$&$3p5s\ ^3P$& $3d_{3/2}5f_{7/2}$&$3d5f\ ^3F$& $3p_{3/2}5d_{3/2}$&$3p5d\ ^3F$& $3p_{3/2}5g_{9/2}$&$3p5g\ ^3H$\\
$3p_{3/2}5g_{7/2}(5)$&$3p5g\ ^3H_5$& $3p_{3/2}5d_{3/2}$&$3p5d\ ^3P$& $3p_{3/2}5s_{1/2}$&$3p5s\ ^3P$& $3p_{3/2}5d_{5/2}$&$3p5d\ ^1F$& $3p_{1/2}5g_{7/2}$&$3p5g\ ^3G$\\
$3p_{3/2}5g_{9/2}(5)$&$3p5g\ ^1H_5$& $3p_{3/2}5d_{5/2}$&$3p5d\ ^1P$& $3p_{3/2}5d_{3/2}$&$3p5d\ ^3F$& $3p_{1/2}5d_{5/2}$&$3p5d\ ^3D$& $3p_{1/2}5g_{9/2}$&$3p5g\ ^1G$\\
$3p_{1/2}5g_{9/2}(5)$&$3p5g\ ^3G_5$& $3p_{1/2}5d_{3/2}$&$3p5d\ ^3D$& $3p_{3/2}5d_{5/2}$&$3p5d\ ^3P$& $3p_{3/2}5g_{7/2}$&$3p5g\ ^3F$& $3s_{1/2}5f_{7/2}$&$3s5f\ ^3F$\\
$3d_{5/2}5f_{7/2}(6)$&$3d5f\ ^3H_6$& $3s_{1/2}5p_{1/2}$&$3s5p\ ^3P$& $3p_{1/2}5d_{3/2}$&$3p5d\ ^3D$& $3p_{3/2}5g_{9/2}$&$3p5g\ ^1F$&              &       \\
$3p_{3/2}5g_{9/2}(6)$&$3p5g\ ^3H_6$& $3s_{1/2}5p_{3/2}$&$3s5p\ ^1P$& $3p_{1/2}5d_{5/2}$&$3p5d\ ^1D$& $3p_{1/2}5g_{7/2}$&$3p5g\ ^3G$&              &       \\
                     &             &                   &           & $3p_{3/2}5g_{7/2}$&$3p5g\ ^3F$& $3s_{1/2}5f_{5/2}$&$3s5f\ ^3F$&              &       \\
                     &             &                   &           & $3s_{1/2}5p_{3/2}$&$3s5p\ ^3P$& $3s_{1/2}5f_{7/2}$&$3s5f\ ^1F$&              &       \\
                     &             &                   &           & $3s_{1/2}5f_{5/2}$&$3s5f\ ^3F$&                   &             &              &         \\
\hline
\multicolumn{10}{c}{even-parity states}\\
\multicolumn{2}{c}{$J$=0,5,6,7}&
\multicolumn{2}{c}{$J$=1}&
\multicolumn{2}{c}{$J$=2}&
\multicolumn{2}{c}{$J$=3}&
\multicolumn{2}{c}{$J$=4\rule{0ex}{2.3ex}}\\
\multicolumn{1}{c}{$jj$~coupl.}&
\multicolumn{1}{c}{$LS$~coupl.}&
\multicolumn{1}{c}{$jj$~coupl.}&
\multicolumn{1}{c}{$LS$~coupl.}&
\multicolumn{1}{c}{$jj$~coupl.}&
\multicolumn{1}{c}{$LS$~coupl.}&
\multicolumn{1}{c}{$jj$~coupl.}&
\multicolumn{1}{c}{$LS$~coupl.}&
\multicolumn{1}{c}{$jj$~coupl.}&
\multicolumn{1}{c}{$LS$~coupl.\rule{0ex}{2.3ex}}\\
\hline
$3d_{5/2}5d_{5/2}(0)$&$3d5d\ ^3P_0$& $3d_{3/2}5s_{1/2}$&$3d5s\ ^3D$& $3d_{5/2}5s_{1/2}$&$3d5s\ ^3D$& $3d_{5/2}5s_{1/2}$&$3d5s\ ^3D$& $3d_{5/2}5d_{3/2}$&$3d5d\ ^3G$ \\
$3d_{3/2}5d_{3/2}(0)$&$3d5d\ ^1S_0$& $3d_{5/2}5d_{3/2}$&$3d5d\ ^3S$& $3d_{3/2}5s_{1/2}$&$3d5s\ ^1D$& $3d_{5/2}5d_{3/2}$&$3d5d\ ^3G$& $3d_{5/2}5d_{5/2}$&$3d5d\ ^1G$ \\
$3p_{3/2}5p_{3/2}(0)$&$3p5p\ ^3P_0$& $3d_{5/2}5d_{5/2}$&$3d5d\ ^1P$& $3d_{5/2}5d_{3/2}$&$3d5d\ ^3P$& $3d_{5/2}5d_{5/2}$&$3d5d\ ^3D$& $3d_{3/2}5d_{5/2}$&$3d5d\ ^3F$ \\
$3p_{1/2}5p_{1/2}(0)$&$3p5p\ ^1S_0$& $3d_{3/2}5d_{3/2}$&$3d5d\ ^3D$& $3d_{5/2}5d_{5/2}$&$3d5d\ ^3D$& $3d_{3/2}5d_{3/2}$&$3d5d\ ^3F$& $3d_{5/2}5g_{7/2}$&$3d5g\ ^3F$ \\
$3s_{1/2}5s_{1/2}(0)$&$3s5s\ ^1S_0$& $3d_{3/2}5d_{5/2}$&$3d5d\ ^3P$& $3d_{3/2}5d_{3/2}$&$3d5d\ ^3F$& $3d_{3/2}5d_{5/2}$&$3d5d\ ^1F$& $3d_{5/2}5g_{9/2}$&$3d5g\ ^3H$ \\
$3d_{5/2}5d_{5/2}(5)$&$3p5f\ ^3G_5$& $3d_{5/2}5g_{7/2}$&$3d5g\ ^3D$& $3d_{3/2}5d_{5/2}$&$3d5d\ ^1D$& $3d_{5/2}5g_{7/2}$&$3d5g\ ^3D$& $3d_{3/2}5g_{7/2}$&$3d5g\ ^3G$ \\
$3d_{5/2}5g_{7/2}(5)$&$3d5g\ ^3I_5$& $3p_{3/2}5p_{1/2}$&$3p5p\ ^3D$& $3d_{5/2}5g_{7/2}$&$3d5g\ ^3D$& $3d_{5/2}5g_{9/2}$&$3d5g\ ^3F$& $3d_{3/2}5g_{9/2}$&$3d5g\ ^1G$ \\
$3d_{5/2}5g_{9/2}(5)$&$3d5g\ ^3H_5$& $3p_{3/2}5p_{3/2}$&$3p5p\ ^3S$& $3d_{5/2}5g_{9/2}$&$3d5g\ ^3F$& $3d_{3/2}5g_{7/2}$&$3d5g\ ^1F$& $3p_{3/2}5f_{5/2}$&$3p5f\ ^3G$ \\
$3d_{3/2}5g_{7/2}(5)$&$3d5g\ ^1H_5$& $3p_{1/2}5p_{1/2}$&$3p5p\ ^1P$& $3d_{3/2}5g_{7/2}$&$3d5g\ ^1D$& $3d_{3/2}5g_{9/2}$&$3d5g\ ^3G$& $3p_{3/2}5f_{7/2}$&$3p5f\ ^1G$ \\
$3d_{3/2}5g_{9/2}(5)$&$3d5g\ ^3G_5$& $3p_{1/2}5p_{3/2}$&$3p5p\ ^3P$& $3p_{3/2}5p_{1/2}$&$3p5p\ ^3D$& $3p_{3/2}5p_{3/2}$&$3p5p\ ^3D$& $3p_{1/2}5f_{7/2}$&$3p5f\ ^3F$ \\
$3p_{3/2}5f_{7/2}(5)$&$3p5f\ ^3G_5$& $3p_{3/2}5f_{5/2}$&$3p5f\ ^3D$& $3p_{3/2}5p_{3/2}$&$3p5p\ ^1D$& $3p_{3/2}5f_{5/2}$&$3p5f\ ^3D$& $3s_{1/2}5g_{7/2}$&$3s5g\ ^3G$ \\
$3s_{1/2}5g_{9/2}(5)$&$3s5g\ ^3G_5$& $3s_{1/2}5s_{1/2}$&$3s5s\ ^3S$& $3p_{1/2}5p_{3/2}$&$3p5p\ ^3P$& $3p_{1/2}5f_{5/2}$&$3p5f\ ^3G$& $3s_{1/2}5g_{9/2}$&$3s5g\ ^1G$ \\
$3d_{5/2}5g_{7/2}(6)$&$3d5g\ ^3I_6$& $3s_{1/2}5d_{3/2}$&$3s5d\ ^3D$& $3p_{3/2}5f_{5/2}$&$3p5f\ ^3D$& $3p_{3/2}5f_{7/2}$&$3p5f\ ^1F$&                   &        \\
$3d_{5/2}5g_{9/2}(6)$&$3d5g\ ^3H_6$&                   &           & $3p_{3/2}5f_{7/2}$&$3p5f\ ^1D$& $3p_{1/2}5f_{7/2}$&$3p5f\ ^3F$&                   &        \\
$3d_{3/2}5g_{9/2}(6)$&$3d5g\ ^1I_6$&                   &           & $3p_{1/2}5f_{5/2}$&$3p5f\ ^3F$& $3s_{1/2}5d_{5/2}$&$3s5d\ ^3D$&                   &        \\
$3d_{5/2}5g_{9/2}(7)$&$3d5g\ ^3I_7$&                   &           & $3s_{1/2}5d_{3/2}$&$3s5d\ ^3D$& $3s_{1/2}5g_{7/2}$&$3s5g\ ^3G$&                   &        \\
                     &             &                   &           & $3s_{1/2}5d_{5/2}$&$3s5d\ ^1D$&                   &           &                   &        \\
\end{tabular}
\end{ruledtabular}
\label{tab0}
\end{table}

\begin{table}
\caption{ Second-order contributions to the energy matrices
(a.u.)\
 for  odd-parity states with
$J$=1 in the case of Ni-like tungsten, $Z=74$. One-body and
two-body second-order Coulomb and Breit-Coulomb contributions are
given in columns labeled $E^{(2)}_{1}$, $E^{(2)}_{2}$,
$B^{(2)}_{1}$, and $B^{(2)}_{2}$, respectively. }
\begin{ruledtabular}
\begin{tabular}{llrrrr}
\multicolumn{2}{c} {}& \multicolumn{2}{l}{Coulomb Interaction:}&
\multicolumn{2}{l}{Breit-Coulomb Correction:} \\
\multicolumn{2}{c}{$3l_1j_1\ 5l_2j_2,3l_3j_3\ 5l_4j_4$} &
\multicolumn{1}{c}{$E^{(2)}_{1}$} &
\multicolumn{1}{c}{$E^{(2)}_{2}$}&
\multicolumn{1}{c}{$B^{(2)}_{1}$} &
\multicolumn{1}{c}{$B^{(2)}_{2}$}\\
\hline
 $3d_{3/2}5p_{1/2}$&$3d_{3/2}5p_{1/2}$&  -0.139676&   0.019347&   0.067240&   0.002331\\
 $3d_{5/2}5f_{5/2}$&$3d_{5/2}5f_{5/2}$&  -0.118423&   0.016129&   0.067375&   0.001401\\
 $3p_{3/2}5s_{1/2}$&$3p_{3/2}5s_{1/2}$&  -0.205416&   0.010692&   0.058375&   0.001411\\
 $3p_{3/2}5d_{3/2}$&$3p_{3/2}5d_{3/2}$&  -0.206632&   0.125851&   0.057029&   0.003750\\
 $3p_{3/2}5d_{5/2}$&$3p_{3/2}5d_{5/2}$&  -0.204386&   0.018147&   0.057244&   0.000973\\
 $3s_{1/2}5p_{1/2}$&$3s_{1/2}5p_{1/2}$&  -0.275748&   0.030048&   0.056515&   0.002185\\
 $3p_{3/2}5d_{3/2}$&$3s_{1/2}5p_{3/2}$&   0.000000&  -0.035615&   0.000000&   0.000660\\
 $3s_{1/2}5p_{3/2}$&$3p_{3/2}5d_{3/2}$&   0.000000&   0.011785&   0.000000&  -0.000227\\
\end{tabular}
\end{ruledtabular}
\label{tab2}
\end{table}

\begin{table}
\caption{ Contributions to the energy matrix
 $E[3l_1j_1\ 5l_2j_2(J),3l_3j_3\ 5l_4j_4(J)]$
=$E^{(0)}$+$E^{(1)}$+$E^{(2)}$+$B^{(1)}_{hf}$+$B^{(2)}$ before
diagonalization. These contributions are given for a hole-particle
ion with a $1s^22s^22p^63s^23p^63d^{10}$ core, in the case of
odd-parity states with $J$=1, and $Z=74$.}
\begin{ruledtabular}\begin{tabular}{llrrrrr}
\multicolumn{2}{c}{$3l_1j_1\ 5l_2j_2,3l_3j_3\ 5l_4j_4$} &
\multicolumn{1}{c}{$E^{(0)}$} & \multicolumn{1}{c}{$E^{(1)}$} &
\multicolumn{1}{c}{$B^{(1)}_{hf}$}& \multicolumn{1}{c}{$E^{(2)}$}&
\multicolumn{1}{c}{$B^{(2)}$} \\
\hline
 $3d_{3/2}5p_{1/2}$&$3d_{3/2}5p_{1/2}$& 100.189697&   -1.830819&  -0.163806&  -0.120330&   0.069571\\
 $3d_{5/2}5f_{5/2}$&$3d_{5/2}5f_{5/2}$& 105.118128&   -1.955034&  -0.151975&  -0.102294&   0.068776\\
 $3p_{3/2}5s_{1/2}$&$3p_{3/2}5s_{1/2}$& 112.576761&   -1.788302&  -0.213595&  -0.194724&   0.059787\\
 $3p_{3/2}5d_{3/2}$&$3p_{3/2}5d_{3/2}$& 118.739109&   -1.829829&  -0.219091&  -0.080782&   0.060779\\
 $3p_{3/2}5d_{5/2}$&$3p_{3/2}5d_{5/2}$& 119.130345&   -1.795553&  -0.231362&  -0.186238&   0.058218\\
 $3s_{1/2}5p_{1/2}$&$3s_{1/2}5p_{1/2}$& 134.581590&   -1.827208&  -0.202851&  -0.245700&   0.058700\\
 $3p_{3/2}5d_{3/2}$&$3s_{1/2}5p_{3/2}$&   0.000000&    0.073512&   0.001228&  -0.035615&   0.000660\\
 $3s_{1/2}5p_{3/2}$&$3p_{3/2}5d_{3/2}$&   0.000000&    0.073512&   0.001228&   0.011785&  -0.000227\\
\end{tabular}
\end{ruledtabular}
\label{tab3}
\end{table}

\begin{table}
\caption{Energies  of Ni-like tungsten for odd-parity states with
$J$=1 relative to the ground state. $E^{(0+1)} \equiv E^{(0)} +
E^{(1)} + B_{hf}^{(1)}$ }
\begin{ruledtabular}\begin{tabular}{llrrrrr}
\multicolumn{1}{c} {$jj$ coupl.} & \multicolumn{1}{c} {$LS$
coupl.} & \multicolumn{1}{c}{$ E^{(0+1)} $} & \multicolumn{1}{c}{$
E^{(2)} $} & \multicolumn{1}{c}{$ B^{(2)} $} &
\multicolumn{1}{c}{$E_{\rm LAMB}$}&
\multicolumn{1}{c}{$E_{\rm tot}$} \\
\hline
$3d_{3/2}5p_{1/2}$&$3d5p\ ^3P$&  97.513893&  -0.106270&   0.067045&  -0.000005&  97.474663\\
$3d_{5/2}5p_{3/2}$&$3d5p\ ^3D$&  98.192405&  -0.117845&   0.069535&   0.005535&  98.149631\\
$3d_{3/2}5p_{3/2}$&$3d5p\ ^1P$&  99.945568&  -0.110473&   0.069717&   0.007671&  99.912484\\
$3d_{5/2}5f_{5/2}$&$3d5f\ ^3P$& 102.970560&  -0.104892&   0.068378&  -0.004300& 102.929746\\
$3d_{5/2}5f_{7/2}$&$3d5f\ ^3D$& 103.514261&  -0.105673&   0.068308&  -0.003590& 103.473306\\
$3d_{3/2}5f_{5/2}$&$3d5f\ ^1P$& 105.804274&  -0.120211&   0.070741&   0.003184& 105.757987\\
$3p_{3/2}5s_{1/2}$&$3p5s\ ^1P$& 110.577497&  -0.195522&   0.059780&   0.001186& 110.442942\\
$3p_{1/2}5s_{1/2}$&$3p5s\ ^3P$& 116.681143&  -0.188040&   0.058691&  -0.026832& 116.524962\\
$3p_{3/2}5d_{3/2}$&$3p5d\ ^3P$& 117.108814&  -0.186591&   0.058643&  -0.025649& 116.955217\\
$3p_{3/2}5d_{5/2}$&$3p5d\ ^1P$& 121.897136&  -0.239625&   0.066817&   0.014855& 121.739183\\
$3p_{1/2}5d_{3/2}$&$3p5d\ ^3D$& 128.067580&  -0.239058&   0.065662&  -0.013400& 127.880785\\
$3s_{1/2}5p_{1/2}$&$3s5p\ ^3P$& 132.564903&  -0.259297&   0.058640&  -0.168955& 132.195290\\
$3s_{1/2}5p_{3/2}$&$3s5p\ ^1P$& 134.381627&  -0.255693&   0.058780&  -0.167081& 134.017633\\
\end{tabular}
\end{ruledtabular}
\label{tab4}
\end{table}

\begin{table}
\caption{E1, E2, and E3 uncoupled reduced matrix elements in length $L$
and velocity $V$ forms for transitions from $av(J)$  states with
$J$=1 into the ground state in W$^{46+}$.}
\begin{ruledtabular}\begin{tabular}{lrrrrrrrr}
\multicolumn{1}{c}{$av(J)$}& \multicolumn{1}{c}{$Z^{(1)}_L$} &
\multicolumn{1}{c}{$Z^{(1)}_V$} & \multicolumn{1}{c}{$Z^{(2)}_L$}
& \multicolumn{1}{c}{$Z^{(2)}_V$} &
\multicolumn{1}{c}{$B^{(2)}_L$} & \multicolumn{1}{c}{$B^{(2)}_V$}
& \multicolumn{1}{c}{$P^{(\rm derv)}_L$} &
\multicolumn{1}{c}{$P^{(\rm derv)}_V$}\\[0.25pc]
\hline
\multicolumn{9}{c}{ E1 uncoupled reduced matrix elements } \\
$3d_{3/2}5p_{1/2}(1)$&   0.019680&  0.018604&  0.001566&  0.001632&  0.000009& -0.000050&  0.019582& -0.000028\\
$3d_{5/2}5f_{5/2}(1)$&   0.027924&  0.026556& -0.001042&  0.000168&  0.000066& -0.000025&  0.028039&  0.000295\\
$3p_{3/2}5s_{1/2}(1)$&  -0.029022& -0.027515& -0.002520& -0.002775& -0.000138& -0.000038& -0.028886& -0.000053\\
$3p_{1/2}5s_{1/2}(1)$&  -0.015155& -0.014409& -0.000913& -0.000950& -0.000091& -0.000029& -0.014932&  0.000152\\
$3p_{3/2}5d_{3/2}(1)$&   0.023649&  0.022517& -0.084802& -0.080260&  0.001803&  0.001642&  0.023594&  0.000129\\
$3s_{1/2}5p_{1/2}(1)$&  -0.025272& -0.024060& -0.001166& -0.001529& -0.000369& -0.000284& -0.025044&  0.000012\\
\multicolumn{9}{c}{ E2 uncoupled reduced matrix elements } \\
$3d_{5/2}5s_{1/2}(2)$&  -0.006168& -0.005738& -0.000223& -0.000251& -0.000016& -0.000004& -0.012377& -0.005790\\
$3d_{5/2}5d_{3/2}(2)$&   0.005121&  0.004825&  0.000148&  0.000208&  0.000012&  0.000003&  0.010268&  0.004875\\
$3d_{5/2}5g_{7/2}(2)$&   0.034683&  0.033504&  0.000921&  0.001359&  0.000149&  0.000078&  0.068911&  0.033144\\
$3p_{3/2}5p_{1/2}(2)$&   0.008878&  0.008378& -0.000119&  0.000002&  0.000030&  0.000010&  0.017789&  0.008438\\
$3p_{3/2}5f_{5/2}(2)$&  -0.008588& -0.008137& -0.001304& -0.001494& -0.000023& -0.000002& -0.017229& -0.008203\\
$3s_{1/2}5d_{5/2}(2)$&   0.014261&  0.013610&  0.000417&  0.000618&  0.000030&  0.000005&  0.028377&  0.013437\\
\multicolumn{9}{c}{ E3 uncoupled reduced matrix elements } \\
$3d_{5/2}5p_{1/2}(3)$&   0.000951&  0.000946&  0.000021&  0.000020&  0.000010&  0.000008&  0.003216&  0.001918\\
$3d_{5/2}5f_{5/2}(3)$&  -0.000665& -0.000717&  0.000048&  0.000015&  0.000000&  0.000002& -0.002439& -0.001461\\
$3p_{3/2}5d_{3/2}(3)$&  -0.002783& -0.002747& -0.000228& -0.000201& -0.000012& -0.000005& -0.008823& -0.005529\\
$3p_{3/2}5g_{7/2}(3)$&  -0.010030& -0.009580& -0.001309& -0.001357& -0.000074& -0.000044& -0.029489& -0.019033\\
$3s_{1/2}5f_{7/2}(3)$&   0.003416&  0.002857&  0.000430&  0.000436& -0.000004& -0.000009&  0.009114&  0.005713\\
\end{tabular}
\end{ruledtabular}
\label{tab-euncop}
\end{table}

\begin{table}
\caption{E1, E2, and E2 coupled reduced matrix elements in length
$L$ and velocity $V$ forms for transitions from $av(J)$  states
into the ground state in W$^{46+}$.}
\begin{ruledtabular}\begin{tabular}{llrrrr}
\multicolumn{2}{c}{}& \multicolumn{2}{c}{First order}&
\multicolumn{2}{c}{RMBPT}\\
\multicolumn{1}{c}{$av(J)$}& \multicolumn{1}{c}{$av(LSJ)$}&
\multicolumn{1}{c}{$L$} & \multicolumn{1}{c}{$V$} &
\multicolumn{1}{c}{$L$} &
\multicolumn{1}{c}{$V$} \\[0.25pc]
\hline
\multicolumn{6}{c}{ E1 coupled reduced matrix elements } \\
$3d_{5/2}5f_{7/2}(1)$&$3d5f\ ^3D_1$&  -0.111022&  -0.105589&  -0.107077&  -0.107826\\
$3d_{5/2}5f_{5/2}(1)$&$3d5f\ ^3P_1$&   0.002220&   0.002113&   0.003493&   0.003475\\
$3p_{3/2}5d_{3/2}(1)$&$3p5d\ ^3P_1$&   0.070874&   0.067571&   0.062757&   0.062619\\
$3p_{3/2}5d_{5/2}(1)$&$3p5d\ ^1P_1$&   0.013898&   0.013212&   0.014071&   0.013734\\
$3s_{1/2}5p_{1/2}(1)$&$3s5p\ ^3P_1$&   0.022564&   0.021480&   0.024998&   0.024575\\
\multicolumn{6}{c}{ E2 coupled reduced matrix elements } \\
$3d_{5/2}5d_{5/2}(2)$&$3d5d\ ^3D_2$&  -0.010513&  -0.009927&  -0.010655&  -0.010415\\
$3d_{5/2}5g_{7/2}(2)$&$3d5g\ ^3D_2$&   0.005729&   0.005528&   0.005828&   0.005803\\
$3p_{3/2}5f_{5/2}(2)$&$3p5f\ ^3D_2$&  -0.023500&  -0.022286&  -0.018038&  -0.017947\\
$3p_{1/2}5f_{5/2}(2)$&$3p5f\ ^3F_2$&   0.017059&   0.016240&   0.017606&   0.017423\\
$3s_{1/2}5d_{3/2}(2)$&$3s5d\ ^3D_2$&  -0.010691&  -0.010194&  -0.010651&  -0.010474\\
$3s_{1/2}5d_{5/2}(2)$&$3s5d\ ^1D_2$&   0.014352&   0.013700&   0.014628&   0.014402\\
\multicolumn{6}{c}{ E3 coupled reduced matrix elements } \\
$3d_{5/2}5p_{1/2}(3)$&$3d5p\ ^3F_3$&  -0.001036&  -0.001027&  -0.001040&  -0.001072\\
$3d_{5/2}5p_{3/2}(3)$&$3d5p\ ^1F_3$&   0.001335&   0.001260&   0.001364&   0.001324\\
$3p_{3/2}5d_{3/2}(3)$&$3p5d\ ^3F_3$&  -0.003072&  -0.003012&  -0.003292&  -0.003279\\
$3p_{3/2}5g_{7/2}(3)$&$3p5g\ ^3F_3$&   0.008368&   0.008151&   0.008444&   0.008468\\
$3p_{1/2}5g_{7/2}(3)$&$3p5g\ ^3G_3$&   0.009924&   0.009479&   0.011368&   0.011112\\
\end{tabular}
\end{ruledtabular}
\label{tab-ecop}
\end{table}

\begin{table}
\caption{ Wavelengths ( ${\lambda}$ in \AA\ ) and multipole (E1,
E2, E3, M1, M2, and M3) transition rates ($A$ in s$^{-1}$) for
Ni-like tungsten with nuclear charge $Z$=74. Numbers in brackets
represent powers of 10.}
\begin{ruledtabular}\begin{tabular}{lrllrllrl}
\multicolumn{1}{c}{level}&
\multicolumn{1}{c}{${\lambda}$}&
\multicolumn{1}{c}{$A^{E1}$}&
\multicolumn{1}{c}{level}&
\multicolumn{1}{c}{${\lambda}$}&
\multicolumn{1}{c}{$A^{M2}$}&
\multicolumn{1}{c}{level}&
\multicolumn{1}{c}{${\lambda}$}&
\multicolumn{1}{c}{$A^{E3}$}\\
\hline
$3d5p\ ^3P_1$&  4.674&   4.758[12]&$3d5p\ ^3F_2$&   4.763&   4.544[06]&$3d5p\ ^3F_3$&   4.763&   8.742[05]\\
$3d5p\ ^3D_1$&  4.642&   4.105[12]&$3d5p\ ^3P_2$&   4.675&   1.409[07]&$3d5p\ ^1F_3$&   4.672&   1.719[06]\\
$3d5p\ ^1P_1$&  4.560&   6.106[11]&$3d5p\ ^1D_2$&   4.645&   5.359[05]&$3d5p\ ^3D_3$&   4.560&   3.615[06]\\
$3d5f\ ^3P_1$&  4.427&   9.500[10]&$3d5p\ ^3D_2$&   4.558&   6.081[03]&$3d5f\ ^3D_3$&   4.421&   1.586[06]\\
$3d5f\ ^3D_1$&  4.403&   9.069[13]&$3d5f\ ^3P_2$&   4.423&   6.767[07]&$3d5f\ ^3G_3$&   4.415&   2.294[06]\\
$3d5f\ ^1P_1$&  4.308&   1.157[14]&$3d5f\ ^3D_2$&   4.419&   8.102[08]&$3d5f\ ^3F_3$&   4.317&   3.151[06]\\
$3p5s\ ^1P_1$&  4.126&   8.472[12]&$3d5f\ ^1D_2$&   4.320&   1.781[07]&$3d5f\ ^1F_3$&   4.314&   8.873[05]\\
$3p5s\ ^3P_1$&  3.910&   5.510[13]&$3d5f\ ^3F_2$&   4.317&   1.884[08]&$3p5d\ ^3F_3$&   3.910&   3.487[07]\\
$3p5d\ ^3P_1$&  3.896&   4.499[13]&$3p5s\ ^3P_2$&   4.127&   3.525[07]&$3p5d\ ^1F_3$&   3.894&   2.029[07]\\
$3p5d\ ^1P_1$&  3.743&   2.551[12]&$3p5d\ ^3F_2$&   3.908&   2.357[06]&$3p5d\ ^3D_3$&   3.552&   2.001[09]\\
$3p5d\ ^3D_1$&  3.563&   3.665[13]&$3p5d\ ^3P_2$&   3.896&   2.299[08]&$3p5g\ ^3F_3$&   3.762&   3.001[08]\\
$3s5p\ ^3P_1$&  3.447&   1.031[13]&$3p5d\ ^3D_2$&   3.564&   1.179[03]&$3p5g\ ^1F_3$&   3.548&   6.245[07]\\
$3s5p\ ^1P_1$&  3.400&   1.064[13]&$3p5d\ ^1D_2$&   3.553&   3.084[06]&$3p5g\ ^3G_3$&   3.440&   1.018[09]\\
             &       &            &$3p5g\ ^3F_2$&   3.548&   5.371[07]&$3s5f\ ^3F_3$&   3.265&   1.676[08]\\
             &       &            &$3s5p\ ^3P_2$&   3.401&   6.153[07]&$3s5f\ ^1F_3$&   3.262&   7.520[07]\\
             &       &            &$3s5f\ ^3F_2$&   3.266&   7.466[04]&             &        &            \\
\hline \multicolumn{1}{c}{level}& \multicolumn{1}{c}{${\lambda}$}&
\multicolumn{1}{c}{$A^{M1}$}& \multicolumn{1}{c}{level}&
\multicolumn{1}{c}{${\lambda}$}& \multicolumn{1}{c}{$A^{E2}$}&
\multicolumn{1}{c}{level}& \multicolumn{1}{c}{${\lambda}$}&
\multicolumn{1}{c}{$A^{M3}$}\\
\hline
$3d5s\ ^3D_1$&  4.727&  1.218[04]&$3d5s\ ^3D_2$&   4.850&   3.430[09]&$3d5s\ ^3D_3$&   4.851&   1.506[04]\\
$3d5d\ ^3S_1$&  4.557&  4.287[07]&$3d5s\ ^1D_2$&   4.727&   2.516[09]&$3d5d\ ^3G_3$&   4.552&   2.151[03]\\
$3d5d\ ^1P_1$&  4.538&  3.668[07]&$3d5d\ ^3P_2$&   4.553&   2.160[09]&$3d5d\ ^3D_3$&   4.534&   1.102[05]\\
$3d5d\ ^3D_1$&  4.446&  1.375[07]&$3d5d\ ^3D_2$&   4.534&   1.328[10]&$3d5d\ ^3F_3$&   4.446&   1.606[03]\\
$3d5d\ ^3P_1$&  4.430&  1.312[07]&$3d5d\ ^3F_2$&   4.442&   7.777[09]&$3d5d\ ^1F_3$&   4.426&   9.004[02]\\
$3d5g\ ^3D_1$&  4.361&  4.381[05]&$3d5d\ ^1D_2$&   4.427&   3.333[09]&$3d5g\ ^3D_3$&   4.358&   1.869[05]\\
$3p5p\ ^3D_1$&  4.064&  3.318[08]&$3d5g\ ^3D_2$&   4.360&   4.831[09]&$3d5g\ ^3F_3$&   4.357&   1.649[06]\\
$3p5p\ ^3S_1$&  4.000&  3.212[07]&$3d5g\ ^3F_2$&   4.357&   2.867[11]&$3d5g\ ^1F_3$&   4.258&   9.997[04]\\
$3p5p\ ^1P_1$&  3.816&  1.449[06]&$3d5g\ ^1D_2$&   4.259&   1.994[11]&$3d5g\ ^3G_3$&   4.257&   5.703[05]\\
$3p5p\ ^3P_1$&  3.691&  3.345[06]&$3p5p\ ^3D_2$&   4.063&   1.539[10]&$3p5p\ ^3D_3$&   3.999&   2.067[05]\\
$3p5f\ ^3D_1$&  3.638&  2.955[08]&$3p5p\ ^1D_2$&   3.997&   2.099[10]&$3p5f\ ^3D_3$&   3.811&   1.402[04]\\
$3s5s\ ^3S_1$&  3.493&  3.933[06]&$3p5p\ ^3P_2$&   3.813&   2.537[09]&$3p5f\ ^3G_3$&   3.809&   1.089[06]\\
$3s5d\ ^3D_1$&  3.336&  1.926[05]&$3p5f\ ^3D_2$&   3.806&   9.129[10]&$3p5f\ ^1F_3$&   3.484&   2.283[04]\\
             &       &           &$3p5f\ ^1D_2$&   3.637&   1.471[10]&$3p5f\ ^3F_3$&   3.480&   6.181[05]\\
             &       &           &$3p5f\ ^3F_2$&   3.481&   1.358[11]&$3s5d\ ^3D_3$&   3.326&   9.078[05]\\
             &       &           &$3s5d\ ^3D_2$&   3.336&   6.152[10]&             &        &            \\
             &       &           &$3s5d\ ^1D_2$&   3.326&   1.178[11]&             &        &           \\
\end{tabular}
\end{ruledtabular}
\label{tab-ar}
\end{table}

\begin{table}
\caption{ Energies (eV)  of odd-parity states relative
to the ground state in Ni-like tungsten. Comparison RMBPT data
with theoretical results from DFS code by Zhang {\it et al.} in
Ref.~\protect\cite{zhang} and from COWAN code~\protect\cite{web}.}
\begin{ruledtabular}\begin{tabular}{llllllllllllll}
\multicolumn{1}{c}{Level}&
\multicolumn{1}{c}{RMBPT}&
\multicolumn{1}{c}{DFS}&
\multicolumn{1}{c}{COWAN}&
\multicolumn{1}{c}{Level}&
\multicolumn{1}{c}{RMBPT}&
\multicolumn{1}{c}{DFS}&
\multicolumn{1}{c}{COWAN}&
\multicolumn{1}{c}{Level}&
\multicolumn{1}{c}{RMBPT}&
\multicolumn{1}{c}{DFS}&
\multicolumn{1}{c}{COWAN}\\
\hline
$3d5p\ ^3P_0$&  2717.2& 2715.5&  2721.9&$3d5p\ ^3F_2$&   2602.9&  2601.6&  2604.0&$3d5p\ ^3F_3$&  2603.3&  2602.9&     2604.4\\
$3d5f\ ^3P_0$&  2799.6& 2798.1&  2801.3&$3d5p\ ^3P_2$&   2652.3&  2650.9&  2655.4&$3d5p\ ^1F_3$&  2653.6&  2652.3&     2656.6\\
$3p5s\ ^3P_0$&  3169.1& 3170.7&  3161.2&$3d5p\ ^1D_2$&   2669.1&  2665.2&  2670.0&$3d5p\ ^3D_3$&  2718.7&  2717.8&     2724.0\\
$3p5d\ ^3P_0$&  3312.1& 3314.2&  3320.5&$3d5p\ ^3D_2$&   2719.9&  2718.6&  2724.7&$3d5f\ ^3D_3$&  2804.2&  2802.7&     2805.4\\
$3s5p\ ^3P_0$&  3596.6& 3602.2&  3598.3&$3d5f\ ^3P_2$&   2802.9&  2801.4&  2804.3&$3d5f\ ^3G_3$&  2808.4&  2806.9&     2809.6\\
$3d5p\ ^3P_1$&  2652.4& 2655.3&  2655.7&$3d5f\ ^3D_2$&   2805.8&  2804.2&  2807.3&$3d5f\ ^3F_3$&  2872.0&  2870.4&     2874.7\\
$3d5p\ ^3D_1$&  2670.8& 2669.5&  2673.4&$3d5f\ ^1D_2$&   2869.7&  2868.1&  2872.8&$3d5f\ ^1F_3$&  2874.2&  2872.5&     2877.2\\
$3d5p\ ^1P_1$&  2718.8& 2717.3&  2723.5&$3d5f\ ^3F_2$&   2871.8&  2870.1&  2874.9&$3p5d\ ^3F_3$&  3171.3&  3172.9&     3163.1\\
$3d5f\ ^3P_1$&  2800.9& 2799.4&  2802.6&$3p5s\ ^3P_2$&   3004.1&  3005.5&  2997.1&$3p5d\ ^1F_3$&  3183.8&  3185.4&     3175.7\\
$3d5f\ ^3D_1$&  2815.7& 2814.9&  2816.6&$3p5d\ ^3F_2$&   3172.6&  3174.3&  3164.2&$3p5d\ ^3D_3$&  3490.2&  3492.6&     3498.0\\
$3d5f\ ^1P_1$&  2877.8& 2877.2&  2880.4&$3p5d\ ^3P_2$&   3182.4&  3184.0&  3174.5&$3s5f\ ^3F_3$&  3797.0&  3802.2&     3799.0\\
$3p5s\ ^1P_1$&  3005.3& 3006.7&  2998.1&$3p5d\ ^3D_2$&   3479.1&  3481.6&  3486.5&$3s5f\ ^1F_3$&  3801.3&  3806.7&     3803.3\\
$3p5s\ ^3P_1$&  3170.8& 3172.5&  3162.7&$3p5d\ ^1D_2$&   3489.8&  3492.3&  3497.6&$3d5p\ ^3F_4$&  2651.5&  2650.1&     2654.8\\
$3p5d\ ^3P_1$&  3182.5& 3184.1&  3174.6&$3s5p\ ^3P_2$&   3645.8&  3651.1&  3649.5&$3d5f\ ^3H_4$&  2804.6&  2801.1&     2805.8\\
$3p5d\ ^1P_1$&  3312.7& 3314.8&  3321.0&$3s5f\ ^3F_2$&   3796.2&  3801.7&  3798.4&$3d5f\ ^3G_4$&  2807.5&  2806.0&     2808.9\\
$3p5d\ ^3D_1$&  3479.8& 3482.4&  3487.3&$3d5f\ ^3H_5$&   2802.4&  2800.9&  2804.0&$3d5f\ ^3F_4$&  2868.9&  2867.3&     2872.1\\
$3s5p\ ^3P_1$&  3597.2& 3602.7&  3598.8&$3d5f\ ^3G_5$&   2808.1&  2806.6&  2809.5&$3d5f\ ^1G_4$&  2874.7&  2873.1&     2877.7\\
$3s5p\ ^1P_1$&  3646.8& 3652.2&  3650.3&$3d5f\ ^1H_5$&   2872.8&  2871.2&  2876.1&$3p5d\ ^3F_4$&  3181.7&  3183.2&     3173.9\\
             &        &       &        &$3d5f\ ^3H_6$&   2805.1&  2803.5&  2806.9&$3s5f\ ^3F_4$&  3799.6&  3804.9&     3801.9\\
\end{tabular}
\end{ruledtabular}
\label{tab-com1}
\end{table}

\begin{table}

\caption{ Wavelengths ($\lambda$ in $\AA$) and oscillator
strengths $f$ for Ni-like ions for odd-parity states with $J$=1.
The RMBPT and RMBPT1 (first-order
approximation) oscillator strengths are compared with theoretical data
from DFS code by
Zhang {\it et al.}
presented in Ref.~\protect\cite{zhang}.}
\begin{ruledtabular}\begin{tabular}{lllllllllllll}
\multicolumn{1}{c}{}&
\multicolumn{2}{c}{RMBPT}&
\multicolumn{1}{c}{RMBPT1}&
\multicolumn{1}{c}{DFS}&
\multicolumn{2}{c}{RMBPT}&
\multicolumn{1}{c}{RMBPT1}&
\multicolumn{1}{c}{DFS}&
\multicolumn{2}{c}{RMBPT}&
\multicolumn{1}{c}{RMBPT1}&
\multicolumn{1}{c}{DFS}\\
\multicolumn{1}{c}{}&
\multicolumn{1}{c}{$\lambda$}&
\multicolumn{1}{c}{$f$}&
\multicolumn{1}{c}{$f$}&
\multicolumn{1}{c}{$f$}&
\multicolumn{1}{c}{$\lambda$}&
\multicolumn{1}{c}{$f$}&
\multicolumn{1}{c}{$f$}&
\multicolumn{1}{c}{$f$}&
\multicolumn{1}{c}{$\lambda$}&
\multicolumn{1}{c}{$f$}&
\multicolumn{1}{c}{$f$}&
\multicolumn{1}{c}{$f$}\\
\multicolumn{1}{c}{$Z$}&
\multicolumn{3}{c}{$3d5f\ ^3D_{1}$}&
\multicolumn{1}{c}{M17+18}&
\multicolumn{3}{c}{$3d5f\ ^1P_{1}$}&
\multicolumn{1}{c}{M19}&
\multicolumn{3}{c}{$3p5d\ ^1P_{1}$}&
\multicolumn{1}{c}{M22}\\
\hline
 60&   8.146&  0.5772&  0.6286&  0.7423&   8.023&  1.2070&  1.3320&  1.4313&   6.990&  0.3922&  0.3691&  0.3821\\
 61&   7.740&  0.5970&  0.6501&  0.8370&   7.620&  1.1870&  1.3070&  1.4734&   6.663&  0.3840&  0.3706&  0.3842\\
 62&   7.364&  0.6541&  0.6709&  0.8814&   7.247&  1.2170&  1.2840&  1.4721&   6.359&  0.6628&  0.3727&  0.3865\\
 63&   7.015&  0.6358&  0.6907&  0.5653&   6.901&  1.1460&  1.2600&  1.4576&   6.075&  0.3982&  0.3749&  0.3890\\
 64&   6.691&  0.6544&  0.7096&  0.6774&   6.579&  1.1270&  1.2380&  1.4370&   5.809&  0.4125&  0.3771&  0.3914\\
 65&   6.389&  0.6695&  0.7275&  0.7304&   6.279&  1.1040&  1.2160&  1.4139&   5.561&  0.4117&  0.3792&  0.3937\\
 66&   6.107&  0.6884&  0.7447&  0.7666&   6.000&  1.0910&  1.1940&  1.3896&   5.328&  0.4297&  0.3813&  0.3958\\
 67&   5.844&  0.7030&  0.7608&  0.7953&   5.739&  1.0710&  1.1740&  1.3650&   5.110&  0.4642&  0.3831&  0.3978\\
 68&   5.598&  0.7121&  0.7761&  0.8196&   5.494&  1.0570&  1.1540&  1.3405&   4.904&  0.0878&  0.3848&  0.3996\\
 69&   5.367&  0.7311&  0.7905&  0.8408&   5.265&  1.0400&  1.1360&  1.3164&   4.711&  0.3487&  0.3864&  0.4012\\
 70&   5.150&  0.7448&  0.8040&  0.8597&   5.050&  1.0240&  1.1170&  1.2930&   4.529&  0.3761&  0.3877&  0.4026\\
 71&   4.946&  0.7575&  0.8168&  0.8765&   4.847&  1.0090&  1.1000&  1.2702&   4.357&  0.3894&  0.3890&  0.4039\\
 72&   4.755&  0.7695&  0.8288&  0.8916&   4.657&  0.9951&  1.0830&  1.2481&   4.195&  0.4006&  0.3901&  0.4051\\
 73&   4.574&  0.7808&  0.8400&  0.9030&   4.478&  0.9817&  1.0670&  1.2268&   4.041&  0.4179&  0.3912&  0.4062\\
 74&   4.403&  0.7909&  0.8506&  0.9169&   4.308&  0.9655&  1.0520&  1.2064&   3.896&  0.3071&  0.3921&  0.4071\\
 75&   4.242&  0.8025&  0.8604&  0.9272&   4.148&  0.9553&  1.0370&  1.1866&   3.758&  0.5088&  0.3929&  0.4079\\
 76&   4.090&  0.8079&  0.8696&  0.9359&   3.997&  0.9446&  1.0230&  1.1677&   3.628&  0.2183&  0.3936&  0.4086\\
 77&   3.946&  0.8183&  0.8782&  0.9429&   3.854&  0.9329&  1.0100&  1.1495&   3.504&  0.3295&  0.3943&  0.4092\\
 78&   3.809&  0.8271&  0.8862&  0.9478&   3.718&  0.9220&  0.9976&  1.1320&   3.386&  0.3707&  0.3949&  0.4097\\
 79&   3.680&  0.8349&  0.8937&  0.9501&   3.589&  0.9118&  0.9855&  1.1152&   3.274&  0.3778&  0.3954&  0.4101\\
 80&   3.557&  0.8422&  0.9005&  0.9485&   3.467&  0.9016&  0.9740&  1.0991&   3.167&  0.3725&  0.3959&  0.4105\\
 81&   3.440&  0.8490&  0.9070&  0.9400&   3.351&  0.8924&  0.9630&  1.0835&   3.065&  0.3820&  0.3962&  0.4108\\
 82&   3.329&  0.8556&  0.9129&  0.9165&   3.240&  0.8834&  0.9526&  1.0585&   2.969&  0.4128&  0.3965&  0.4110\\
 83&   3.223&  0.8613&  0.9185&  0.8435&   3.135&  0.8752&  0.9426&  1.0540&   2.876&  0.3864&  0.3968&  0.4111\\
 84&   3.122&  0.8667&  0.9235&  0.7220&   3.035&  0.8672&  0.9331&  1.0398&   2.788&  0.3880&  0.3969&  0.4112\\
 85&   3.026&  0.8721&  0.9283&  1.1741&   2.939&  0.8584&  0.9240&  1.0260&   2.704&  0.3893&  0.3971&  0.4112\\
 86&   2.934&  0.8767&  0.9326&  1.1235&   2.848&  0.8518&  0.9154&  0.0124&   2.623&  0.3903&  0.3971&  0.4112\\
 87&   2.846&  0.8811&  0.9366&  1.0954&   2.761&  0.8452&  0.9072&  0.9988&   2.546&  0.3910&  0.3972&  0.4111\\
 88&   2.762&  0.8853&  0.9402&  1.0806&   2.678&  0.8386&  0.8994&  0.9850&   2.472&  0.3917&  0.3973&  0.4110\\
 89&   2.682&  0.8897&  0.9435&  1.0704&   2.598&  0.8330&  0.8922&  0.9704&   2.401&  0.3939&  0.3971&  0.4108\\
 90&   2.606&  0.8924&  0.9465&  1.0606&   2.522&  0.8265&  0.8853&  0.9540&   2.333&  0.3918&  0.3970&  0.4106\\
 91&   2.532&  0.8953&  0.9491&  1.0456&   2.449&  0.8208&  0.8789&  0.9334&   2.268&  0.3885&  0.3969&  0.4103\\
 92&   2.462&  0.8985&  0.9516&  1.0018&   2.380&  0.8170&  0.8733&  0.9013&   2.206&  0.3981&  0.3968&  0.4099\\
\end{tabular}
\end{ruledtabular}
\label{tab-com3}
\end{table}

\begin{table}

\caption{ Wavelengths ($\lambda$ in $\AA$) and transition rates ($A$ in s$^{-1}$
  for odd-parity states with $J$=1 in Ni-like xenon, $Z$=54.
The RMBPT and RMBPT1 (first-order
approximation) results are compared with theoretical data
from HULLAC code by
Skobelev {\it et al.}
presented in Ref.~\protect\cite{skobelev}.}
\begin{ruledtabular}\begin{tabular}{llrrrlll}
\multicolumn{2}{c}{}&
\multicolumn{1}{c}{RMBPT}&
\multicolumn{1}{c}{RMBPT1}&
\multicolumn{1}{c}{HULLAC}&
\multicolumn{1}{c}{RMBPT}&
\multicolumn{1}{c}{RMBPT1}&
\multicolumn{1}{c}{HULLAC}\\
\multicolumn{1}{c}{$LS$-coupl.}&
\multicolumn{1}{c}{$jj$-coupl.}&
\multicolumn{1}{c}{$\lambda$}&
\multicolumn{1}{c}{$\lambda$}&
\multicolumn{1}{c}{$\lambda$}&
\multicolumn{1}{c}{$A$}&
\multicolumn{1}{c}{$A$}&
\multicolumn{1}{c}{$A$}\\
\hline
$3d5p\ ^3P_1$&$3d_{3/2}5p_{1/2}$&  12.418&    12.390& 12.417&   7.650E+11&  5.949E+11&  3.98E+11\\
$3d5p\ ^3D_1$&$3d_{5/2}5p_{3/2}$&  12.353&    12.320& 12.351&   7.365E+11&  5.875E+11&  4.32E+11\\
$3d5p\ ^1P_1$&$3d_{3/2}5p_{3/2}$&  12.254&    12.230& 12.243&   1.005E+11&  7.710E+10&  5.10E+10\\
$3d5f\ ^3P_1$&$3d_{5/2}5f_{5/2}$&  11.502&    11.480& 11.407&   6.184E+10&  6.748E+10&  7.00E+10\\
$3d5f\ ^3D_1$&$3d_{5/2}5f_{7/2}$&  11.444&    11.420& 11.445&   7.541E+12&  8.277E+12&  7.13E+12\\
$3d5f\ ^1P_1$&$3d_{3/2}5f_{5/2}$&  11.292&    11.270& 11.286&   2.319E+13&  2.585E+13&  2.77E+13\\
$3p5s\ ^1P_1$&$3p_{3/2}5s_{1/2}$&  10.160&    10.120& 10.136&   1.096E+12&  1.093E+12&  1.07E+12\\
$3p5s\ ^3P_1$&$3p_{1/2}5s_{1/2}$&   9.653&     9.614&  9.626&   6.262E+11&  7.007E+11&  4.74E+11\\
$3p5d\ ^3P_1$&$3p_{3/2}5d_{3/2}$&   9.591&     9.555&  9.568&   3.045E+10&  2.424E+10&  3.00E+09\\
$3p5d\ ^1P_1$&$3p_{3/2}5d_{5/2}$&   9.572&     9.537&  9.548&   7.979E+12&  7.590E+12&  8.19E+12\\
$3p5d\ ^3D_1$&$3p_{1/2}5d_{3/2}$&   9.129&     9.093&  9.104&   4.229E+12&  3.791E+12&  3.90E+12\\
$3s5p\ ^3P_1$&$3s_{1/2}5p_{1/2}$&   8.543&     8.498&  8.513&   1.039E+12&  1.039E+12&  1.24E+12\\
$3s5p\ ^1P_1$&$3s_{1/2}5p_{3/2}$&   8.489&     8.446&  8.459&   2.405E+12&  1.932E+12&  2.39E+12\\
\end{tabular}
\end{ruledtabular}
\label{tab-com4}
\end{table}

 \begin{table}
\caption{ The $3d_{5/2}\ -5f_{7/2}$ and $3d_{3/2}\ -5f_{5/2}$
transition energies ($E$ in eV) and transition rates ($A$ in
10$^{13}$~s$^{-1}$) for Ni=like ions with $Z$ = 56--92. Comparison
our RMBPT calculations with experimental measurements presented in
$a$--Ref.~\protect\cite{doron}, $b$--Ref.~\protect\cite{zigler},
$c$--Ref.~\protect\cite{doron-58},
$d$--Ref.~\protect\cite{elliot}, and
$f$--Ref.~\protect\cite{au-03}.}
\begin{ruledtabular}
\begin{tabular}{lllllll}
\multicolumn{1}{c}{}&
\multicolumn{3}{c}{$3d_{5/2}\ -5f_{7/2}$ }&
\multicolumn{3}{c}{$3d_{3/2}\ -5f_{5/2}$ }\\
\multicolumn{1}{c}{}&
\multicolumn{1}{c}{$E$, eV}&
\multicolumn{1}{c}{$E$, eV}&
\multicolumn{1}{c}{$A$, 10$^{13}$~s$^{-1}$}&
\multicolumn{1}{c}{$E$, eV}&
\multicolumn{1}{c}{$E$, eV}&
\multicolumn{1}{c}{$A$, 10$^{13}$~s$^{-1}$}\\
\multicolumn{1}{c}{$Z$}&
\multicolumn{1}{c}{RMBPT}&
\multicolumn{1}{c}{Expt.}&
\multicolumn{1}{c}{RMBPT}&
\multicolumn{1}{c}{RMBPT}&
\multicolumn{1}{c}{Expt.}&
\multicolumn{1}{c}{RMBPT}\\
\hline
 56&  1221.87&  1222.1$^a$        &  1.058& 1238.98&           1240.0$^a$&   2.864\\
 57&  1294.01&  1296.0$^b$        &  1.240& 1312.50&           1314.3$^b$&   3.159\\
 58&  1368.07&  1370.7$^c$        &  1.448& 1388.12&                     &   3.486\\
 59&  1444.07&  1440.2$^b$        &  1.676& 1465.72&           1465.5$^b$&   3.813\\
 64&  1853.04&1853.20$\pm$0.30$^d$&  3.250& 1884.43&   1885.1$\pm$0.3    &   5.791\\
 70&  2407.44&                    &  6.243& 2455.25& 2455.55$\pm$0.05$^d$&   8.930\\
 73&  2710.70&                    &  8.298& 2768.99& 2769.31$\pm$0.11$^d$&   10.89\\
 77&  3142.11& 3142.2$\pm$0.2$^d$ &  11.69& 3217.18&                     &   13.97\\
 79&  3369.43& 3370.6$\pm$0.5$^f$ &  13.71& 3454.22&   3458.3$\pm$0.5$^f$&   15.74\\
 90&  4758.25&4758.36$\pm$0.05$^d$&  29.22& 4915.86&                     &   28.89\\
 92&  5036.05&                    &  32.96& 5210.48& 5210.85$\pm$0.05$^d$&   32.08\\
\end{tabular}
\end{ruledtabular}
\label{tab-com5}
\end{table}


\end{document}